# Positron emission in elastic collisions of fully ionized high-Z heavy ions

Theo de Reus
*Member of Walter Greiner Gesellschaft, Frankfurt am Main*

Recent theoretical investigations of pair-production probabilities for given sets of nuclear trajectories predict the observability of spontaneous positron emission in elastic collisions of heavy ions [1, 2]. These calculations are extended to the investigation of ratios $P_y(\eta)/P_y(1)$ for partial probabilities of positron emission at small positron energies. The results suggest a possible observability of spontaneous positron emission in elastic collisions of bare heavy ions by analyzing the slope of $P_y(\eta)/P_y(1)$ for values of $R_{min} \lesssim 0.75\ R_{cr}$. In addition calculations of cross sections for positron emission in U+Cm and Cm+Cm collisions of fully ionized heavy ions are compared with former results with partly ionized projectiles and show an increase by factors of 57 to 72.

## 1 Introduction

### 1.1 Review

Within the current standard model of Quantum Electrodynamics (QED), a transition from the neutral vacuum into a charged vacuum is predicted in the presence of strong external electromagnetic fields [3–12]. Spontaneous $e^+e^-$ pair creation indicates this phase transition [13–15]. The required strong external electromagnetic fields - also called supercritical fields - can be generated in collisions of heavy ions with a combined nuclear charge $Z$ of projectile and target $Z_P + Z_T > 173$ for internuclear distances smaller than a critical value ($R < R_{cr}$) for the considered charge $Z$, when the binding energy of the $1s\sigma$- state exceeds twice the electron rest mass as in Fig. 1. An experimental evidence for the predicted emission of spontaneous positrons in heavy ion collisions would prove the fundamental theoretical framework [16, 17]. Structures were measured in positron spectra [18–29], but didn't show the theoretical predicted peak positions as function of $Z$ for the emission of spontaneous positrons. Since these structures also emerged in subcritical collisions, the spontaneous mechanism c) in Fig. 2 could not be the reason. Experiments conducted later also could not verify spontaneous positron emission, but hint at lines originating from internal conversion processes in the colliding nuclei [30–33].
For supercritical charges Z > 173 and beam energies in the vicinity of the Coulomb barrier, a vacant molecular $1s\sigma$- level can join the lower Dirac continuum at $-m_ec^2$ and stimulate spontaneous positron emission (Fig. 1). The theoretical description for QED in strong fields, the solution of the two centre Dirac equation in supercritical fields, the calculation of excitation amplitudes of inner shells and the decay of the neutral vacuum in supercritical collisions by spontaneous positron emission are described comprehensively in [36]. Spontaneous positron emission could be detected as a sharp line in the spectra of emitted positrons, if long living nuclear molecules existed [16, 37, 38].

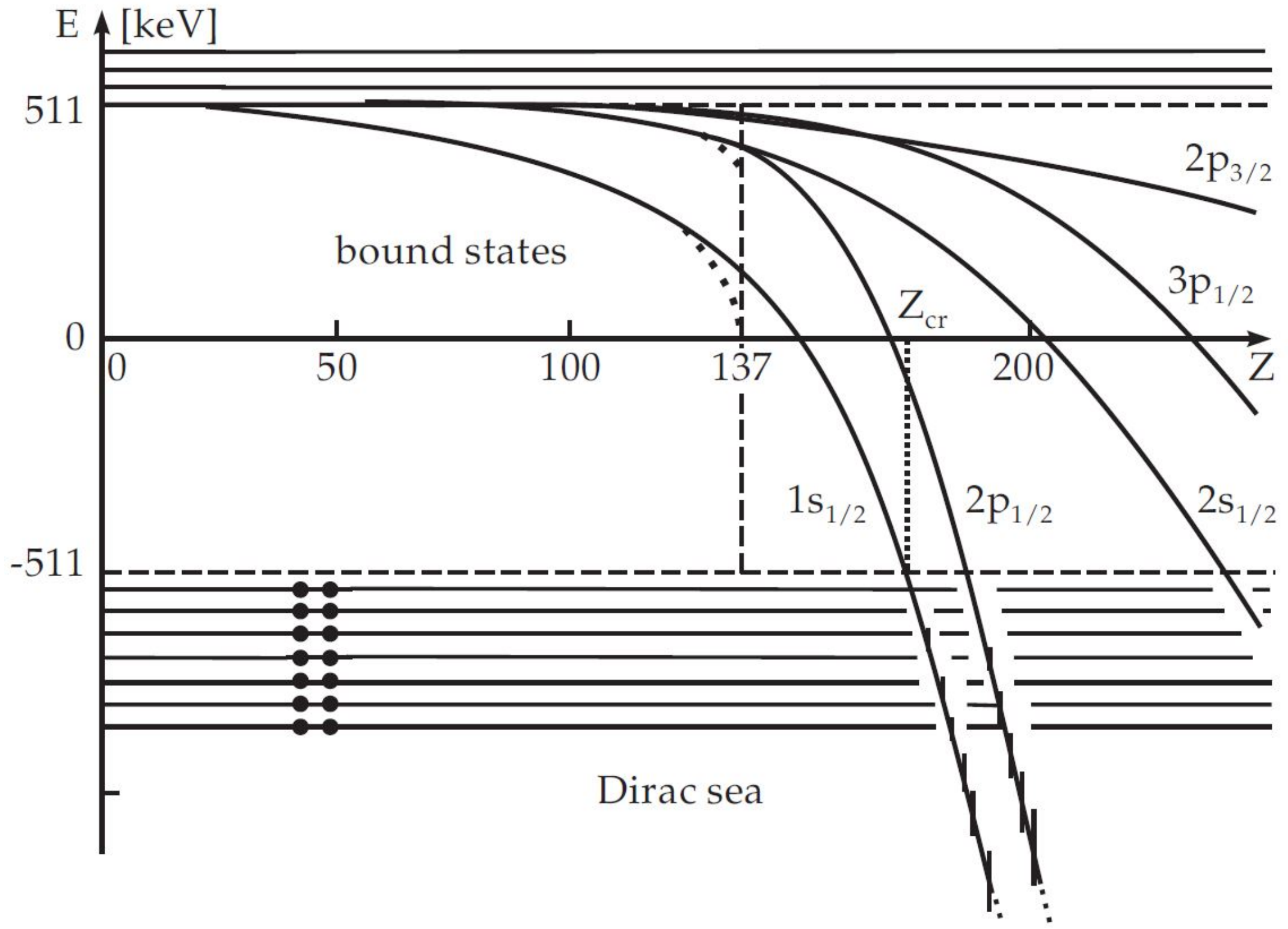

Figure 1: *Solutions of the Dirac equation as a function of the total charge Z [34, 35]. The bound states are located within the energy gap between* $-m_ec^2$ *and* $+m_ec^2$*, which separates the continuum solutions for negative and positive frequencies. Using the approximation of point like nuclei, the Sommerfeld fine structure formula for* $1s_{\frac{1}{2}}$*- und* $2p_{\frac{1}{2}}$*-binding energies is obtained, indicated by the dotted curves ending at* $Z = 137$*. For* $Z > 1/\alpha \simeq 137$ *the energy eigenvalues for the Dirac quantum numbers* $\kappa = \pm 1$*, however, become imaginary, so that right of the vertical dashed line at* $Z = 137$ *no bound states exist. Considering nuclei with finite radii, inner shell binding energies can be calculated continuously to the negative continuum at* $E = -m_ec^2$*. The* $1s_{\frac{1}{2}}$*- level dives into the Dirac see for charges above* $Z_{cr} = 173$ *where it resembles a resonance with finite decay width. The corresponding resonance widths - magnified by a factor 10 - are indicated by vertical bars.*

Another way to infer the emission of spontaneous positrons provide deep inelastic collisions, if production rates associated with an impact parameter range $b = 0$ - $4\ fm$ could be measured with sufficient statistics [39]. In both cases the time derivative of the internuclear distance $R$ (internuclear velocity $\dot{R}$) either needs to vanish at the distance of closest approach $R_{min}$ (in the case of long living nuclear molecules) or to be reduced significantly (in deep inelastic collisions) for a sufficient period of time to reduce the dynamically induced positron emission adequately to make spontaneous positron emission detectable (cf. process c) in Fig. 2).

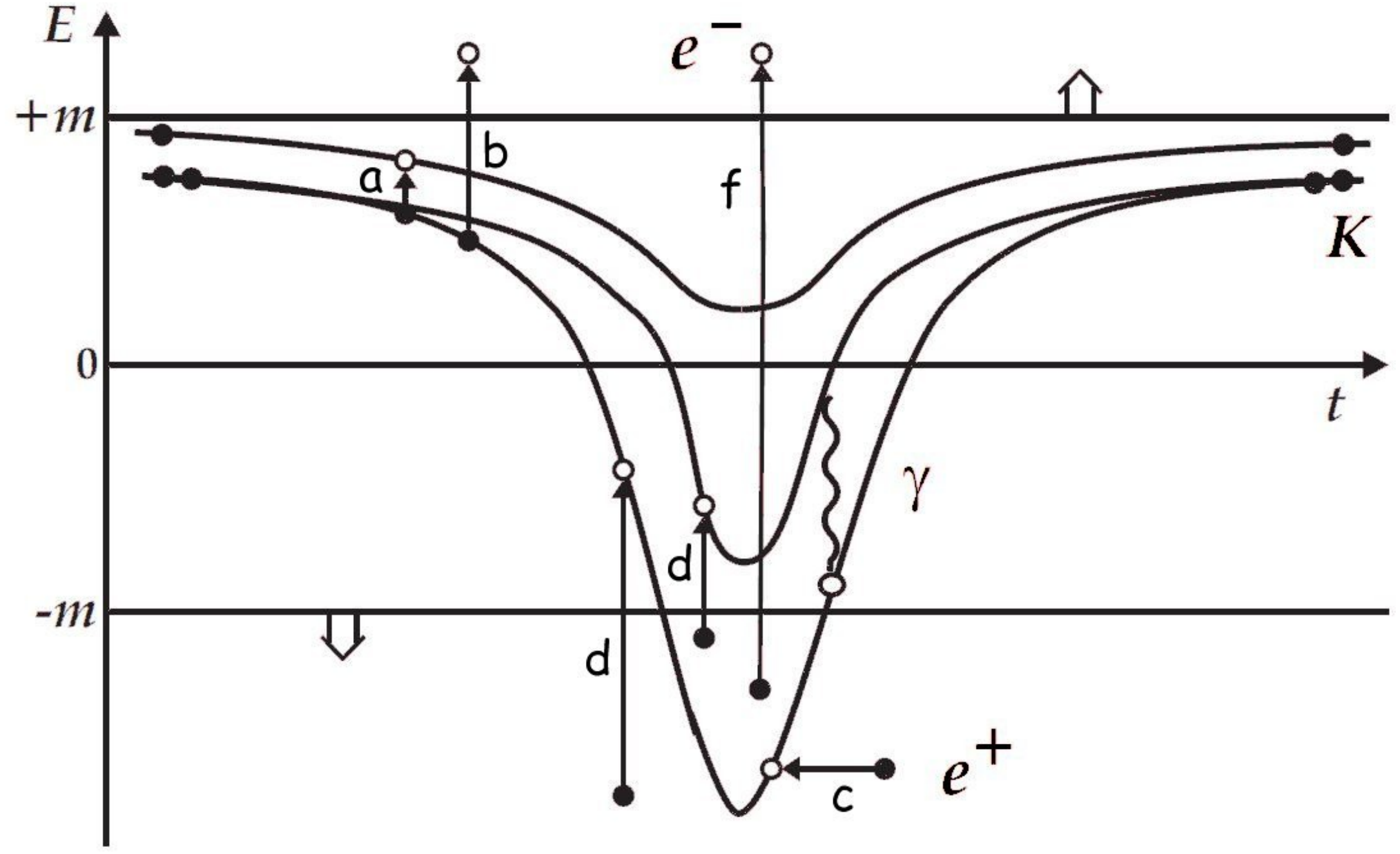

Figure 2: *Energy eigenvalues of the strongest bound $1s_{\frac{1}{2}}$, $2p_{\frac{1}{2}}$ und $2s_{\frac{1}{2}}$- states as a function of the collision time t in supercritical heavy ion collisions ($Z > Z_{cr}$) [40]. The $1s_{\frac{1}{2}}$-level dives into the negative energy continuum for a period of time $\Delta t$ and experiences the strongest binding energy at the distance of closest approach $R(t) = R_{min}$. Electronic transitions being induced by the collision dynamics are indicated by vertical arrows. An initially occupied $1s_{\frac{1}{2}}$- level can be partially emptied by transitions into higher lying vacant states (a) or by ionization into the upper continuum (b). These processes contribute to the formation of K- vacancies and δ- electron emission. The transition of continuum electrons with negative energy into vacant bound states (d) plus pair creation (f) correspond to induced positron creation by the radial velocity $\dot{R}$. The required energy for these transitions is transferred by the collision dynamics. Process (c) in contrast takes place spontaneously, i.e. independent of the collision dynamics of the two centre coordinate R, without any required transition energy. The occupation of an empty submerged $1s_{\frac{1}{2}}$-state causes spontaneous positron emission, which signals the change of the vacuum state in supercritical fields.*

Predicted oscillations in electron and positron spectra being caused by time delays of $3 \cdot 10^{-21}$ s to $10^{-20}$ s in inelastic heavy ion collisions with beam energies up to $E_{Lab}$ = 10 MeV/n could not be detected experimentally up to now.
Beyond this the slope of the high energetic part (up to $E_{e^\pm}$= 50 MeV) in the emission spectra for electrons and positrons in deep inelastic collisions at intermediate energies ($E_{Lab}$ = 20 - 100 MeV/n) can theoretically serve as 'atomic clock' to detect nuclear stopping or reaction times on time scales down to 2 fm/c, resp. $6.7 \cdot 10^{-24}$ s [41–46].

### 1.2 Scope

The calculations presented here, focus on the $(e^+e^-)$- pair production in elastic collisions of fully ionized, high-Z, heavy ions for sub- $(Z_1 + Z_2 < 174)$ and supercritical $(Z_1 + Z_2 > 173)$ systems. The coupled channel code developed and employed by the Atomic Physics Group of the Institute for Theoretical Physics at the J.W. von Goethe University Frankfurt, between end of the 1970s until end of the 90s, was used for the calculations (c.f. [36, 47]). First measurements of dynamical positron emission in heavy ion collisions were conducted at the UNILAC of GSI/ Darmstadt in the late 70s [48, 49]. Before discussing the results of the current calculations, we start with an excerpt of the theoretical framework in section 2, which is fully described in [36].
In section 3 the continuous transition from sub- to supercritical collisions is discussed in 3.1, followed by the presentation of positron spectra as a function of the kinetic parameter $\eta$, as proposed by [1, 2] in subsection 3.2.
Positron spectra and cross sections for the systems U+Cm and Cm+Cm in section 4 are more of theoretical interest and show the characteristic differences between emission rates of partly ionized projectiles onto neutral targets versus those from fully ionized heavy ion collisions. Since the colliding nuclei undergo Coulomb excitation up to high-lying nuclear levels, the nuclear internal pair creation also contributes to the positron spectra. Such contributions, however, will not be considered here, and should be addressed in a forthcoming paper.
The effects of spontaneous positron emission in elastic collisions in section 5 are investigated with respect to the contribution of positrons with small kinetic energies in 5.1, whereas subsection 5.2 deals with a possible visibility of spontaneous positron emission by analyzing the slope of $P_y(\eta)/P_y(1)$ at small values of $R_{min}$. Section 6 contains the conclusion.

## 2 Theoretical background

The dynamical behaviour of the electron- positron- field is described within a semi classical model, i.e. the internuclear trajectory is treated classically for velocities $< 0.2$c. At small internuclear distances $R$ the strongest bound electrons move in the combined field of both nuclei. In the adiabatic picture, where the electrons are in their ground states at each $R$, single particle wave functions and binding energies for the two centre Dirac equation were first calculated by B. Müller [50, 51]. The two centre Dirac equation reads:

$$\left[\vec{\alpha} \cdot \vec{p} + \beta m_e + V_{TC} - E_n\left(\vec{R}\,(t)\right)\right] \varphi_n\left(\vec{r}, \vec{R}\right) = 0\ . \tag{1}$$

The two centre potential $V_{TC}$ acting on the electrons is composed of the potential of two extended nuclei with charge densities $\rho_1(\vec{r}_1)$ and $\rho_2(\vec{r}_2)$:

$$V_{TC}(\vec{r}, \vec{R}) = -\int\limits_{V_1} \frac{\rho_1(\vec{r}_1^{\,\prime})}{|\vec{r} - \vec{r}_1^{\,\prime}|} d^3 r_1^{\,\prime} - \int\limits_{V_2} \frac{\rho_2(\vec{r}_2^{\,\prime})}{|\vec{r} - \vec{r}_2^{\,\prime}|} d^3 \vec{r}_2^{\,\prime} \,. \tag{2}$$

Instead of solving (1) with the exact two centre potential (2), a multipole expansion of the potential in equation (2) is used, according to $V_{TC}(\vec{r}, \vec{R}) = \Sigma_\ell V_\ell(r, R) P_\ell(cos\theta)$, where only the leading monopole term ($\ell = 0$) is used [52–54]. A similar ansatz was employed by [55]. The monopole approximation of two homogenously charged spheres with radii $a_i$ and charges $Z_i$ reads:

$$\begin{aligned}
\bar{V}_o(r, R) &= \sum_{i=1,2} -\frac{Z_i e}{S_i} && \text{for} \quad r \leq S_i - a_i, \\
\bar{V}_o(r, R) &= \sum_{i=1,2} -\frac{3}{2}\frac{Z_i e}{a_i^3}\left[\frac{a_i^3}{3}\left(\frac{1}{r} + \frac{1}{S_i}\right) + \frac{1}{24}\left(\frac{S_i^3}{r} + \frac{r^3}{S_i}\right) - \frac{a_i^2}{4}\left(\frac{S_i}{r} + \frac{r}{S_i}\right) - \right. \\
&\qquad \left. \frac{1}{6}\left(r^2 + S_i^2\right) + \frac{a_i^2}{2} + \frac{r S_i}{4} - \frac{a_i^4}{8 r S_i}\right] \\
& && \text{for} \quad S_i - a_i < r < S_i + a_i, \\
\bar{V}_o(r, R) &= \sum_{i=1,2} -\frac{Z_i e}{r} && \text{for} \quad r \geq S_i + a_i.
\end{aligned} \tag{3}$$

$S_i$ denotes the distance between the centre of charge and the centres of the nuclei $Z_i$, with $S_1 = \frac{Z_2 R}{Z_1 + Z_2}$ and $S_2 = \frac{Z_1 R}{Z_1 + Z_2}$. For internuclear distances $R < 500$ fm the monopole approximation describes binding energies and matrix elements within deviations of about 2% as demonstrated for the systems Pb + Pb and U + U by comparison with calculations using higher multipole terms of the two centre potential [56].
A time-dependent non-perturbative approach, including the full multipole expansion [57] is required to describe resulting K- and L-shell ionization probabilities at large distances $R$. By use of the monopole approximation the stationary two centre Dirac equation can be solved numerically for different two centre distances, provided the binding energy of the $1s\sigma$- state does not exceed the limit of twice the electron rest mass $2m_e c^2$. Equation (1) reduces to two coupled differential equations:

$$\begin{aligned}
\frac{d}{dr} u_1 &= -\frac{\kappa}{r} u_1 + (E + m - V_o)\, u_2 \\
\frac{d}{dr} u_2 &= -(E - m - V_o)\, u_1 + \frac{\kappa}{r} u_2
\end{aligned} \tag{4}$$

with $u_1(r) = r g(r)$ and $u_2(r) = r f(r)$, where $g(r)$ and $f(r)$ are the radial wave functions [56], $\kappa = \ell$ for $j = \ell - \frac{1}{2}$ resp. $\kappa = -\ell - 1$ for $j = \ell + \frac{1}{2}$. For collision systems with a total charge $Z > 173$ and sufficiently small two centre distances $R < R_{cr}$ the $1s\sigma$- binding energy exceeds the value of $2m_ec^2$, causing the $1s\sigma$- state to dive into the lower Dirac continuum $\varphi_{E_p}$ in form of a resonance $\Phi_R$. As described in [17] it is advantageous to construct modified continuum states $|\tilde{\varphi}_{E_p}>$, with the condition of being orthogonal to $|\Phi_R>$, the remaining bound states and all states in the upper Dirac continuum. This is achieved by employing the following projection operators:

$$\hat{Q} = \sum\!\!\!\!\!\!\int |\varphi_\alpha >< \varphi_\alpha| + |\Phi_R >< \Phi_R| \tag{5}$$

and

$$\hat{P} = 1 - \hat{Q} = \int dE_p |\tilde{\phi}_{E_p} >< \tilde{\phi}_{E_p}|, \tag{6}$$

where the index $\alpha$ spans bound states plus the upper Dirac continuum. The modified continuum states are no longer eigenstates of the Hamilton- operator in equation (1) but eigenstates of the projected Hamilton operator $\hat{P}\hat{H}_{TCD}\hat{P}$.

$$\hat{P}\hat{H}_{TCD}\ \hat{P}\ |\tilde{\varphi}_{E_p}> = E_p\ |\tilde{\varphi}_{E_p}> \tag{7}$$

or

$$(E_p - \hat{H}_{TCD})|\tilde{\varphi}_{E_p}> = - < \Phi_R|\hat{H}_{TCD}|\tilde{\varphi}_{E_p}> |\tilde{\varphi}_{E_p}> \ . \tag{8}$$

Thus $|\tilde{\varphi}_{E_p}>$ are solutions of equation (8), which differs from the original Dirac equation for $|\varphi_{E_p}>$ by an additional $|\Phi_R>$ dependent potential coupling. The square of the absolute value of this matrix element corresponds to the decay width of the resonance:

$$\Gamma(E_p) = 2\pi| < \tilde{\varphi}_{E_p}|\hat{H}_{TCD}\ |\Phi_R> |^2. \tag{9}$$

The modified continuum states $|\tilde{\varphi}_{E_p}>$ are calculated by numerical solution of equation (8). The wave function $|\Phi_R>$ is similar to the one of the $1s\sigma$- state, except for an oscillating tail caused by the tunneling process through the particle- antiparticle gap. The construction principle of Wang and Shakin [58] is employed for $|\Phi_R>$ by cutting off its oscillating tail. This is achieved by keeping the potential constant for $r > r_c$:

$$\tilde{V}(r) = \Theta(r_c - r)\ \ V(r) + \Theta(r - r_c)\ \ V(r_c). \tag{10}$$

The value of the cut- off- radius $r_c$ is located within the particle- antiparticle- gap and the resonance wave function is constructed by starting with a continuum wave function in the lower Dirac continuum for a continuum state $E_p = E_{Res}$ according to

$$\Phi_R(r) = C\ \Phi_{E_{Res}}(r)\ \Theta(r_c - r), \tag{11}$$

where C is a normalization constant [17] and $r_c$ is defined by $E_R - V(r_c) = -\gamma\ m_e c^2$ with a value of $\gamma = 0.9$ chosen in the numerical calculations to determine $|\Phi_R>$ and the resonance energy $E_R$.
A different approach to circumvent a computational basis with limited continuum energies using the construction principle described above, is published in [59], where a mapped Fourier grid matrix representation was employed, also using the monopole approximation. These calculations yield smaller values for correlated spectra in the (2s$e^-$, free $e^+$) channel.

In a next step the time dependent wave function for an electron with index $j$ is developed in a basis of Born- Oppenheimer states $\varphi_k$, given by the stationary molecular eigenstates of the Hamiltonian in equation (1), resp. (8).

$$\Psi_j(t) = \sum\!\!\!\!\!\!\int_k a_{jk}(t)\varphi_k(R(t))e^{-i\chi_k(t)}, \tag{12}$$

with $\chi_k(t) = \int^t <\varphi_k|\hat{H}_{TCD}|\varphi_k> dt' = \int^t E_k(t')dt'$, where the index $k$ spans subcritical states ($Z < 174$) fulfilling equation (1). Inserting equation (12) into the time dependent two centre Dirac equation

$$\hat{H}_{TCD}\Psi_j(t) = [\vec{\alpha}\cdot\vec{p} + \beta m_e + V_o\,(r, R\,(t))]\,\Psi_j(t) = i\frac{\partial}{\partial t}\Psi_j(t) \tag{13}$$

and projecting with stationary eigenstates of equ. (1) delivers a set of first order coupled differential equations for the occupation amplitudes $a_{ij}(t)$, also denoted as 'coupled channel equations':

$$\dot{a}_{ij}(t) = -\sum\!\!\!\!\!\!\int_{k\neq j} a_{ik} <\varphi_j|\frac{\partial}{\partial t} + i\hat{H}_{TCD}|\varphi_k> \; e^{i(\chi_j - \chi_k)}. \tag{14}$$

If all $\varphi_i$ in (14) are eigenstates of the Hamiltonian $\hat{H}_{TCD}$, the potential couplings caused by $i\hat{H}_{TCD}$ vanish and only the induced couplings originating from the $\partial/\partial t$- operator remain. For supercritical collision systems $Z > Z_{cr}$ and sufficiently small two centre distances $R < R_{cr}$, however, the modified continuum states $|\tilde{\varphi}_{E_p}>$ according to (8) are no eigenstates of $\hat{H}_{TCD}$, but eigenstates of the modified Hamilton operator $\hat{P}\hat{H}_{TCD}\hat{P}$, so that additional potential couplings occur in equation (14), which have to be added coherently to the time dependent couplings.
The $\partial/\partial t$- operator can be split into a radial and angular velocity dependent term according to:

$$\frac{\partial}{\partial t} = \dot{R}\frac{\partial}{\partial R} - i\vec{\omega}\cdot\vec{j}. \tag{15}$$

The second term, describing the rotational or Coriolis coupling, is neglected for the considered collision systems $Z > 1/\alpha$ where the radial coupling and the behaviour of the wave functions at internuclear distances $R < 500$ fm play the dominant role [52, 53, 60]. Equation (14) is fully equivalent to the time dependent Dirac equation (13) except for the following approximations:

- use of the monopole approximation
- neglection of Coriolis coupling
- considering a finite number of basis wave functions

Having solved the coupled channel equations (14), the number of particles in a state $p$ above the Fermi level can be calculated by [61, 62]:

$$N_p = \sum\!\!\!\!\!\!\int_{r<F} |a_{rp}(t \to +\infty)|^2 \qquad (p > F), \tag{16}$$

and the number of holes in level $q < F$:

$$\bar{N}_q = \sum\!\!\!\!\!\!\int_{r>F} |a_{rq}(t \to +\infty)|^2 \qquad (q < F). \tag{17}$$

F denotes the Fermi- level, i.e. F = 3 indicates that the first 3 shells of bound states are occupied. The sum in equations (16 - 17) applies for bound states. If one of the indices runs over continuum states, the integration symbol applies. Equation (17) describes both: holes, or inner shell vacancies plus positrons, which are represented by holes in the negative energy continuum.

The number of correlated particle- hole- pairs $N_{pq}$ is obtained by the following expression [60, 63]:

$$N_{pq} = N_p \bar{N}_q + |\sum_{r<F} a^*_{rp}(t \to +\infty) a_{rq}(t \to +\infty)|^2. \tag{18}$$

This expression is applied in the analysis of experiments, in which coincidences between electrons and 1s- vacancies are measured. Coincidence measurements reveal e.g., which parts of the detected $\delta$- electrons are emitted from the molecular $1s\sigma$- level. For the calculation of particle- particle or hole- hole- correlations, the sign of the second term in equation (18) has to be inverted.

In collisions of bare nuclei, where $F = 0$, electron capture into bound states is the dominant channel ($> 98\%$) for pair creation. Therefore the bound-free pair production can be calculated in good approximation by the total positron emission rate, which is given by:

$$P_{e^+} = \int \bar{N}_q(F = 0) dE. \tag{19}$$

## 3 Transition from sub- to supercritical collisions

### 3.1 Continuous transition of matrix elements and amplitudes

Equation 14 describes the dynamics of inner-shell excitation processes in sub- and supercritical ($Z > Z_{cr}, R < R_{cr}$) collisions of heavy ions. It contains two separate coupling operators in case the binding energy of the $1s\sigma$- or $2p_{\frac{1}{2}}\sigma$- state exceeds $2m_ec^2$:

$$< \tilde{\varphi}_{E_p}|\frac{\partial}{\partial t}|\Phi_R > +i < \tilde{\varphi}_{E_p}|\hat{H}_{TCD}|\Phi_R > \quad . \tag{20}$$

The first term (called induced, dynamical or radial coupling) is caused by the collision dynamics, resp. the dynamics of the two centre distance R as a function of time. This coupling causes the transitions indicated by the vertical arrows a, b, d and f in Fig. 2. Induced couplings are present in any sub- or supercritical heavy ion collision and cause also 'dynamical' positron emission in systems with a combined charge $< 174$.

The second term in equation (20) being denoted as potential or spontaneous coupling, however, is nonzero only in case of supercritical ($Z > Z_{cr}$, $R < R_{cr}$) collisions, where the binding energy of the $1s\sigma$- state (or for $Z > 188$ in addition the $2p_{\frac{1}{2}}\sigma$- state) exceeds $2m_ec^2$ and enters the lower Dirac continuum, being described as a resonance state $|\Phi_R >$. The spontaneous coupling causes the transition indicated by arrow c in Fig. 2. It is most important to note, that the magnitude of spontaneous coupling depends on the two centre distance $R$, but is independent of $\dot{R}$. In Fig. 3 the dependence of the spontaneous coupling between the resonance state $|\Phi_R>$ and the continuum state $|\tilde{\varphi}_{E_{e^+}}> = -2.1 m_e c^2$ is shown for U+Cm at different values of $R_{min}$.

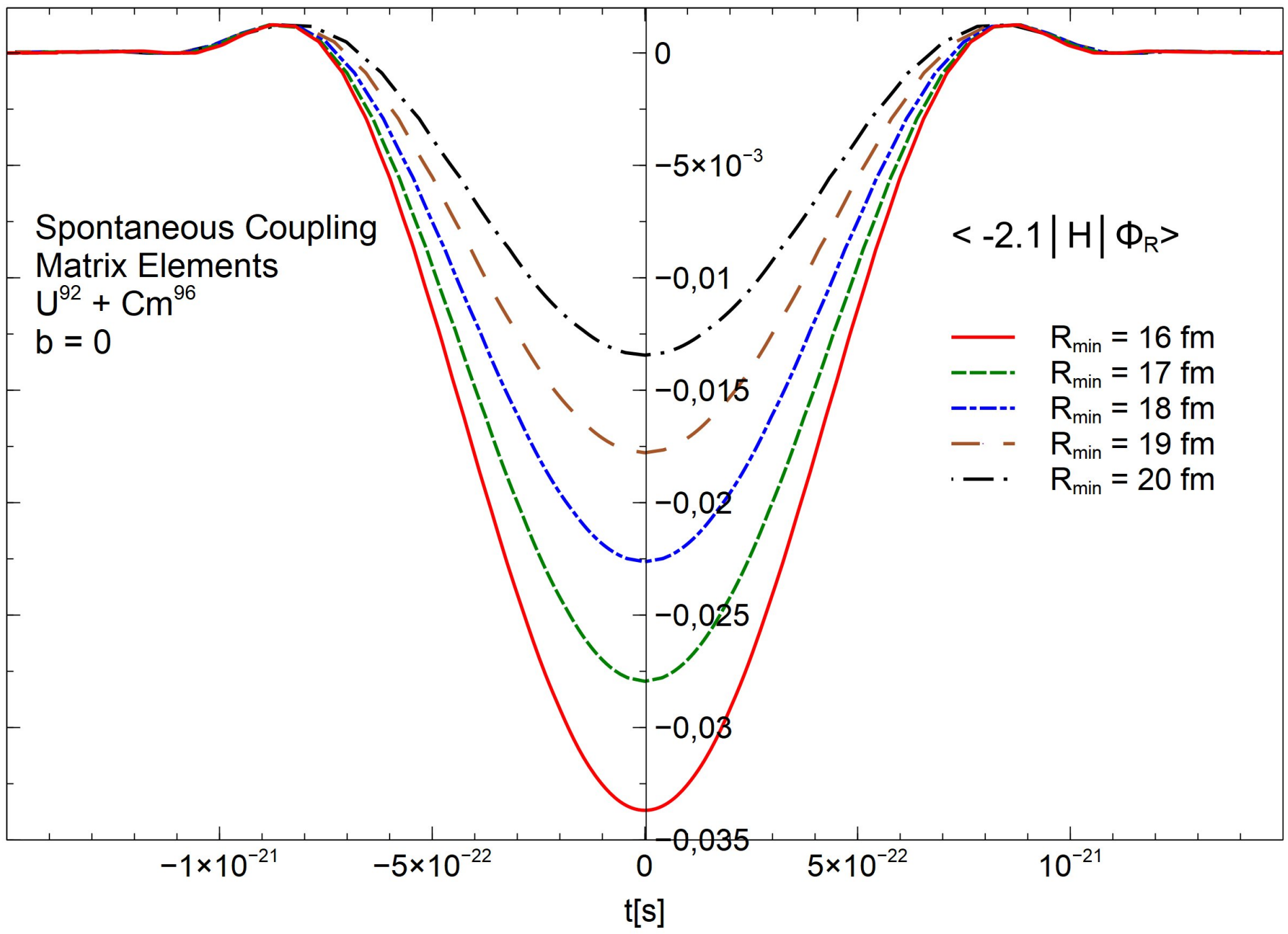

Figure 3: *Spontaneous coupling matrix element* $< \tilde{\varphi}_{E_p}|\hat{H}_{TCD}|\Phi_R >$ *for* $E_p$ = *-2.1*$m_ec^2$ *in central* $U^{92} + Cm^{96}$ *collisions for different values of* $R_{min}$ = *2a (b = 0).*

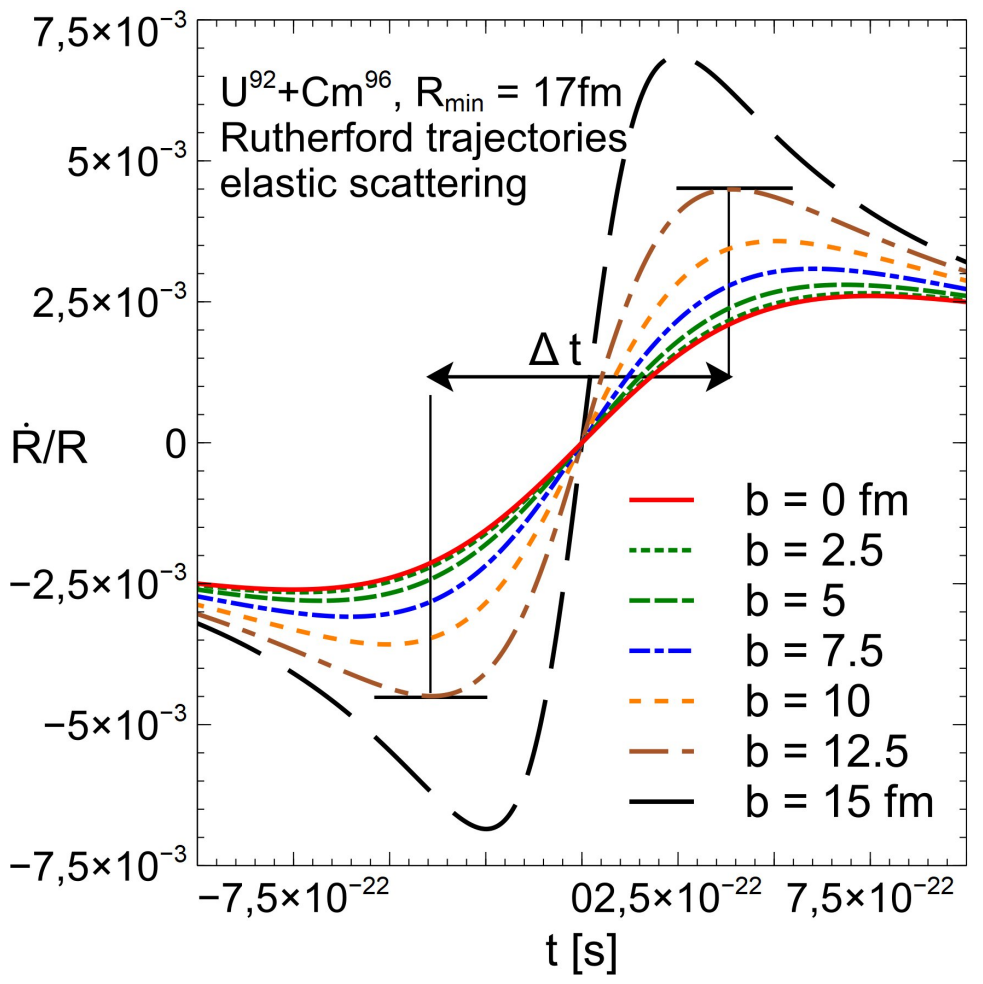

Figure 4: Definition of the collision time in a $U^{92}+Cm^{96}$ collision along classical Rutherford trajectories at a bombarding energy defined by 2a = 17fm ($E_{Lab} \simeq$ 6.16 MeV/n) for different impact parameters b.

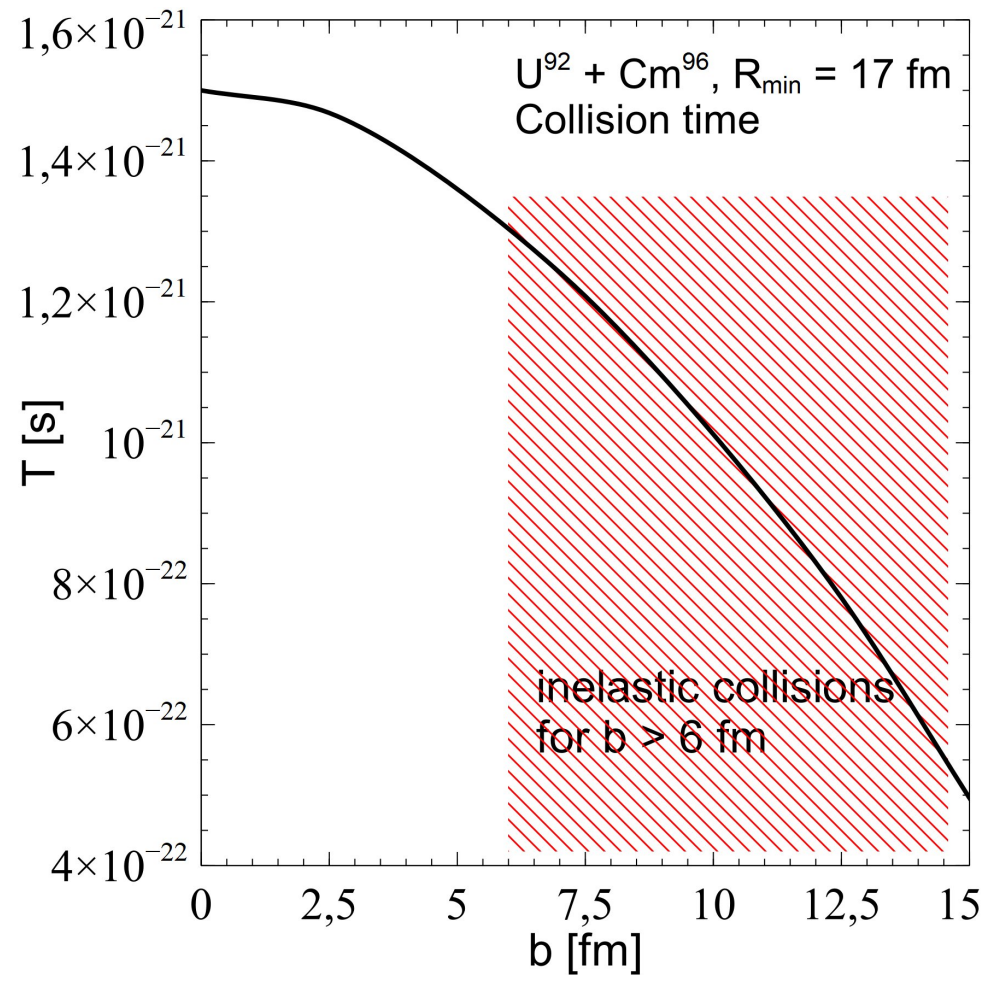

Figure 5: Collision time as a function of the impact parameter b for the same system and bombarding energy as in Fig. 4. The shaded area indicates the range, where inelastic collisions occur.

In the theoretical case, where we could freeze the two centre distance in a U+Cm collision at a fixed value of R = 15 fm for an arbitrary long period of time, the spontaneous coupling would cause the appearance of a sharp line in the emitted positron spectrum at the position $T_{e^+}$ = 476 keV (corresponding to the resonance energy of $|\Phi_R>$) with a width of $\Gamma$ = 6.6 keV, according to equation (9).

As described in [17] there is no threshold effect in the transition regime between sub- and supercritical collisions due to the collision dynamics. This can be seen by considering the collision time $\tau$ characterizing the essential interaction time.

In Fig. 4 the collision time is defined as the time interval $\tau = \Delta t$ between the minimum and maximum of $\dot{R}/R$ (c.f. [64] for a similar definition). In the regime of elastic collisions the value of $\tau$ e.g. varies between 1.5 and $1.3 \cdot 10^{-21}$s as shown in Fig. 5 for U+Cm collisions with a distance of closest approach at $R_{min}$ = 17 fm. According to the uncertainty principle the shortness of $\tau$ prevents the emergence of a sharp structure caused by spontaneous positron emission, which is broadened according to $\Gamma_{dyn} \simeq \hbar/\tau$. For the collision times in our example this yields widths between 440 and 510 keV.

Another explanation for the continuous transition of positron emission rates without threshold effect when crossing the border to the supercritical regime can be obtained by looking at the coupling matrix elements as a function of time or $\dot{R}$. The matrix elements $< \tilde{\varphi}_{E_p}|\dot{R}\frac{\partial}{\partial R}|\Phi_R >$ and $< \tilde{\varphi}_{E_p}|\hat{H}_{TCD}|\Phi_R >$ for $E_p$ = -2.1$m_e c^2$ are shown in Fig. 6 for a central U+Cm collision at $R_m$ = 17 fm. When compared with the induced matrix

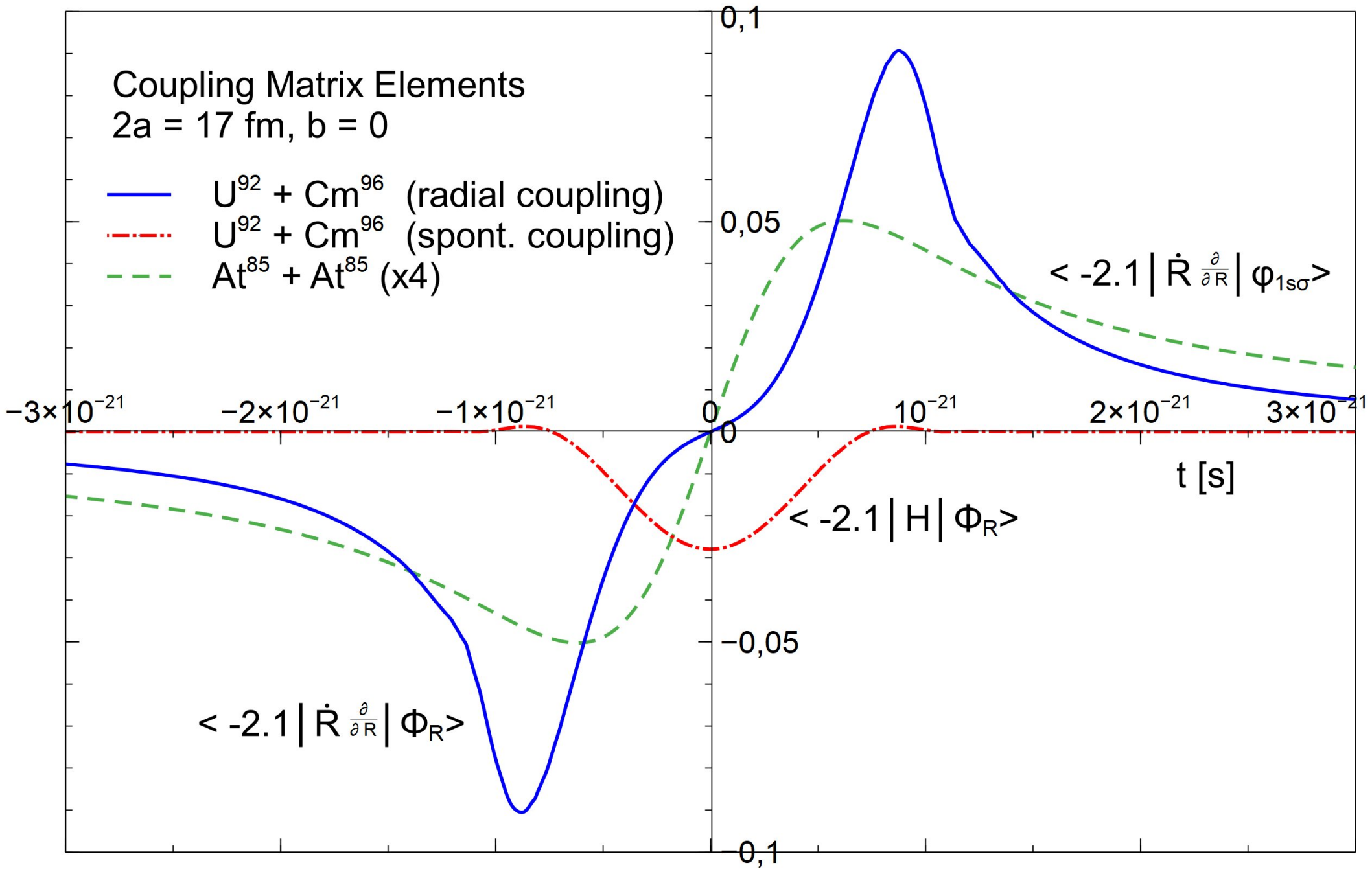

Figure 6: *Matrix elements of induced and spontaneous coupling as a function of the collision time in central collisions for the system $U^{92} + Cm^{96}$ at a bombarding energy $E_{Lab} \simeq$ 6.16 MeV/n (2a = 17 fm). The dashed curve shows the induced coupling for the subcritical system $At^{85} + At^{85}$ (multiplied by 4) for comparison with the same kinematics (2a = 17 fm).*

element of the subcritical System At+At, the radial matrix element of U+Cm shows a different shape, slope and magnitude around t=0, where the $1s\sigma$- state dives into the negative continuum and causes spontaneous positron emission i.a. described by the matrix element $< \tilde{\varphi}_{E_p}|\hat{H}_{TCD}|\Phi_R >$. The superposition of both couplings leads to a continuous progression of the transition amplitudes in equation (14) by coherent addition of both terms and prevents threshold effects when crossing the border from the sub- to the supercritical regime.

### 3.2 Positron spectra as a function of the kinematic parameter $\eta$

A very useful parameter $\eta$ was introduced by [1] to select elastic collisions with different dwell times for the submerged $1s\sigma$- state in supercritical quasi-molecules. This idea and the question, how current results compare with those obtained from the coupled channel code used in earlier calculations, triggered this paper. Furthermore the kinematic param-eter $\eta$ helps to visualize the transition from the sub- to the supercritical regime in elastic collisions of heavy ions [2]. It is defined by the ratio of the bombarding energies leading to predefined $R_{min}$- values for two centre distances $R$, which are located in the range $R_{cont} < R_{min} < R_{cr}$, where $R_{cont}$ denotes the contact point of the two colliding nuclei and $R_{cr}$ the two centre distance, where the level of the $1s\sigma$- state reaches the lower Dirac continuum for combined nuclear charges $Z > 173$. The parameter $\eta$ is defined as $\eta = E_0/E$, with $E_0$ leading to $R_{min}$ in central collisions, i.e. impact parameter b = 0 and bombarding energies $E > E_0$ with b > 0 leading to the same value of $R_{min}$. For the system U+Cm classical Rutherford trajectories are show in Fig. 7 for different values of $\eta$ leading to the same two centre distance $R_{min} = 18$ fm. The horizontal line at $R \simeq 40$ fm denotes $R_{cr}$ for the fully ionized system $U^{92+} + Cm^{96+}$ on the left side of the figure. Similar to Fig. 4 the dwell time of the resonance state $|\Phi_R>$ is maximal in central collisions (b = 0). The significant difference in Fig. 7 is the constant distance of closest approach $R_{min}$ for the considered combinations of E and b. Therefore the parameter $\eta$ is inversely proportional to the dwell time of the resonance state in the supercritical regime. It has to be kept in mind, that for increasing values of $\eta$ also inelastic collisions occur. For the system U+Cm with parameters considered above, this occurs for $\eta > 1.2$, when assuming spherical nuclei with radii $r_i = 1.2fm\ A_i^{\frac{1}{3}}$. In reality the Uranium nucleus shows a hexadecapole defor-mation, i.e. the internuclear distance for contact between the nuclei will be larger in the longitudinal U- orientation and shorter at contact along its short axis. I.e. we expect a distribution around an average value of $\bar{r_i}$. For more precise values of nuclear radii, resp. charge and mass radii c.f. [65–71].

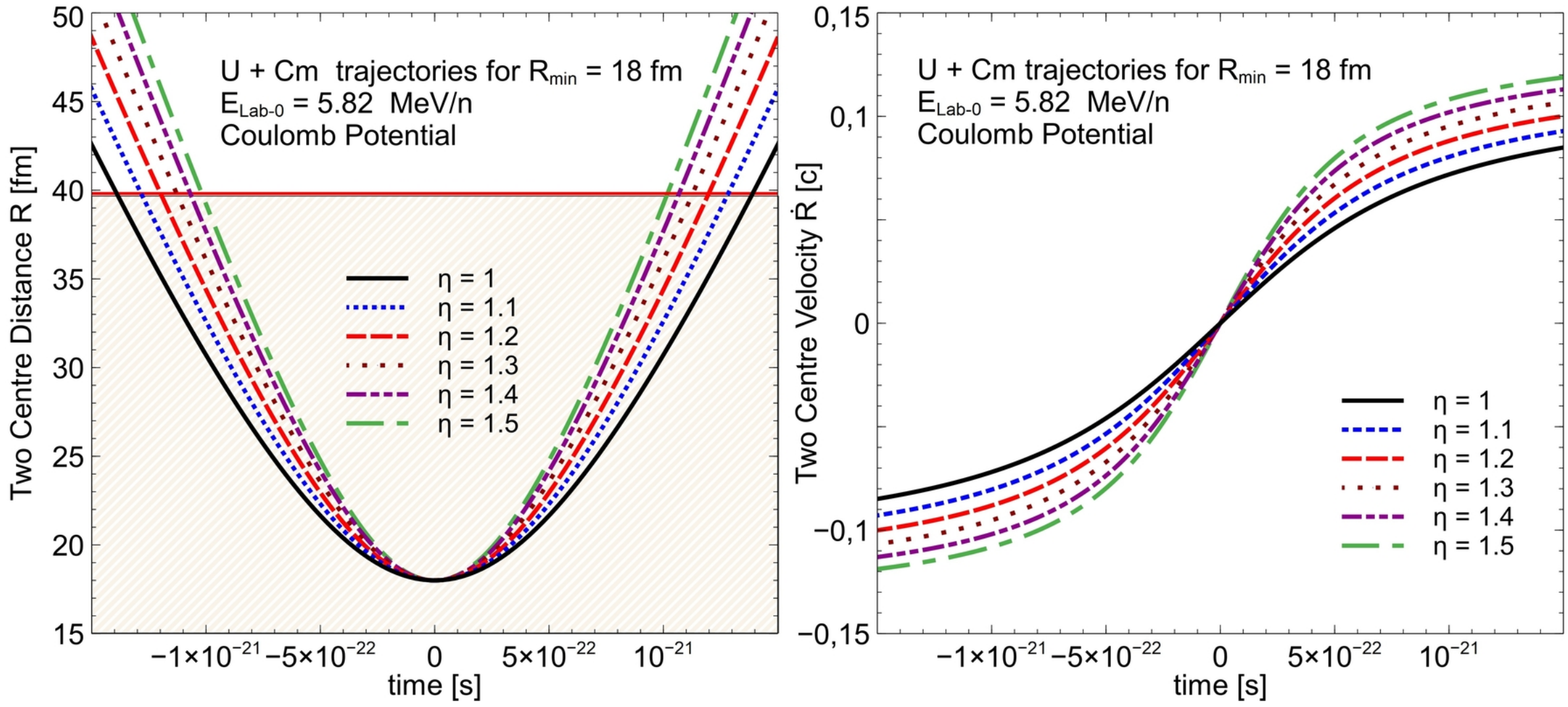

Figure 7: *a) Trajectories for elastic scattering in $U^{92+}$ + $Cm^{96+}$- collisions leading to a fixed distance of closest approach of $R_{min} = 18$ fm (Coulomb- potential for two fully ionized atoms). The trajectories for $\eta = 1 \ldots 1.5$ belong to the following combinations of bombarding energies $E_{Lab}$ [MeV/n] and impact parameters b [fm]: (5.82, 0), (6.4, 5.43), (6.98, 7.35), (7.56, 8.65), 8.15, 9.62) and (8.73, 10.39). At the critical distance $R_{cr} = 40$ fm (horizontal line) the binding energy of the $1s\sigma$- level exceeds the value of $2m_ec^2$, enabling spontaneous positron emission (shaded area). b) Associated values of the internuclear relative velocity of the collision partners $\dot{R}$ versus time $t$ ($R(t = 0) = R_{min}$).*

At the right side of Fig. 7 the two centre velocity $\dot{R}$ is plotted versus time. The velocity increases with increasing bombarding energies and so do the radial coupling matrix elements in equation (6) near the distance of closest approach, which can be transformed by the Hellmann Feynman theorem [72, 73] according to:

$$< \tilde{\varphi}_{E_p} | \frac{\partial}{\partial t} | \Phi_R >= \frac{1}{E_j - E_i} < \tilde{\varphi}_{E_p} | \dot{R} \, \partial V_o / \partial R | \Phi_R > \quad . \tag{21}$$

By calculating positron spectra as a function of $\eta$, the transition from sub- to the supercritical collisions can be visualized quite clearly, as Figs. 8 and 9 demonstrate.

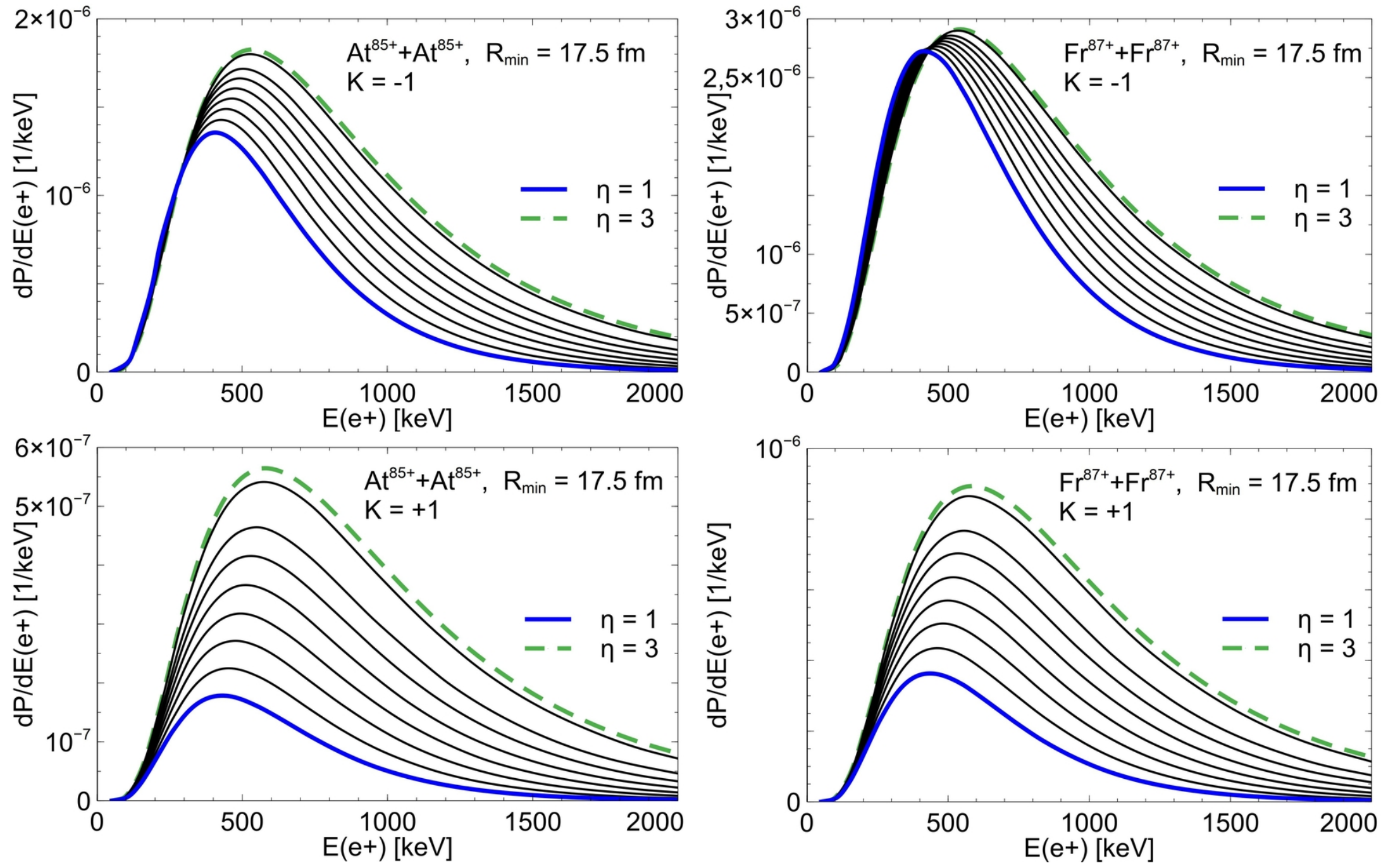

Figure 8: *Differential emission probability for positrons $dP_{e^+}/dE_{e^+}$ as a function of their energy $E_{e^+}$ for the subcritical systems $At^{85+} + At^{85+}$ and $Fr^{87+} + Fr^{87+}$. $E_0$ is chosen to fulfill 2a = 17.5 fm. The following parameters $\eta = E/E_0$ were chosen: $\eta$ = 1, 1.15, 1,32, 1,51, 1,73, 1,99, 2,28, 2,82, 3. Top: $s\sigma$- channel, bottom: $p_{\frac{1}{2}}\sigma$- channel.*

In collisions of partly ionized heavy ion projectiles on neutral targets (conventional collisions), the inner shells ($1s\sigma$ and $2p_{\frac{1}{2}}\sigma$) of the quasi-molecules are occupied. Due to this Pauli blocking, the probability for inner shell vacancies at the distance of closest approach amounts only a few percent [74] and reduces the transitions of electrons from the lower Dirac continuum into vacant $1s\sigma$- and $2p_{\frac{1}{2}}\sigma$- states significantly.

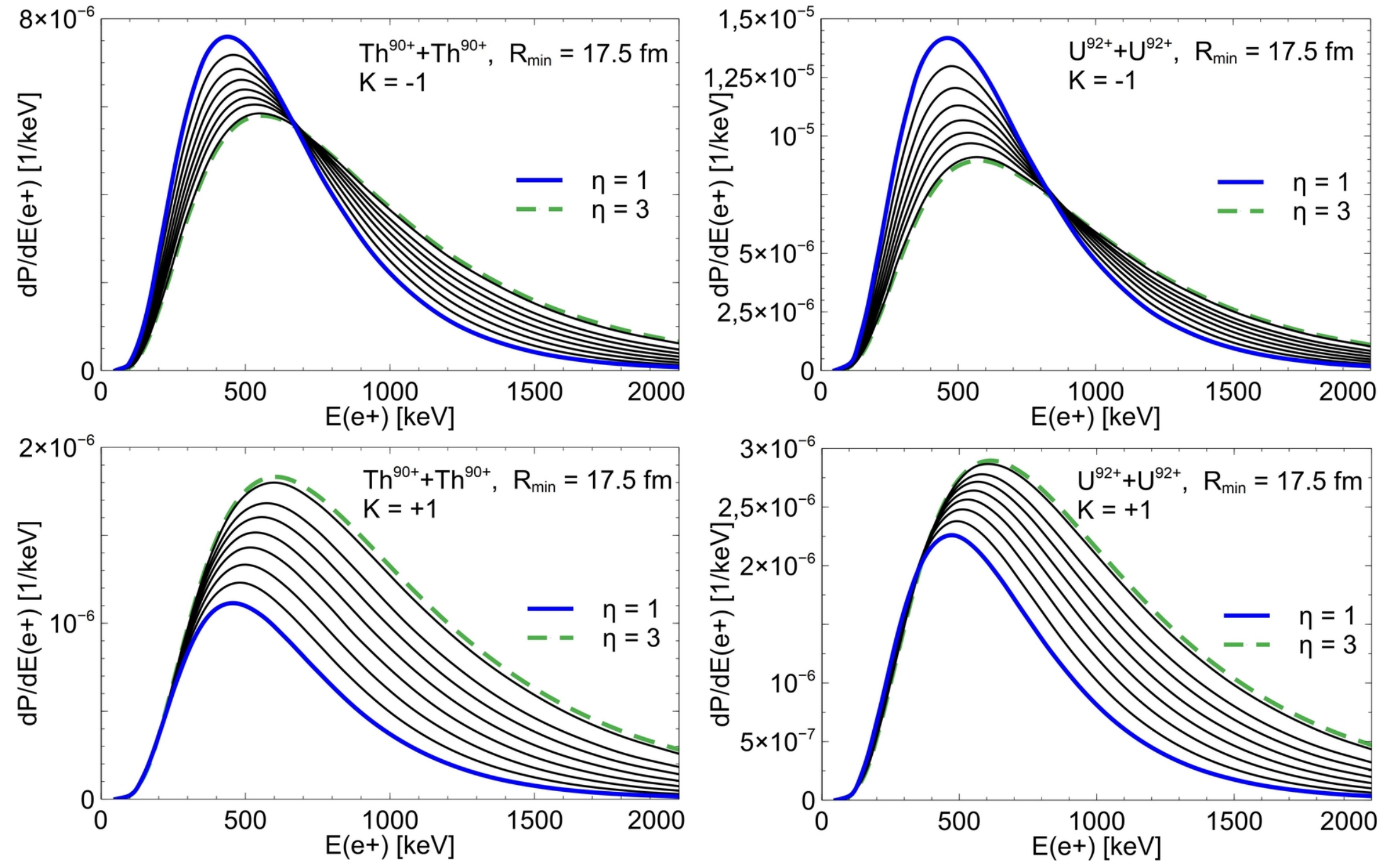

Figure 9: *Same as in Fig. 8 for the supercritical systems* $Th^{90+} + Th^{90+}$ *and* $U^{92+} + U^{92+}$. $E_0$ *again is chosen to fulfill 2a = 17.5 fm and the same parameters for* $\eta = E/E_0$ *were used as in Fig. 8. Top: results for* $\kappa = -1$*, bottom:* $\kappa = +1$.

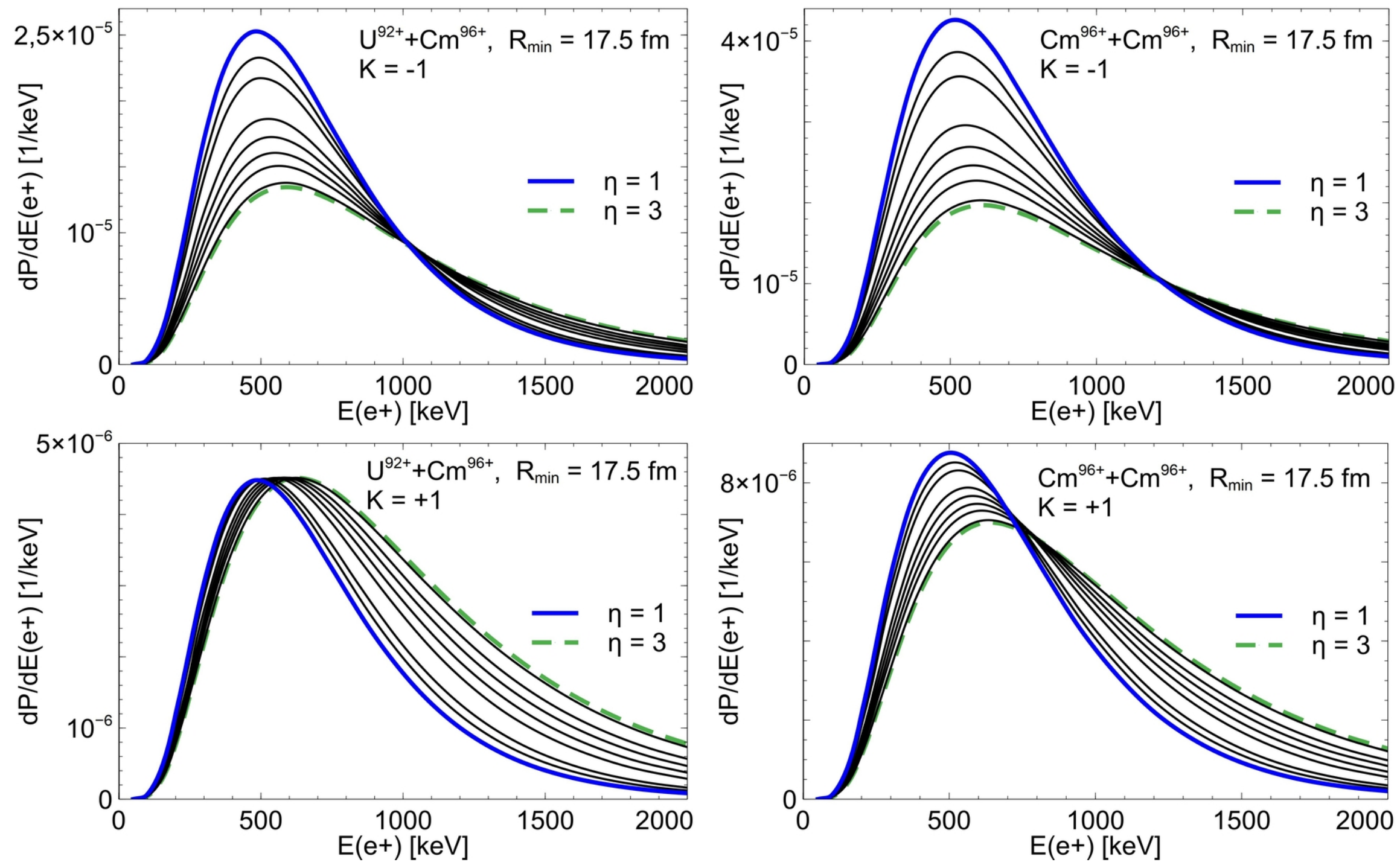

Figure 10: *For the heaviest supercritical systems available,* $U^{92+} + Cm^{96+}$ *and* $Cm^{96+} + Cm^{96+}$*, the* $\kappa = -1$*- spectra increase significantly for* $\eta = 1$ *compared with higher values for* $\eta$*. The kinematical parameters are the same as in Figs. 8 and 9 (2a = 17.5 fm), the values for* $\eta$ *are:* $\eta =$ *1, 1.1, 1,2, 1,51, 1,73, 1,99, 2,28, 2,82, 3.*

Experiments with fully ionized heavy ions would circumvent this problem and offer a unique chance to increase our knowledge of the relation between induced and spontaneous positron emission by significantly increased emission rates when compared with conventional collisions [38, 74, 75]. It is emphasized again, that the following calculations apply for fully ionized projectile- target combinations unless indicated otherwise.
Fig. 8 shows the differential emission probability for positrons versus their energy for two subcritical systems At+At and Fr+Fr, where only induced positron emission occurs. For increasing values of $\eta$ resp. bombarding energies $E$, the induced matrix elements increase, which is also reflected in the spectra.
The situation changes, when considering spectra for combined nuclear charges of the collision partners $Z > 173$, where the $1s\sigma$- state dives into the negative Dirac continuum for two centre distances $R < R_{cr}$ as evident in figures 9 and 10 for the systems Th+Th, U+U, U+Cm and Cm+Cm. The radial or induced matrix elements cause the same effects on the spectra as described before, but exhibit an earlier and stronger decrease around $R_{min}$ (similar as in Fig. 6). In addition the second term in equation (21) describing the spontaneous transitions contributes to the spectra, resp. total emission rates $P_{e^+}$. For Th+Th, U+U and U+Cm the $1s\sigma$- level dives into the lower Dirac continuum. The longer the dwell time in the supercritical region $R < R_{cr}$, being indicated by small values of $\eta$, the bigger the contribution of the second term in equation 21. A reversal of the $\eta$-dependency is observed, i.e. the spectra for the $s\sigma$- states have their maxima at $\eta = 1$, whereas the spectra for the $p_{\frac{1}{2}}\sigma$- states ($\kappa = +1$) keep their $\eta$- dependence as before. For the system Cm+Cm also the $2p_{\frac{1}{2}}\sigma$ level reaches the lower Dirac continuum for two centre distances $R < R_{cr} \simeq 19.5$ fm. Therefore the spectra for $\kappa = \pm\, 1$ show a similar dependence on $\eta$.

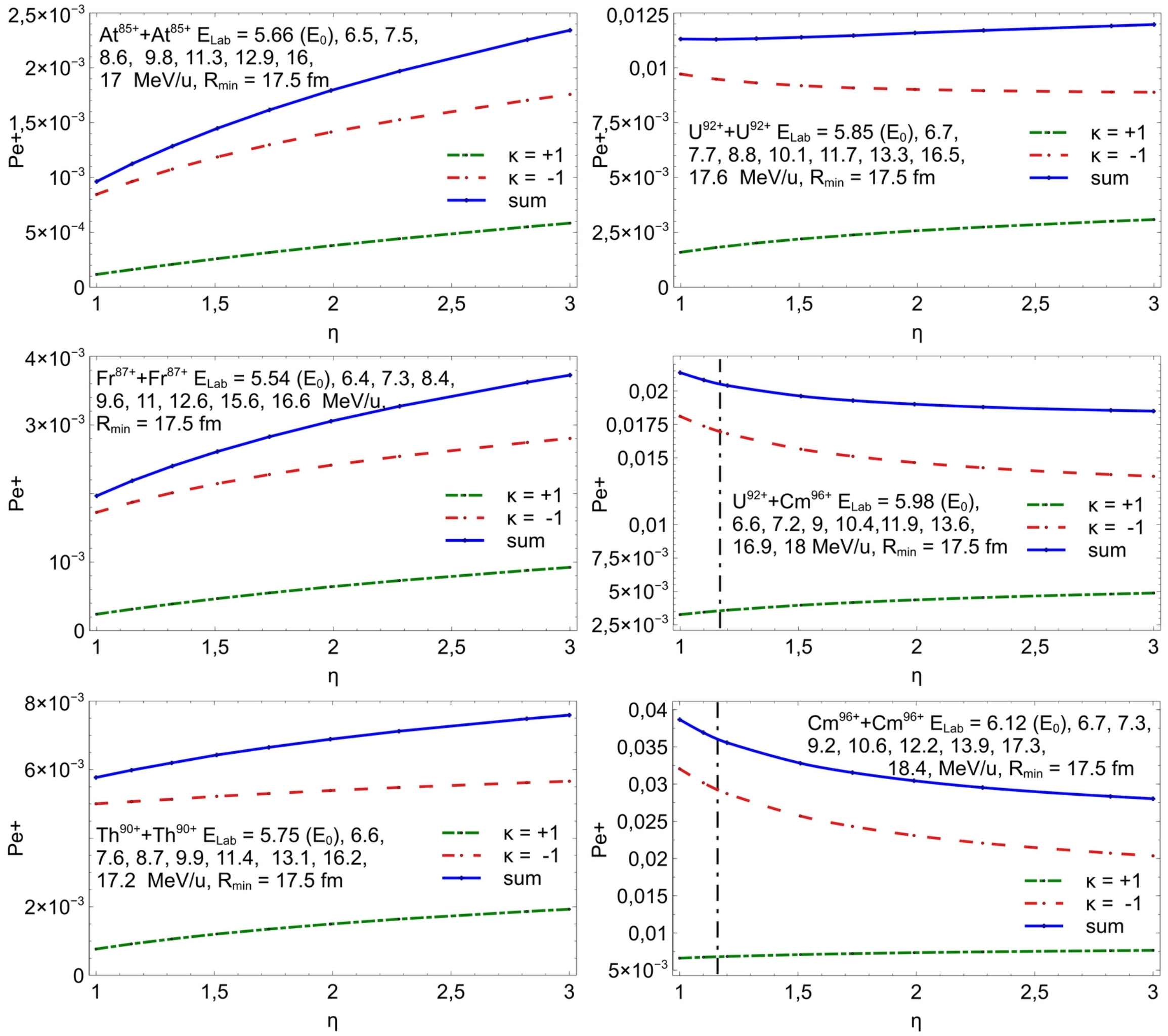

Figure 11: *Total positron emission rates $P_{e^+}$ as a function of Z for the systems $At^{85+} + At^{85+}$ ($Z = 170$) to $Cm^{96+} + Cm^{96+}$ ($Z = 192$). For supercritical systems the slope of $dP_{e^+}/d\eta$ turns negative. Right of the vertical dashed-dotted lines also inelastic collisions can occur.*

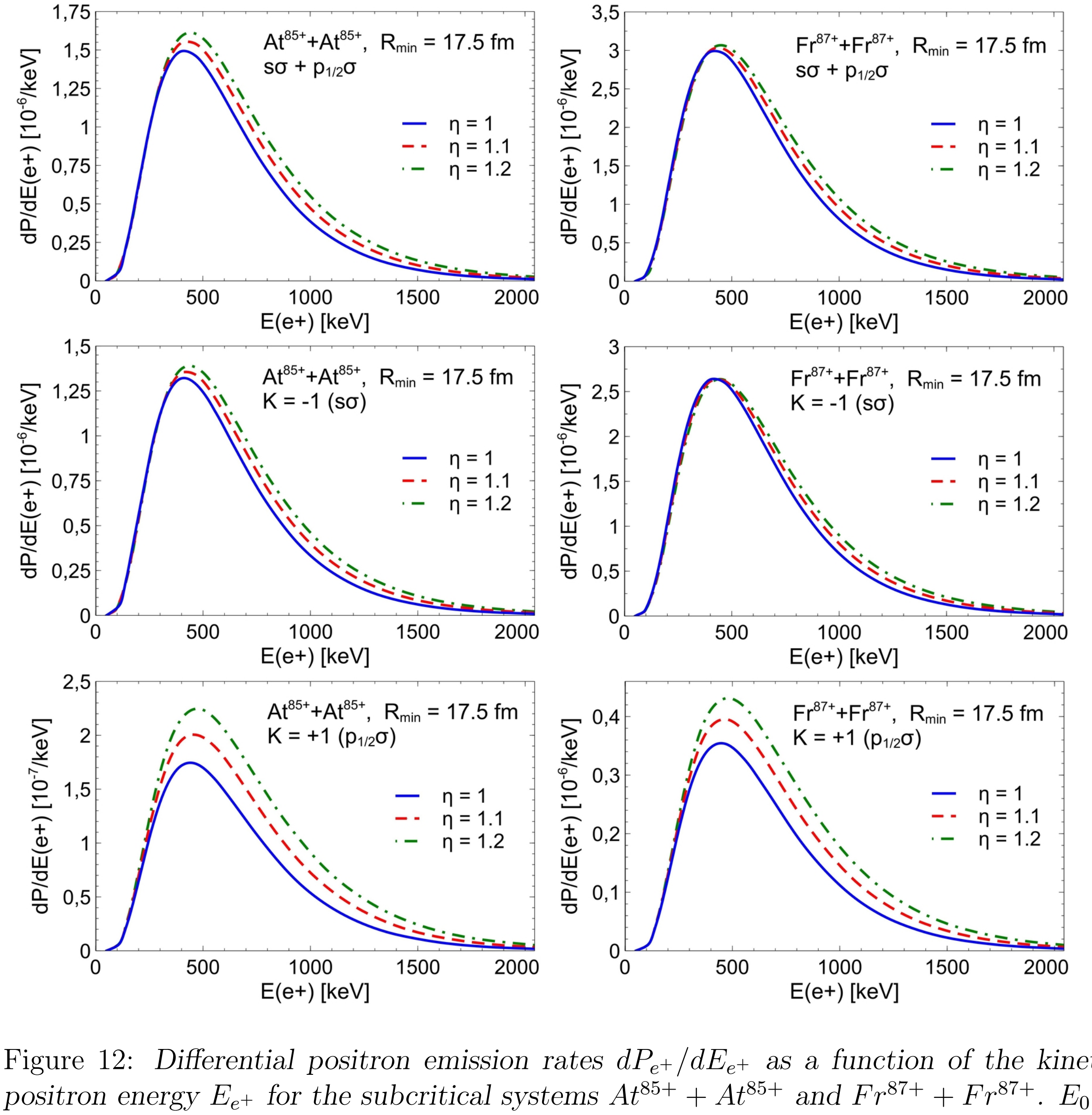

Figure 12: *Differential positron emission rates* $dP_{e^+}/dE_{e^+}$ *as a function of the kinetic positron energy* $E_{e^+}$ *for the subcritical systems* $At^{85+} + At^{85+}$ *and* $Fr^{87+} + Fr^{87+}$. $E_0$ *is chosen to fulfill 2a* $=$ *17.5 fm. The parameters* $\eta = E/E_0 = 1$, *1.1 and 1.2 are considered Top: Total spectrum (s+p), middle:* $s\sigma$- *spectrum, bottom:* $p_{\frac{1}{2}}\sigma$- *spectrum.*

The total positron emission rate $P_{e^+} = \int \bar{N}_q(F=0)dE$ reflects the area under the curves discussed before. Its dependence on $\eta$ is shown in Fig. 11 and the results are in good agreement with [2]. This figure summarizes the previous discussion in terms of the slope of $dP_{e^+}/d\eta$, which should focus on values for $\eta < 1.2$ for elastic collisions. While the slope of $\eta$ is positive for subcritical systems, it turns negative for supercritical systems, where the $1s\sigma$- state dives into the negative Dirac continuum for two centre distances $R < R_{cr}$. Fig. 12 shows positron spectra for the two subcritical systems At+At and Fr+Fr for values $\eta \leq 1.2$ with increasing total emission rates (addition of $\kappa = \pm 1$) for increasing values of $\eta$. Figs. 13 and 14 visualize the reversal of the $\eta$- dependency in the supercritical systems Pb+Cf, U+U, U+Cm and Cm+Cm: For head on collisions with the longest dwell time within $R < R_{cr}$ ($\eta = 1$), radial and spontaneous positron emission rates deliver their maximal contribution to the spectrum (c.f. Fig. 3), which decreases with increasing $\eta$. Therefore, the slope of $\eta$ in the total spectra turns negative in contrast to the $\kappa = +1$ spectra. This is different for the system Cm+Cm, where also the $2p_{\frac{1}{2}}\sigma$- state enters the negative continuum and thus the dependence of the $\kappa = +1$ spectrum shows the same $\eta$-dependence as the $\kappa = -1$ spectrum.

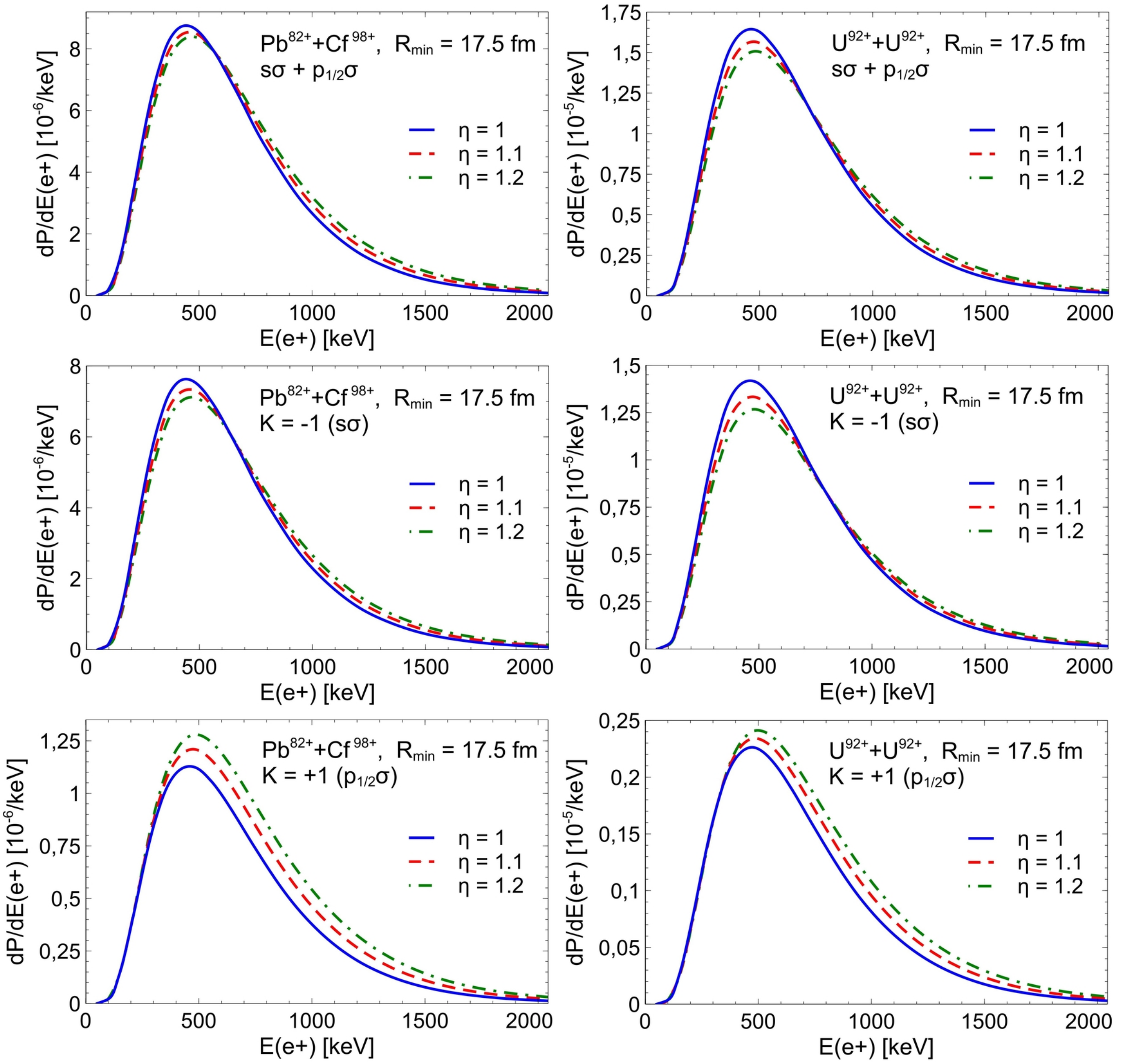

Figure 13: *Same as in Fig. 12 for the systems $Pb^{82+} + Cf^{98+}$ and $U^{92+} + U^{92+}$ indicating the transition from the sub- to the supercritical regime.*

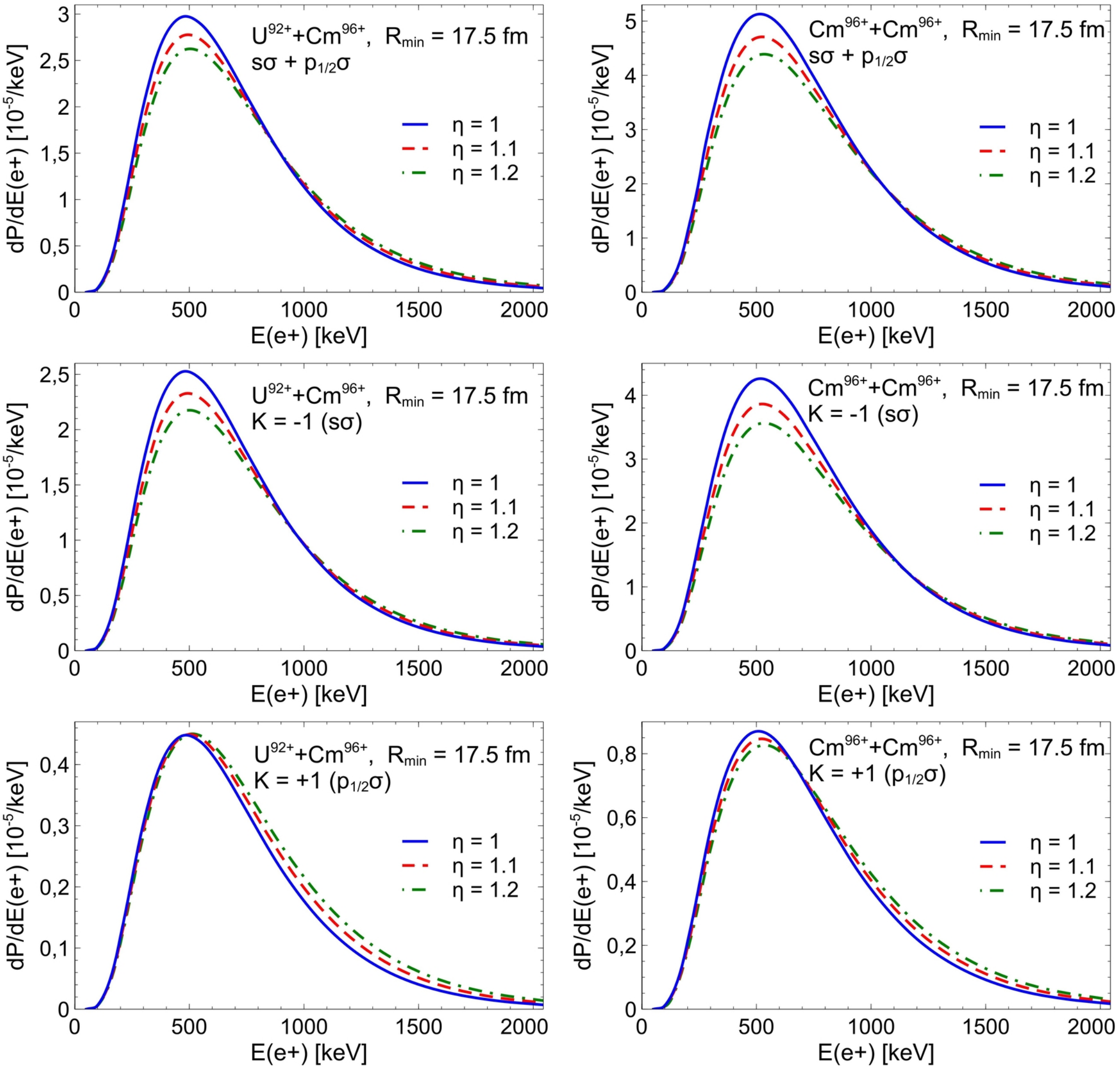

Figure 14: *Differential positron emission rates $dP_{e^+}/dE_{e^+}$ for the two supercritical systems $U^{92+}$+$Cm^{96+}$ and $Cm^{96+}$+$Cm^{96+}$. $E_0$ again is chosen to fulfill 2a = 17.5 fm. For the total spectrum (s+p) the systematics concerning $\eta$ is reversed compared with the subcritical systems in Fig. 12: For $\eta = 1$ $P_{e^+}$ is maximal.*

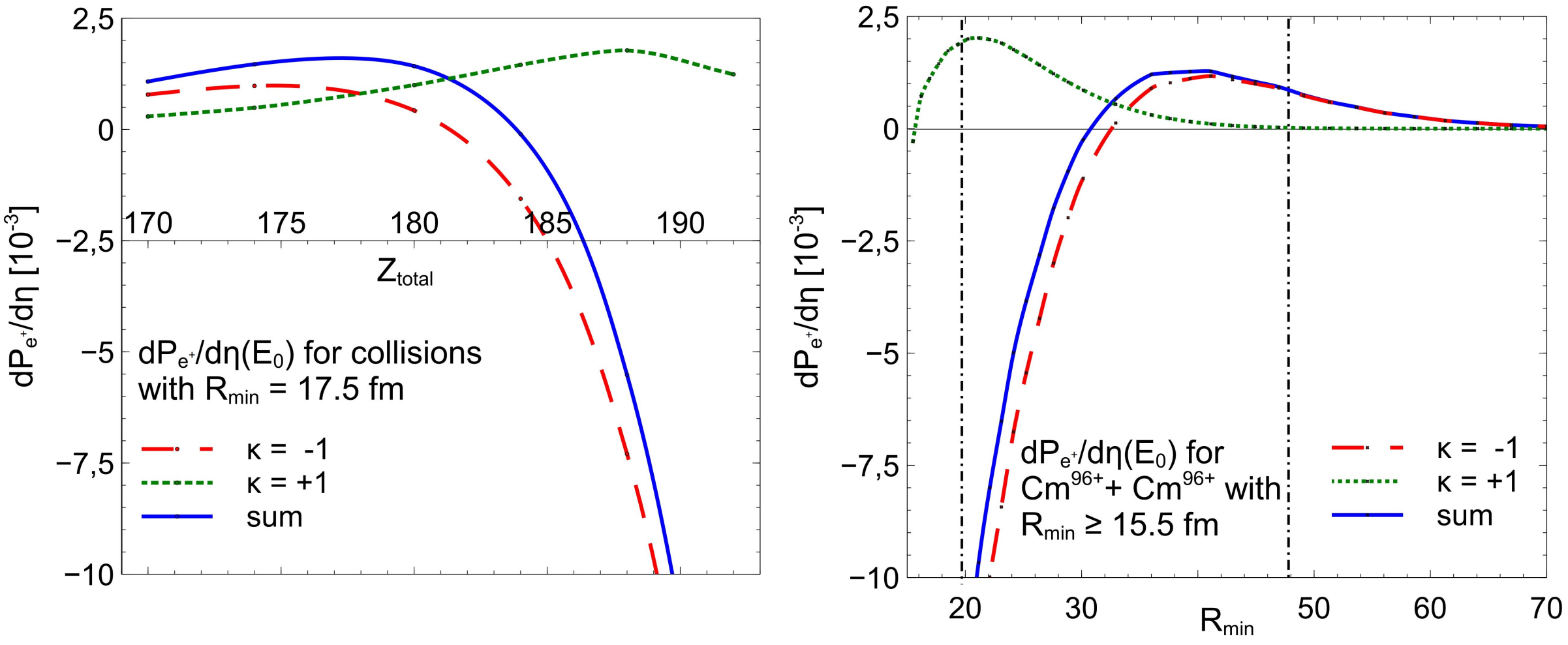

Figure 15: *Left: Derivative of the total positron emission rate $dP_{e^+}/d\eta$ according to equation (19) at $\eta = 1$ as a function of the combined nuclear charge of projectile and target. Combinations of bombarding energy an impact parameters leading to $R_{min}$ = 17.5 fm are chosen. Dashed: $\kappa$ = –1, dotted: $\kappa$ = +1, full curve: sum (s+p). Right: Same as on the left for the system $Cm^{96+} + Cm^{96+}$ as a function of $R_{min}$. The dashed vertical lines indicate the critical two centre distances for this system $R_{cr}^{1s\sigma} \simeq$ 48 fm and $R_{cr}^{2p_{1/2}\sigma} \simeq$ 20 fm, where the $1s\sigma$-, resp. the $2p_{\frac{1}{2}}\sigma$- state dives into the negative energy continuum.*

The dependence of the derivative of total positron emission $dP_{e^+}/d\eta$ at $E_0$ as a function on the combined nuclear charge $Z_{total}$ is plotted in the left part of Fig. 15 for the fixed value $R_{min}$ = 17.5 fm. The curves for the $\kappa = \pm$ 1- states show maxima slightly above the critical charges $Z_{cr} \simeq$ 174 for the $1s\sigma$-, resp. $Z_{cr} \simeq$ 188 for the $2p_{\frac{1}{2}}\sigma$- state. The total value of $dP_{e^+}/d\eta$ (sum of $\kappa = -1$ and $\kappa = +1$ contribution) has a maximum at $\simeq$ 178. This dependence has been demonstrated earlier for $R_{min}$ = 16.5 fm in [1].
Similarly the derivative $dP_{e^+}/d\eta$ at $\eta = 1$ can be plotted as a function of $R_{min}$ for a fixed value of $Z_{total}$ = 192 (e.g. Cm+Cm) in the right part of Fig. 15 to visualize its dependence on the distance of closest approach $R_{min}$. For the critical two centre distances $R_{cr}$ $\simeq$ 48 fm resp. $R_{cr} \simeq$ 20 fm (indicated as vertical dashed-dotted lines) the $1s\sigma$- and $2p_{\frac{1}{2}}\sigma$-levels dive into the antiparticle continuum. For values of $R_{min}$ smaller $\simeq$ 30 fm the total value of $dP_{e^+}/d\eta$ (full line) turns negative, thus indicating an upper bound for the choice of $R_{min}$ to observe the $\eta$- dependence of the spectra for this system. The curves in this figure plus the previous spectra shown in Figs. 12 - 14 are in good agreement with the calculations presented in [2].
**Annotation:** The complete notations of the elements used in the calculations including the nucleon numbers are: ${}^{208}_{82}Pb$, ${}^{210}_{85}At$, ${}^{223}_{87}Fr$, ${}^{232}_{90}Th$, ${}^{238}_{92}U$, ${}^{248}_{96}Cm$, ${}^{251}_{98}Cf$. For the sake of simplicity, fully ionized atoms, like e.g. Uranium, are denoted as $U^{92+}$, instead of ${}^{238}_{92}U^{92+}$.

# 4 Positron spectra and cross sections for U+Cm and Cm+Cm

Former calculations for positron- $\delta$- electron emission or K-vacancy formation [38, 76–80] targeted at 'conventional collisions', i.e. a partly ionized projectile was bombarded onto a neutral target. These systems were described with sufficient precision by assuming the first three inner s$\sigma$- shells (1s$\sigma$-3s$\sigma$) plus the first three p$\sigma$- shells ($2p_{\frac{1}{2}}\sigma$ - $4p_{\frac{1}{2}}\sigma$) to be occupied by electrons and the remaining shells to be empty - indicated by a Fermi level F=3.

For a full agreement between experimental measurements and theory, however, the consideration of screening effects was necessary, e.g. by employing the adiabatic time dependent Hartree- Fock (ATDHF) approximation. Solving the time dependent Dirac equation self consistently within the Hartree-Fock-Slater approximation, delivered partially excellent agreement between experimental data and theory [78, 81–85] in conventional collisions.

Positron emission rates and hence related cross sections, however, are comparatively small in conventional collisions with bombarding energies $E_{Lab}$ = 4-6 MeV/n, assuming a Fermi level of F=3. Considering a U+U collision at $E_{Lab}$ = 5.9 MeV/n the differential emission probability $dP_{e^+}/dE_{e^+}$ at b=0 is about $4.5 \cdot 10^{-7}$ [1/keV] and the cross section $\sigma^{e^+} \simeq 2.5$ mb. From the experimental point of view these small quantities are challenging to measure with the additional difficulty to subtract undesirable (nuclear) background effects, deteriorating the spectra.

For two of the heaviest collision systems, U+Cm and Cm+Cm, the calculated emission spectra for positrons versus their kinetic energy $E_{e^+}$ are shown for different impact parameters b. The results for conventional collisions (F=3) are compared with those for the fully ionized systems. Fig. 16 shows spectra for the system U+Cm at an energy of $E_{c.m.}$ $\simeq$ 748 MeV (2a = 17fm). The results for the differential positron emission spectra for $1s\sigma$- and $p_{\frac{1}{2}}\sigma$- states are in the same order of magnitude, with about 16% higher maxima values for $p_{\frac{1}{2}}\sigma$- states. This changes significantly when considering the spectra for the corresponding fully ionized system in Fig. 17: The s-channel becomes dominant and the maximum values in central collisions for s-states are about a factor 5.3 larger than for the p-channel. For the total spectrum (sum of s- and p-spectra in Fig. 18) the maximum in central collisions for the fully ionized system exceeds the one for conventional collisions by a factor of $\simeq$ 36. Total positron emission rates $P_{e^+}$ as a function of the impact parameter b are listed in Table 1. The emission rates for the fully ionized system in column 2 increases by factors 36 (b = 0) to 185 (b = 40 fm) versus those in column 4. The total cross section for positron emission $\sigma(P_{e^+})$ amounts 203 mb for the fully ionized system, versus 3.6 mb in conventional systems. The increase of the total cross section for positron emission by about a factor of 57 in collisions of bare nuclei is caused by the absence of Pauli- blocking: In conventional collisions the $1s\sigma$- shell is occupied or blocked, so that direct transitions from states in the negative Dirac continuum into a vacant s-state is strongly suppressed. Transitions of the $1s\sigma$- electrons into higher bound states or to the upper Dirac continuum (free electrons) are required, before notable occupation of the 1s-level by electrons from the lower Dirac continuum can occur. Table 1 shows another feature in collisions of bare nuclei: The dominant channel for positron emission is the capture of negative continuum states into bound states in column 3 (more precisely into the $1s\sigma$- state, as we will see later). These emission rates are denoted as $P_{e^+}^{bound}$ and omit the emission rates from direct transitions between the negative and positive Dirac continuum. The effects in collisions of bare nuclei versus partly ionized ones are even more pronounced in the System Cm+Cm at $E_{c.m.} \simeq 781$ MeV (2a = 17 fm), where the $2p_{\frac{1}{2}}\sigma$- state also dives into the negative Dirac continuum at two centre distances $R < 20$ fm. Figs. 19 to 21 show the corresponding spectra for s- and p- states plus total spectra for conventional and fully ionized collisions. Total pair-creation emission rates $P_{e^+}$ for this system are listed for different impact parameters in Tab. 2. The total cross section $\sigma(P_{e^+})$ for positron emission is determined as 383 mb in collisions of bare nuclei, versus 5.3 mb in conventional ones, which is factor of about 72 smaller.

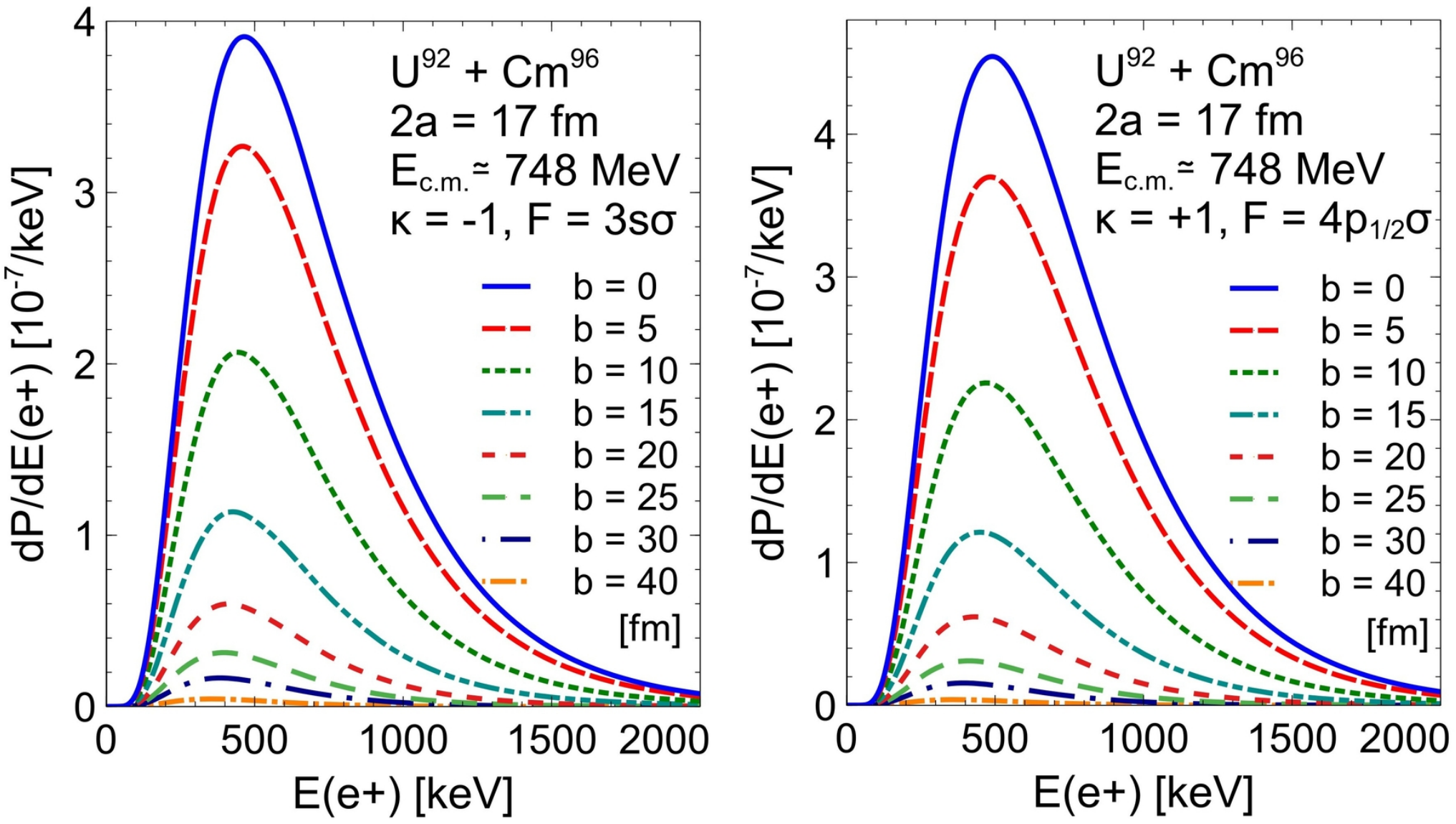

Figure 16: *Positron spectra for $\kappa = -1$ (left) and $\kappa = +1$ in conventional $U^{92} + Cm^{96}$ collisions at $E_{c.m.} \simeq 748$ MeV (2a = 17 fm), assuming a Fermi-Level of F = 3. Different curves show the results for different impact parameters b.*

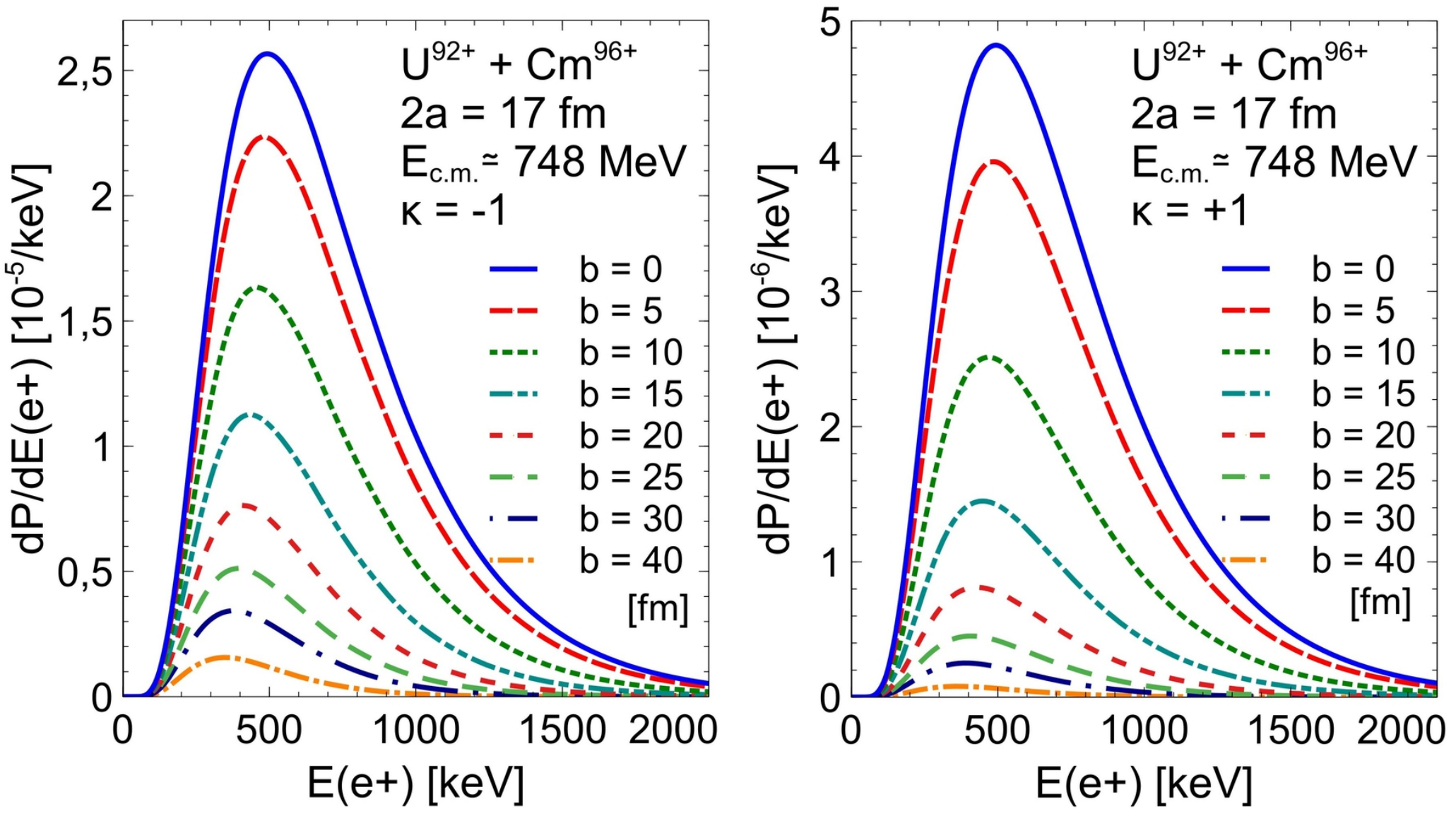

Figure 17: *Same as in Fig. 16, however, assuming fully ionized* $U^{92+} + Cm^{96+}$ *collisions* *(*$F = 0$*).*

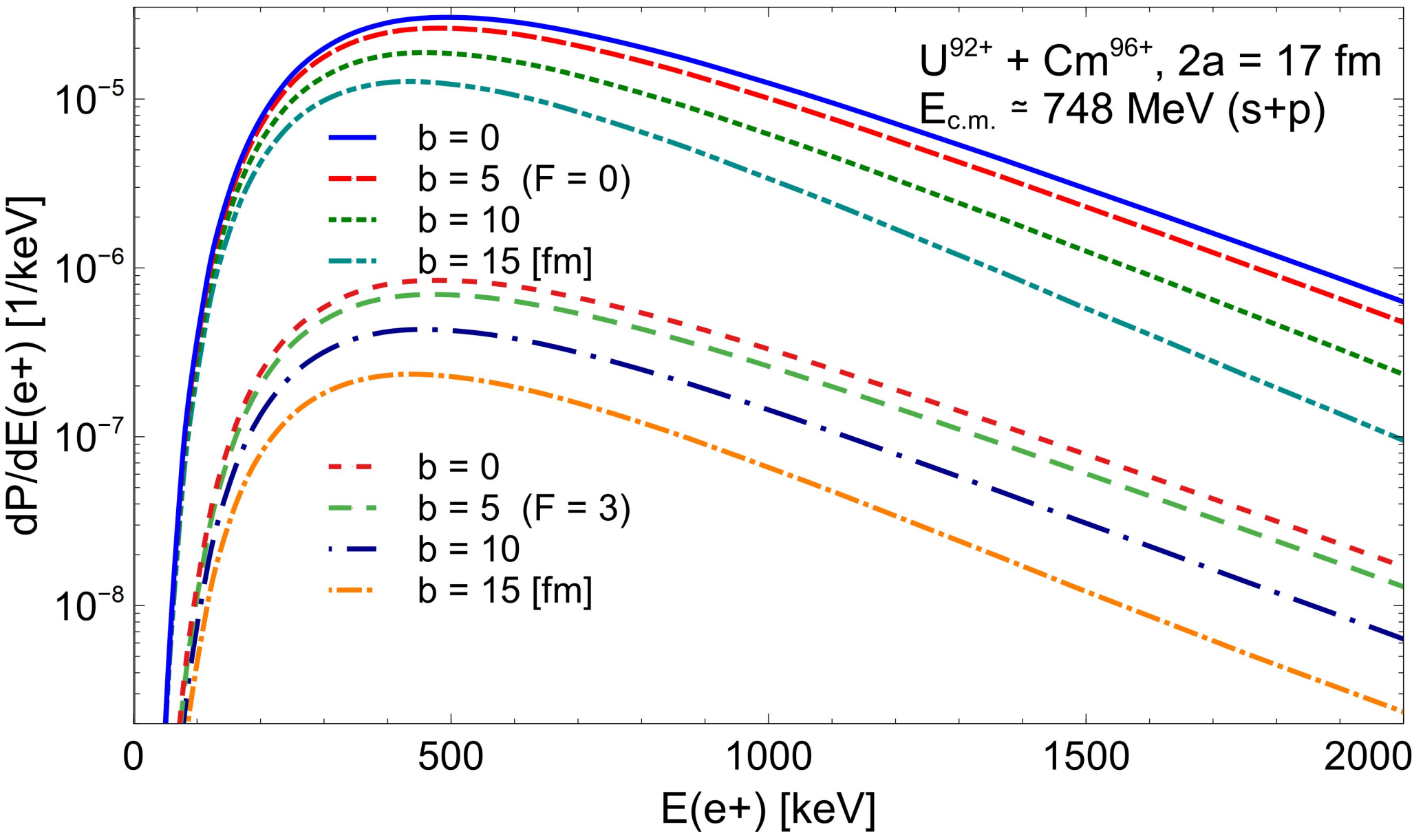

Figure 18: *Total positron spectra in U + Cm collisions (s+p) at* $E_{c.m.} \simeq 748$ *MeV (2a = 17 fm). Upper curves: A Fermi-Level of* $F = 0$ *is chosen, lower curves:* $F = 3$*.*

| $b$ [fm] | $P_{e^+}(F=0)$ | $P_{e^+}^{bound}(F=0)$ | $P_{e^+}(F=3s\sigma,4p_{1/2}\sigma)$ |
|---|---|---|---|
| 0 | $2.26 \cdot 10^{-2}$ | $2.21 \cdot 10^{-2}$ | $6.21 \cdot 10^{-4}$ |
| 5 | $1.90 \cdot 10^{-2}$ | $1.87 \cdot 10^{-2}$ | $5.03 \cdot 10^{-4}$ |
| 10 | $1.29 \cdot 10^{-2}$ | $1.26 \cdot 10^{-2}$ | $2.99 \cdot 10^{-4}$ |
| 15 | $8.05 \cdot 10^{-3}$ | $7.95 \cdot 10^{-3}$ | $1.52 \cdot 10^{-4}$ |
| 20 | $4.95 \cdot 10^{-3}$ | $4.90 \cdot 10^{-3}$ | $7.35 \cdot 10^{-5}$ |
| 25 | $3.05 \cdot 10^{-3}$ | $3.03 \cdot 10^{-3}$ | $3.54 \cdot 10^{-5}$ |
| 30 | $1.90 \cdot 10^{-3}$ | $1.88 \cdot 10^{-3}$ | $1.73 \cdot 10^{-5}$ |
| 40 | $7.54 \cdot 10^{-4}$ | $7.51 \cdot 10^{-4}$ | $4.02 \cdot 10^{-6}$ |

Table 1: *Total positron emission rates $P_{e^+}$ and bound free pair creation rates $P_{e^+}^{bound}$ in $U^{92+}+Cm^{96+}$- collisions at $E_{c.m.} \simeq$ 748 MeV (2a = 17 fm) for different impact parameters b. The total cross sections for this impact parameter range are determined as $\sigma(P_{e^+})$ = 203 mb and $\sigma(P_{e^+}^{bound})$ = 201 mb (F=0). Column 4 displays $P_{e^+}$ for a partly ionized projectile being bombarded onto a neutral target (F=3). The total cross section for $\sigma(P_{e^+})$ (F=3) amounts 3.56 mb and is about a factor 57 smaller than in collisions of fully ionized collision partners. For bound free pair creation the result obtained is $\sigma(P_{e^+}^{bound})$ = 1.13 mb.*

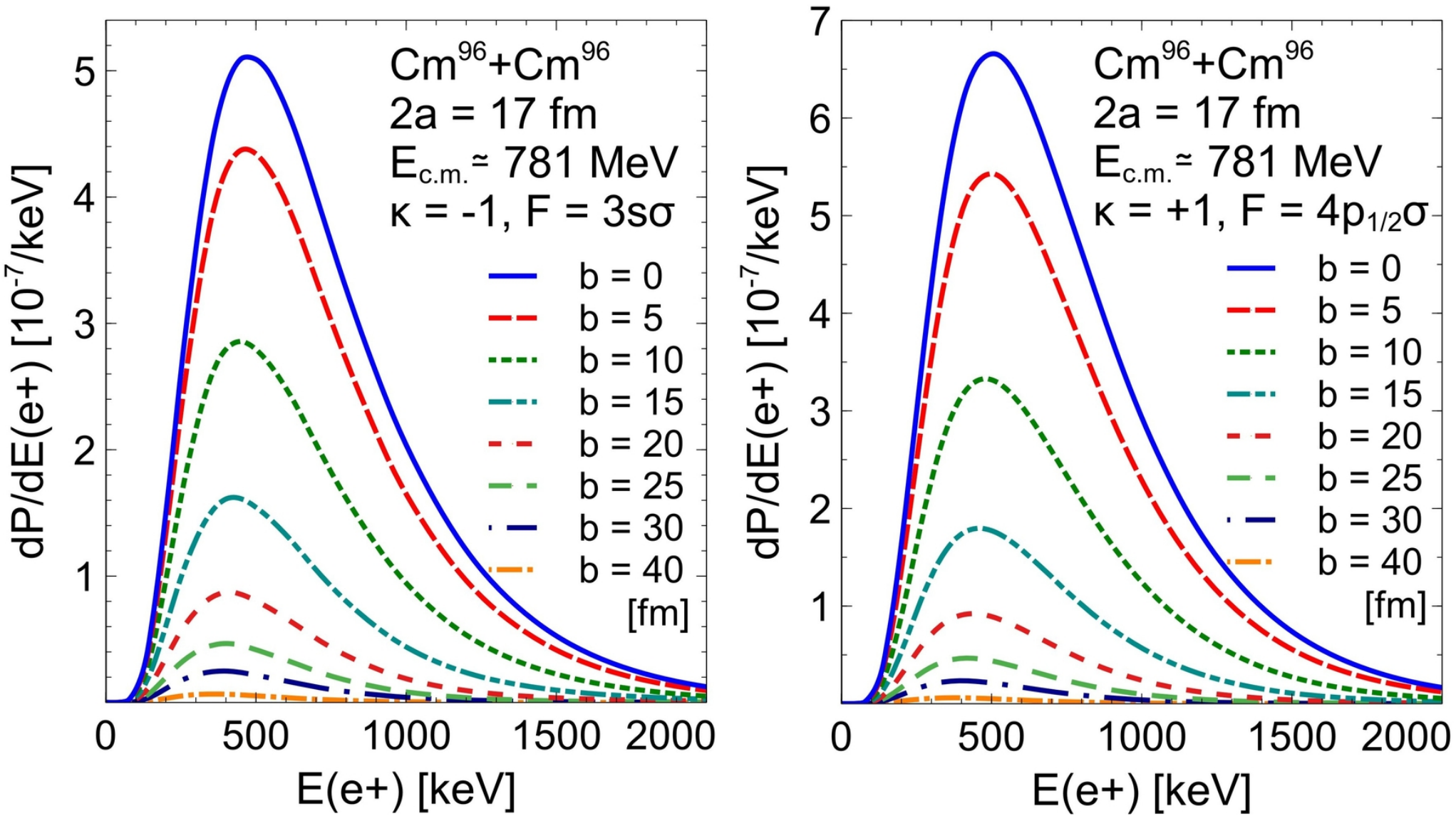

Figure 19: *Positron spectra for $\kappa = -1$ (left) and $\kappa = +1$ in conventional $Cm^{96} + Cm^{96}$ collisions at $E_{c.m.} \simeq 781$ MeV (2a = 17 fm), assuming a Fermi-Level of $F = 3$. Different curves show the results for different impact parameters b.*

To detect spontaneous positron emission in regular elastic heavy ion collisions was deemed to be not feasible [17]. Therefore scenarios where clear signatures for spontaneous positron emission could theoretically be detected were proposed, such as a sufficiently long living ($> 10^{-21}$s) giant quasi-molecule, triggering spontaneous positron emission visible as a sharp structure in the spectra [16, 37, 38]. If this scenario happened as assumed, the spontaneous positron emission would be proved by a 'first order effect', being highly desirable. Alternatively inelastic collisions using classical friction trajectories with nuclear delay times up to $3{\cdot}10^{-21}$s were investigated [39,86]. The channel of spontaneous positron emission would not cause a 'first order effect', but theoretically be visible by a narrower shape in positron spectra and a steeper exponential decay at higher $E_{e^+}$- energies - being a more indirect proof of its existence.

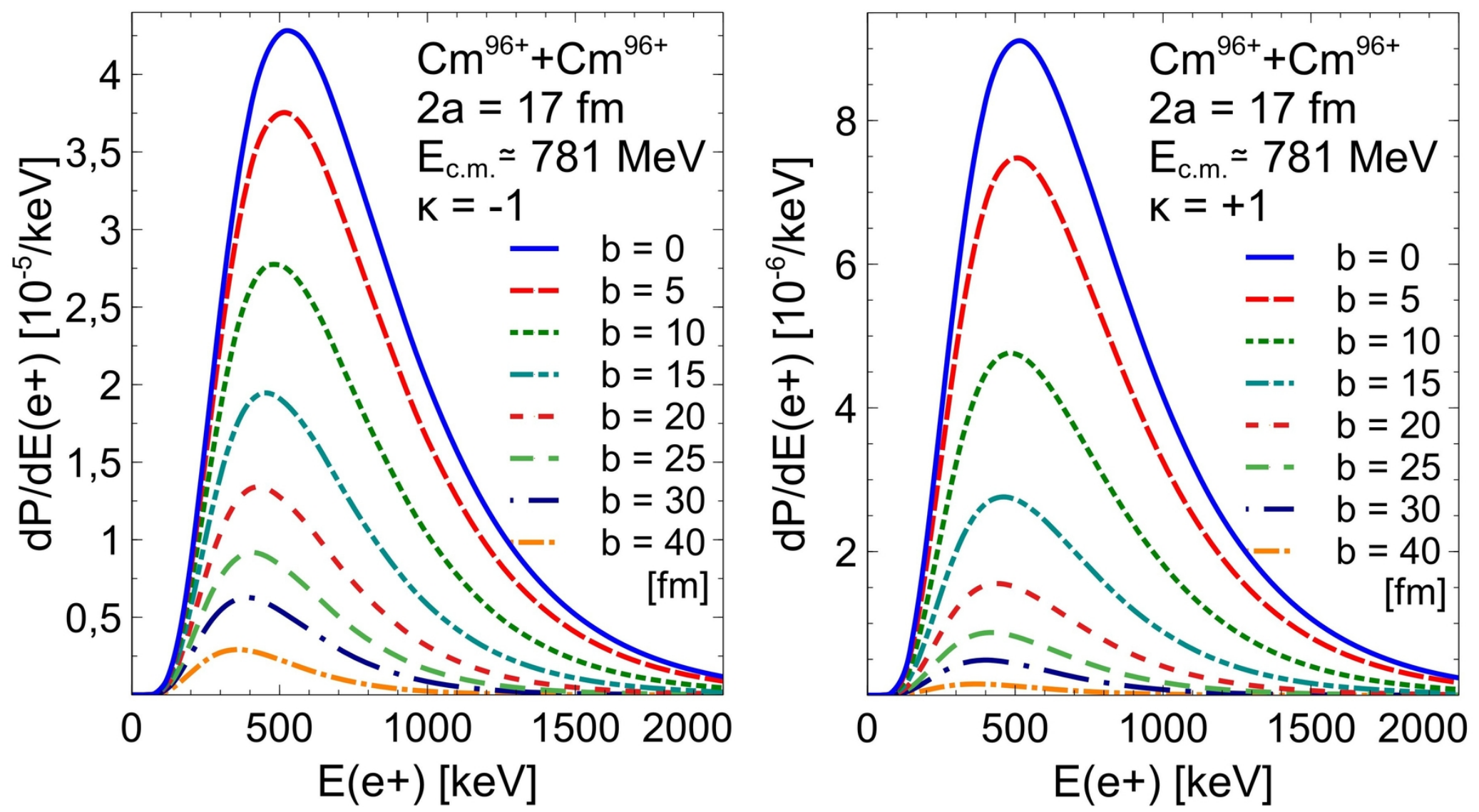

Figure 20: *Same as in Fig. 19, however, assuming fully ionized* $Cm^{96+} + Cm^{96+}$ *collisions (F = 0).*

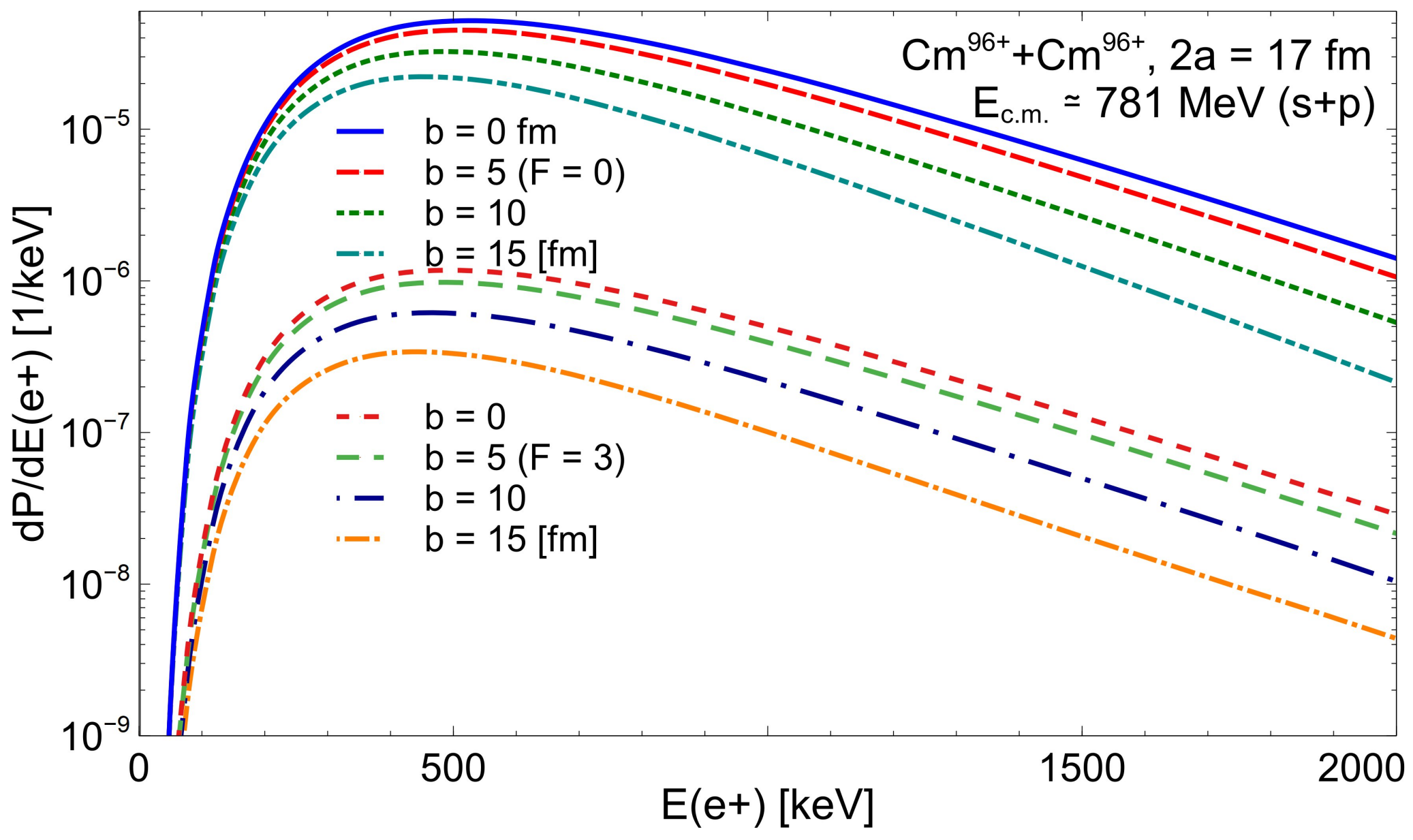

Figure 21: *Total positron spectra in Cm + Cm collisions (s+p) at* $E_{c.m.} \simeq$ *781 MeV (2a = 17 fm). Upper curves: A Fermi-Level of F = 0 is chosen, lower curves: F = 3.*

| $b$ [fm] | $P_{e^+}(F=0)$ | $P_{e^+}^{bound}(F=0)$ | $P_{e^+}(F=3s\sigma,4p_{1/2}\sigma)$ |
|---|---|---|---|
| 0 | $4.04 \cdot 10^{-2}$ | $3.97 \cdot 10^{-2}$ | $8.92 \cdot 10^{-4}$ |
| 5 | $3.41 \cdot 10^{-2}$ | $3.36 \cdot 10^{-2}$ | $7.28 \cdot 10^{-4}$ |
| 10 | $2.32 \cdot 10^{-2}$ | $2.29 \cdot 10^{-2}$ | $4.37 \cdot 10^{-4}$ |
| 15 | $1.46 \cdot 10^{-2}$ | $1.45 \cdot 10^{-2}$ | $2.26 \cdot 10^{-4}$ |
| 20 | $9.09 \cdot 10^{-3}$ | $9.01 \cdot 10^{-3}$ | $1.11 \cdot 10^{-4}$ |
| 25 | $5.65 \cdot 10^{-3}$ | $5.62 \cdot 10^{-3}$ | $5.40 \cdot 10^{-5}$ |
| 30 | $3.55 \cdot 10^{-3}$ | $3.53 \cdot 10^{-3}$ | $2.67 \cdot 10^{-5}$ |
| 40 | $1.44 \cdot 10^{-3}$ | $1.43 \cdot 10^{-3}$ | $6.13 \cdot 10^{-5}$ |

Table 2: *Same as in Table 1 for $Cm^{96+}$ + $Cm^{96+}$- collisions at $E_{c.m.} \simeq$ 781 MeV (2a = 17 fm). The total cross sections for this impact parameter range are determined as $\sigma(P_{e^+})$ = 383 mb and $\sigma(P_{e^+}^{bound})$ = 379 mb (F=0). The total cross section for $\sigma(P_{e^+})$ (F=3) amounts 5.28 mb and is about a factor 72 smaller than in collisions of fully ionized collision partners. For bound free pair creation the result reads $\sigma(P_{e^+}^{bound})$ = 1.75 mb.*

# 5 Effects of spontaneous positron emission in elastic collisions

## 5.1 Contribution of spontaneous positrons with small kinetic energies

The question is, if elastic collisions of fully ionized ions facilitate the detection of spontaneous positron emission as described in [2], i.e. that 'any increase in the pair-production probability at $E/E_0 \to 1$ for a given $R_{min}$ should indicate the effect of spontaneous pair creation'. Before we cope with this question, ratios of the partial probabilities $P_x$ being proposed in the aforementioned reference were also calculated and are shown in Fig. 22. Partial total positron emission or pair-creation probabilities $P_x$ restrict the integration of the $dP_{e^+}/dE_{e^+}$ spectra to energies $E_{e^+} < dP_{e^+}/dE_{e^+} = (1-\text{x})\ (dP_{e^+}/dE_{e^+})_{max}$ right of the maximum in $dP_{e^+}/dE_{e^+}$ for $\eta = 1$. The motivation for the partial spectra stems from the interest to study the changes of interest in the spectra for energies $E_{e^+}$ around the maximum or below, to focus on the energy range of emitted spontaneous positrons.
In equation (14) for the time derivative of the occupation amplitudes, the two components for radial and potential couplings add up coherently to the solution for $a_{ij}(t)$ with a relative phase factor $i$ in the supercritical domain. The problem is, that the final solutions of the occupation amplitudes $a_{ij}(t)$ cannot be disentangled concerning their radial and spontaneous components resp. origin. A simple phenomenological ansatz is to switch off the

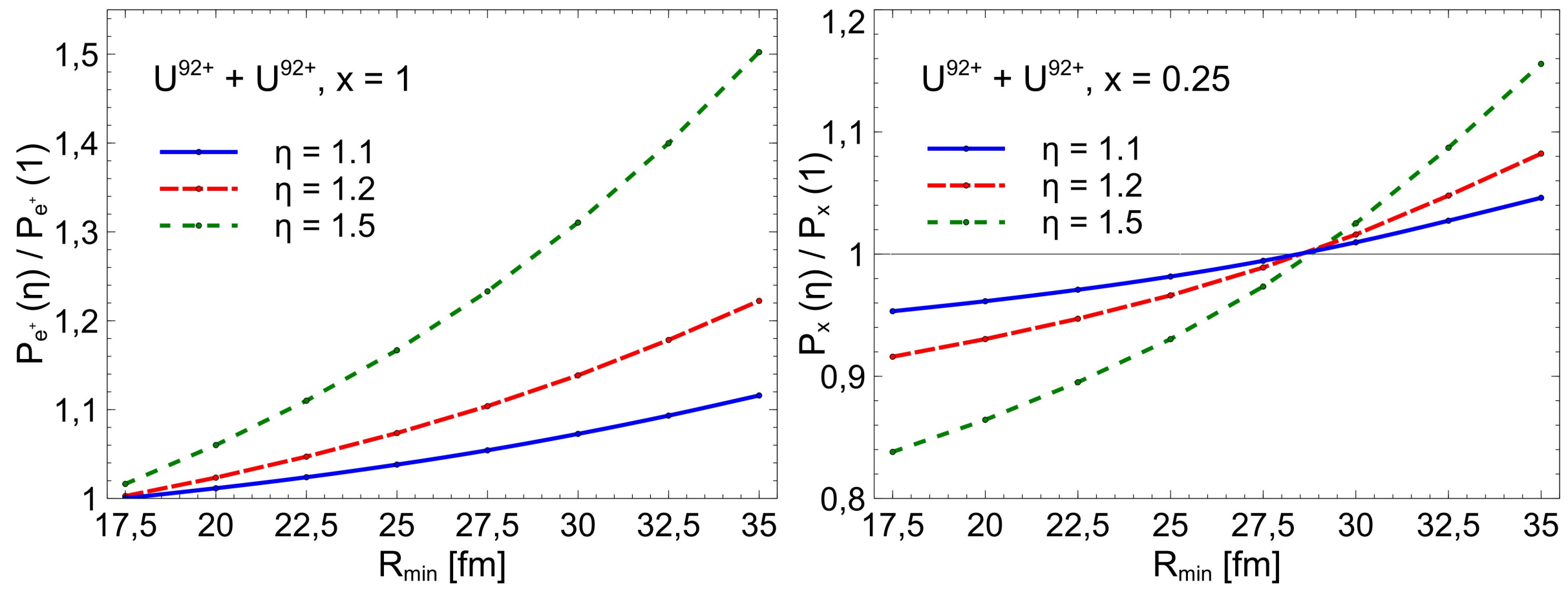

Figure 22: *Left: Ratio of total positron emission for $\eta$ >1 divided by the result for central collisions of U+U versus distance of closest approach $R_{min}$. Right: Same ratio for the partial spectrum with x=0.25.*

potential coupling by setting the corresponding matrix elements $| < \tilde{\varphi}_{E_p} | \hat{H}_{TCD} \, | \Phi_R > |$ to zero. The result is shown in Fig. 23. Although no exact result for this method can be expected, the spectra as a function of $\eta$ look quite similar to those in Fig. 9, i.e. we see an increase of the pair-production probability at $E/E_0 \to 1$ for a given $R_{min}$ even

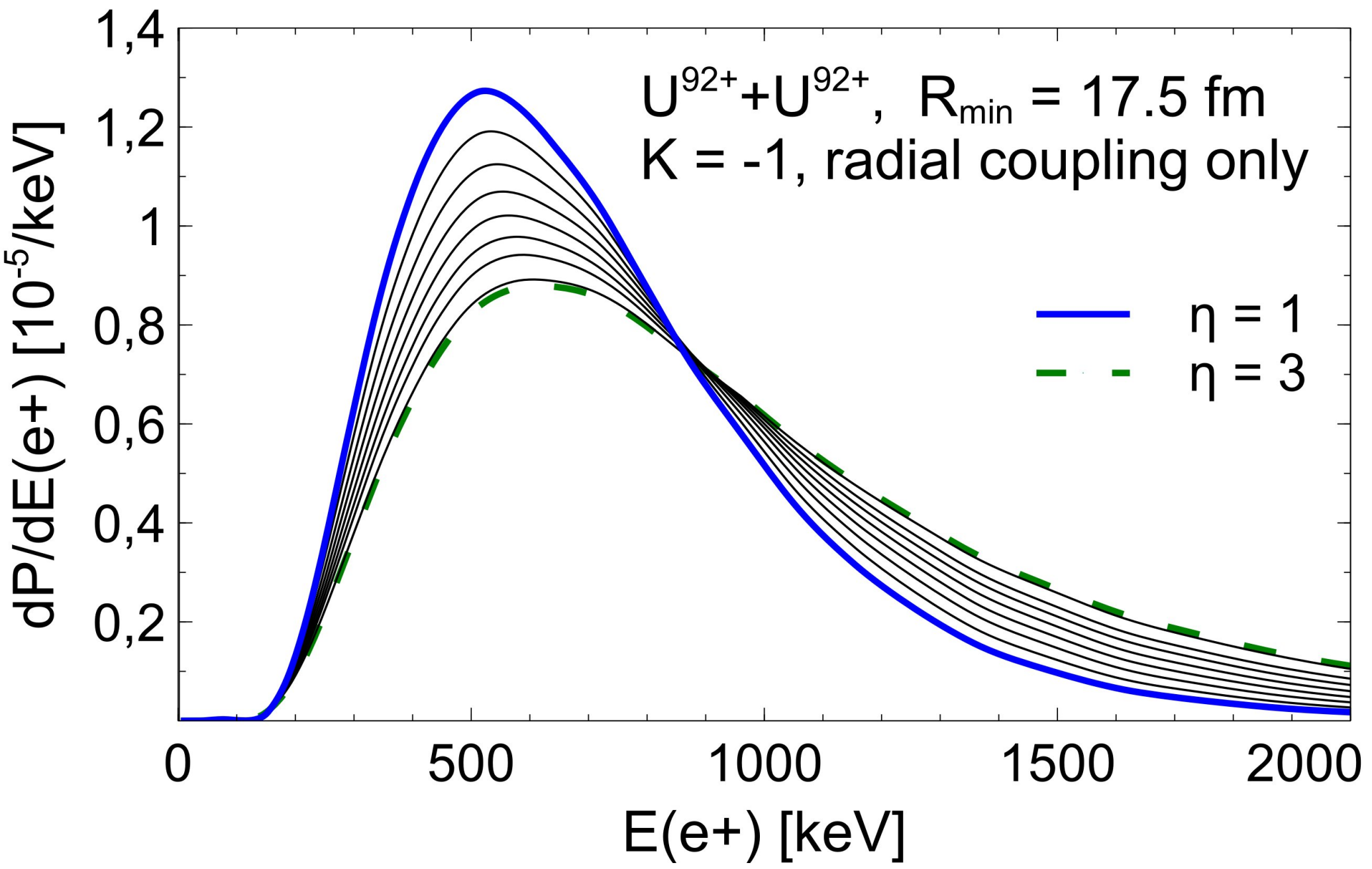

Figure 23: *Differential emission probability for positrons $dP_{e^+}/dE_{e^+}$ as a function of their energy $E_{e^+}$ for the system $U^{92+} + U^{92+}$. Like in Fig. 9 $E_0$ is chosen to fulfill 2a = 17.5 fm. In this spectra for the angular momentum quantum number $\kappa$ = –1, however, the spontaneous coupling matrix elements have been switched off. The following parameters $\eta = E/E_0$ were chosen: $\eta$ = 1, 1.15, 1.32, 1.51, 1.73, 1.99, 2.28, 2.82 and 3.*

without spontaneous couplings being considered. Calculation of the rations for the partial pair-creation probabilities in the same setting for U+U with x = 0.25 are shown in Fig. 24. The curves look quite similar, except for a flattening of the slope for values of $R_{min} < 24$ fm. From this phenomenological consideration neither the increase of the pair-production probability at $E/E_0 \rightarrow 1$ for a given $R_{min}$, nor the crossing of the ratios for the emission probabilities below 1 are clear indicators for spontaneous positron emission. Another ansatz to analyze the relation between radial and spontaneous couplings is to study the transition strengths or matrix elements between the $1s\sigma$- resp. resonance state $|\Phi_R>$ and continuum energies in the negative Dirac continuum $E_{e^+}$ along the trajectory. Of special interest are those matrix elements, which exhibit their largest transition strength, i.e. absolute values of their extrema for the kinematic parameters $R_{min}$ and $\eta$. For the potential or spontaneous coupling matrix element the extremum is located at $R_{min}$, whereas the extrema for radial or induced coupling matrix elements are located between $\approx$ 20 and 30 fm for $R_{min}$ = 17.5 - 25 fm (c.f. Fig. 6). After having selected the extrema, the associated continuum energies $E_{e^+}$ were identified, which are in the range of 10 - 194 keV for spontaneous and 194 - 286 keV for radial coupling matrix elements for s-states. The resonance energies of $|\Phi_R>$ for distances $R_{min}$ = 17.5 - 30 fm vary between 262 and 33 keV, i.e. the strongest couplings occur for continuum energies $E_{e^+}$ in the vicinity of the resonance. The transition strengths of radial couplings in Fig. 25 depend on the kinematic parameters, which is not the case for spontaneous couplings which solely depend on the internuclear distance $R$, so that one curve is obtained for different values of $\eta$. The figure also shows that the maximum transitions strengths of radial couplings are about a factor of 6 -8 larger for kinematic parameters $R_{min} = 17.5$ fm and $\eta = 1.1$ - 1.5 compared with those of spontaneous couplings. The right side of Fig. 25 shows matrix elements for p-states for which no diving of the $2p_{\frac{1}{2}}\sigma$- state in U+U-collisions occurs. Maximal transition strengths are notably smaller in the logarithmic plot in Fig. 26, explaining the significant difference of the emission rates $dP_{e^+}/dE_{e^+}$ between spectra for s- and p-states in the right column of Fig. 13. In addition the transition strengths are maximal for different continuum energies $E_{e^+}$, varying in the range of 450 - 715 keV. In Fig. 27 the ratio of matrix elements is plotted similarly to Fig. 22. In summary the analysis of matrix elements shows that radial couplings are dominant for s-states, maximal transitions strengths occur for continuum energies $E_{e^+}$ adjacent to the resonance energy and that only the radial coupling matrix elements depend on the specific kinematic parameters $R_{min}$ and $\eta$. Of course the spontaneous matrix elements depend on the dwell time in the supercritical regime. This cannot be assessed when analyzing the matrix elements along the trajectory, but only by solving the coupled channel equations (14). In view of the resonance decay time ($\approx 10^{-19}$s), the differences for the spontaneous coupling strength of trajectories varying between $\eta = 1$ and 1.2 for elastic collisions is assumed to be small. To deal with the dynamic aspect of spontaneous positron emission, the occupation amplitudes $a_{1s\sigma,E_{e+}}(t)$, resp. its time derivative $\dot{a}_{1s\sigma,E_{e+}}(t)$ are considered, since $|a_{1s\sigma,E_{e+}}|^2$ delivers the dominant contribution to the emission rates as can be seen in Table 3. For two centre distances $R < R_{cr}$ the $1s\sigma$- state is replaced by $|\Phi_R>$. Fig. 28 shows the time derivative of $a_{1s\sigma,E_{e+}}(t)$ in the complex plane as a function of t according to equation (14) for the system $U^{92+} + U^{92+}$ (F=0). The time interval was chosen as $[-3.3 \cdot 10^{-21}s;\ 0\ ; 3.3 \cdot 10^{-21}s]$, which corresponds to an interval [100 fm; $R_{min}$ ;100 fm] for the two centre distance $R$. For distances $R \gtrsim 100$ fm the radial part of $\dot{a}_{1s\sigma,E_{e+}}(t)$ (first term in equ. (14)) starts to oscillate (suggestion of a circle in the middle and bottom right plots of Fig. 28). The 'radial components' exhibit extrema on the incoming and outgoing branch of the trajectory and run through zero at $R_{min}$, reflecting the corresponding matrix elements (c.f. Fig. 6). The 'spontaneous component' of $\dot{a}_{\Phi_R,E_{e+}}(t)$, i.e. the second term in equ. (14), is shown on the top right and bottom left position of Fig. 28. It starts at zero for $R < R_{cr}$ and has its extremum at $R_{min}$. Two sets of continuum energies $E_{e^+}$ were chosen: a) smaller energies for the rising edge of the differential emission spectra $dP_{e^+}/dE_{e^+}$ in Fig. 13 middle right, and b) energies around its maximum. The addition of the radial and spontaneous or potential coupling components reflect the complete result for $\dot{a}_{1s\sigma,E_{e+}}(t)$ on the left side of equation (14). For smaller energies $E_{e^+}$ the addition of the component caused by the potential coupling pushes the curves for radial couplings away from zero. In addition the plot on the middle left in Fig. 28 exhibits that the curves are twisted an exhibit a knot on the in- and outgoing branch of the trajectory. For higher energies $E_{e^+}$ the potential couplings also push the curves for radial couplings away from zero (at $R_{min}$ in the bottom right of Fig. 28) but without additional impact. Generally the component caused by the potential coupling is about an order of magnitude smaller than the radial one and decreases with higher energies $E_{e^+}$. Qualitatively the potential coupling component seems to have a stronger impact on the total values for $\dot{a}_{1s\sigma,E_{e+}}(t)$ at smaller energies compared to higher ones. Since an analysis of $|a_{1s\sigma,E_{e+}}|^2$ concerning contributions from radial and potential couplings is not feasible retrospectively after having solved the full set of coupled channel equations (14), another approximation is considered by direct numerical integration of $\dot{a}_{1s\sigma,E_{e+}}(t)$. This may be viewed as a 'quasi-perturbative' approach, because the multi channel dependencies for this component are omitted. Figs. 29 and 30 show the results of this approximation. The same time interval $[-3.3 \cdot 10^{-21}s;\ 0\ ; 3.3 \cdot 10^{-21}s]$ for integration was chosen as in Fig. 28, assuring to pick up all contributions from spontaneous couplings for values $R < R_{cr}$ and the dominant ones for the radial couplings. The resulting occupation amplitudes are denoted as $\bar{a}_{1s\sigma,E_{e+}}$, their real and imaginary parts are plotted for different trajectories of U + U with $R_{min}$ = 17.5 fm and $\eta$ = 1, 1.1, 1.2 and 1.5. The full symbols abbreviated e.g. by <1s|R|-1.3> stand for $-\sum_{k\neq j} a_{1s,k} < \varphi_{E_{e+}}|\frac{\partial}{\partial t}|\varphi_k > \ e^{i(\chi_{E_{e+}} - \chi_k)}$, with $E_{e^+}$ = -1.3 $m_e c^2$ and open symbols for <1s|H|-1.3> e.g. denote $-i\sum_{k\neq j} a_{1s,k} < \varphi_{E_{e+}}|\hat{H}_{TCD}|\varphi_k > \ e^{i(\chi_{E_{e+}} - \chi_k)}$. The energies $E_{e^+}$ are the same as in Fig. 28. In the top of Fig. 29 notable contributions of the potential coupling to $dP_{e^+}/dE_{e^+} \simeq |\bar{a}_{1s\sigma,E_{e+}}|^2$ at small energies $E_{e^+}$ are visible for $\eta$ = 1 and 1.1, which is in line with the consideration for matrix elements in Figs. 25 and 26. For increasing values of $\eta$ = 1.2 to 1.5 the contribution of spontaneous positrons decreases. Fig. 30 shows the same as before, but for energies $E_{e^+}$ located around the maximum of $dP_{e^+}/dE_{e^+}$. Real and imaginary parts of the potential coupling components (open symbols) are located around zero and the radial coupling components represented by the full symbols clearly dominate the result for $|\bar{a}_{1s\sigma,E_{e+}}|^2$. In this approximation spontaneous positron emission is expected to impact the positron spectra in elastic collisions mainly for small energies $E_{e^+}$ on the rising edge of the spectra, but only marginally in its maximum being dominated by radial coupling.

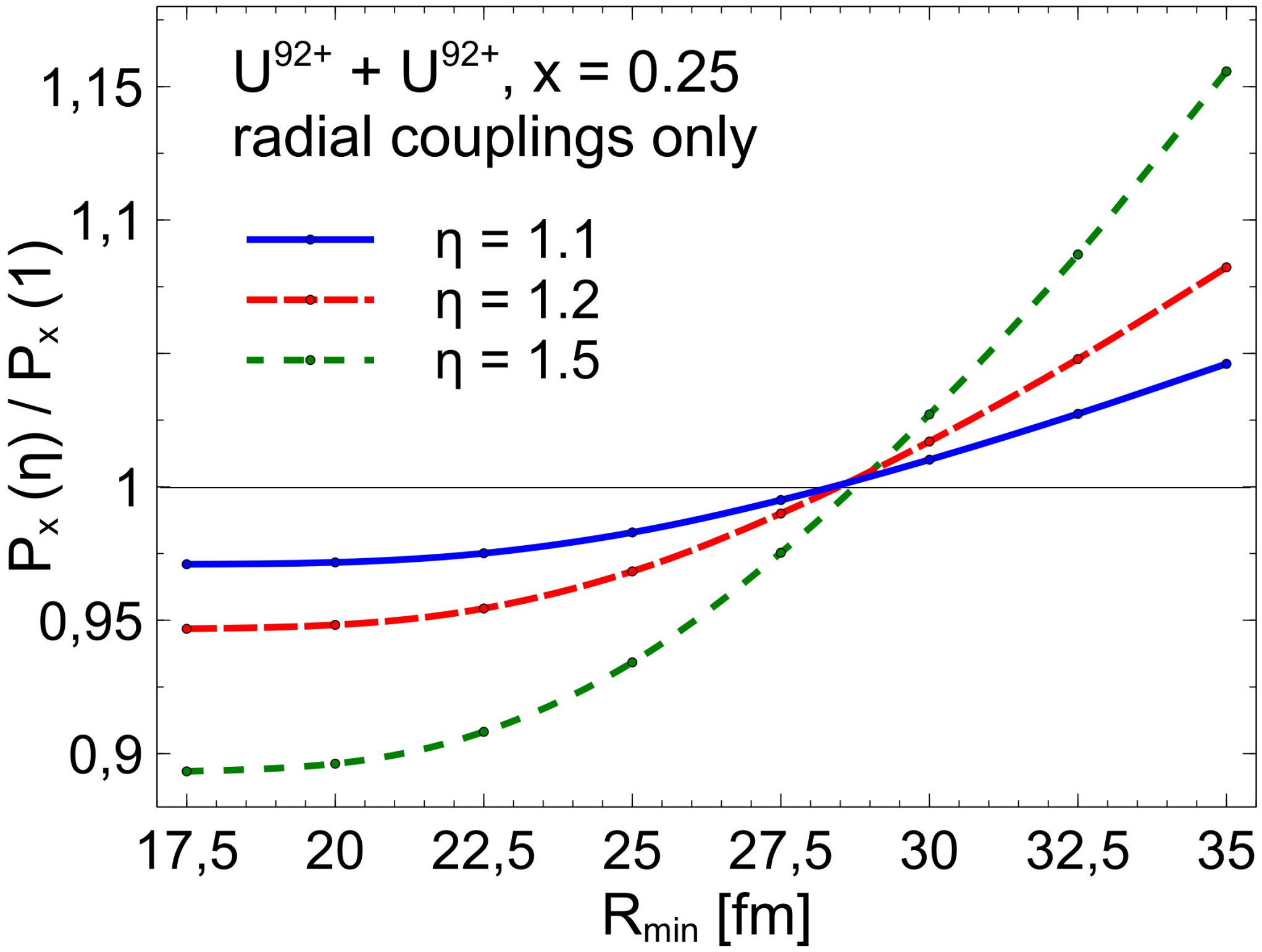

Figure 24: *Similar plot of ratios for emission probabilities like in Fig. 22 for different values of $\eta$, however, considering only radial couplings, i.e. the potential coupling was switched of, similarly to Fig. 23.*

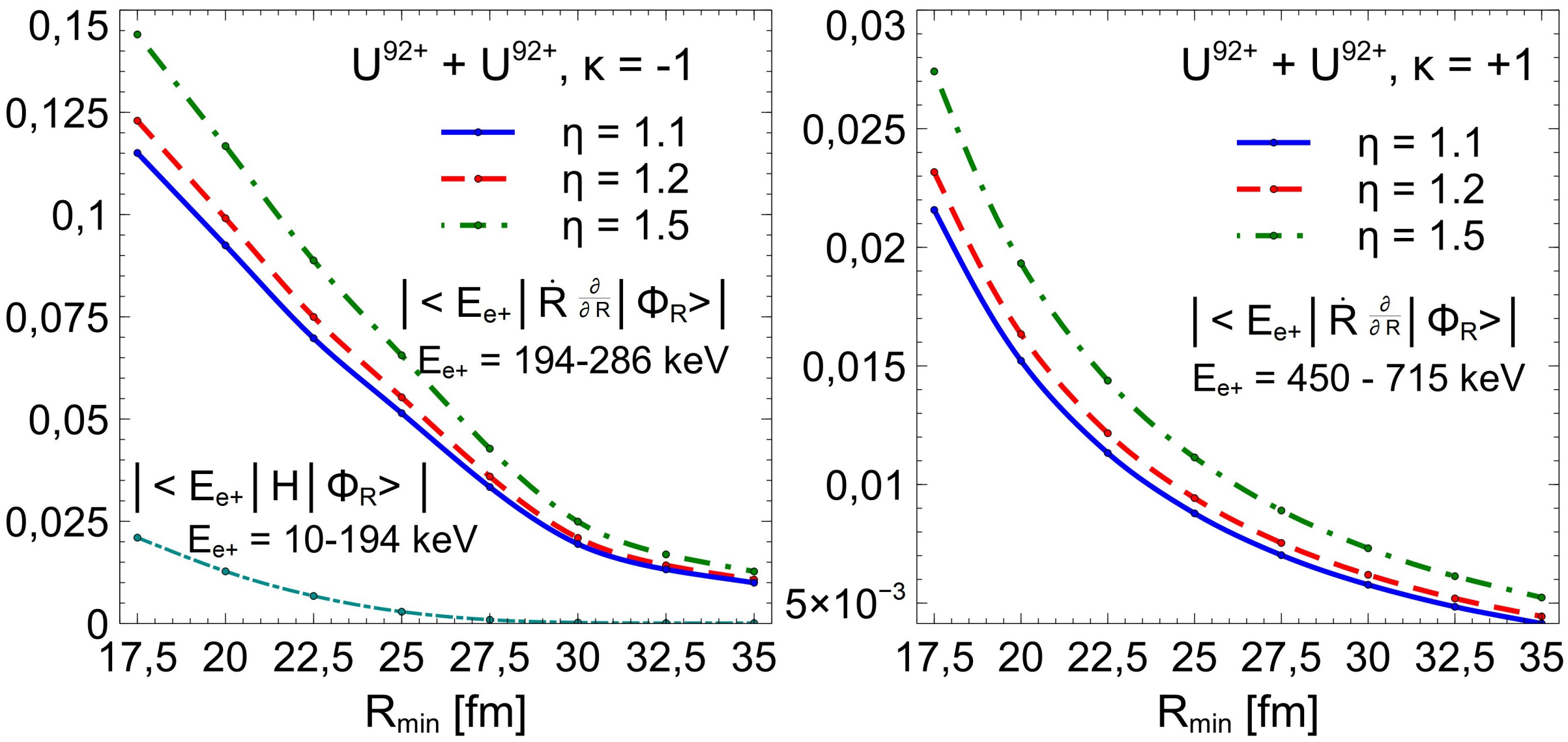

Figure 25: *Left: Analysis of transition matrix elements for radial and potential coupling ($\kappa = -1$) between continuum states in the negative Dirac continuum and the $1s\sigma$- state, resp. $\mid \Phi_R >$ during elastic Rutherford scattering. Only matrix elements with maximum absolute values at corresponding $R_{min}$ values are selected . Right: Same for p-states ($\kappa = +1$), where only radial couplings occur.*

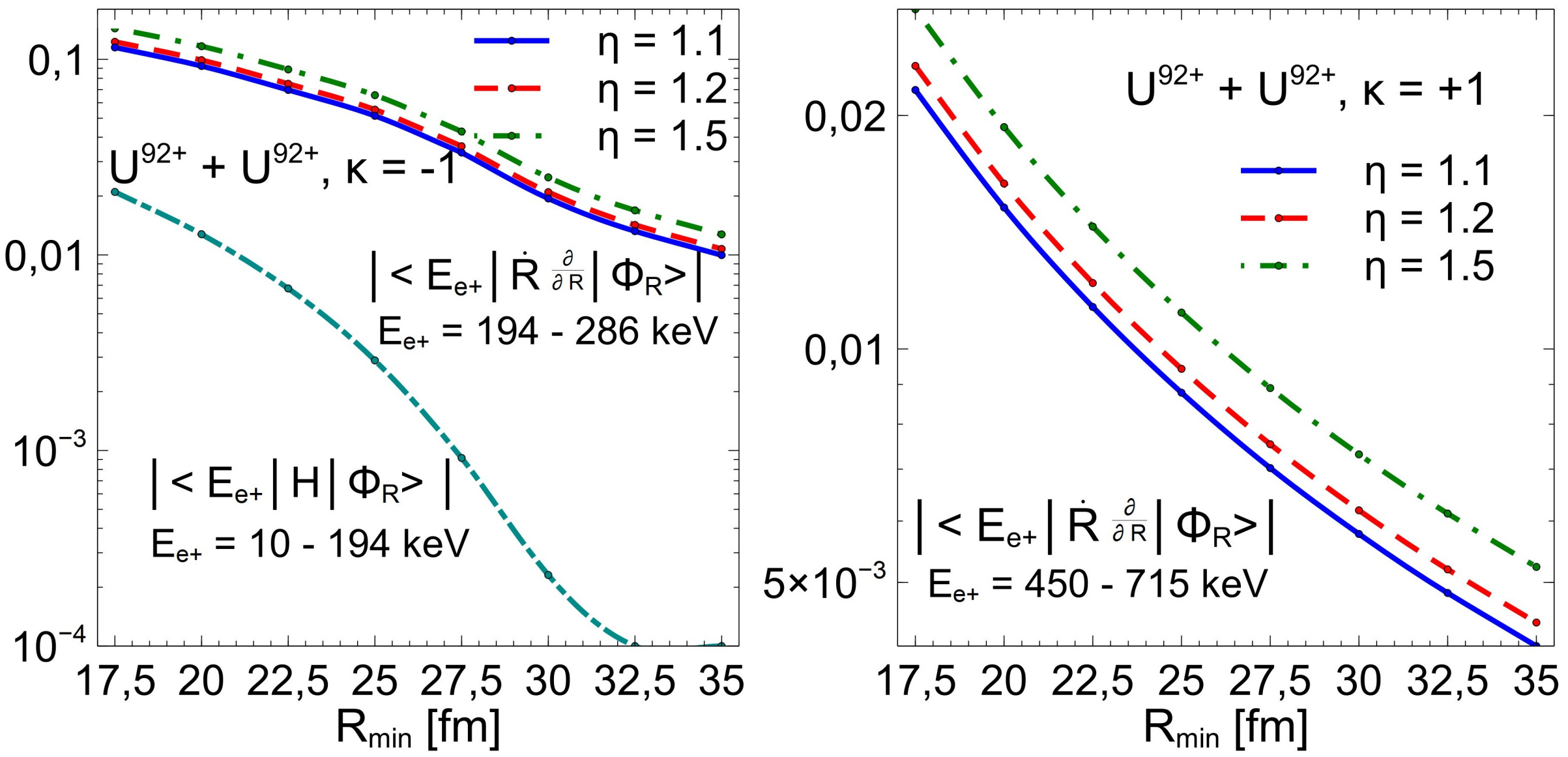

Figure 26: *Same as in Fig. 23 on a logarithmic scale, visualizing the steeper increase of the matrix elements for p-states and the difference in scale between radial and potential coupling for s-states.*

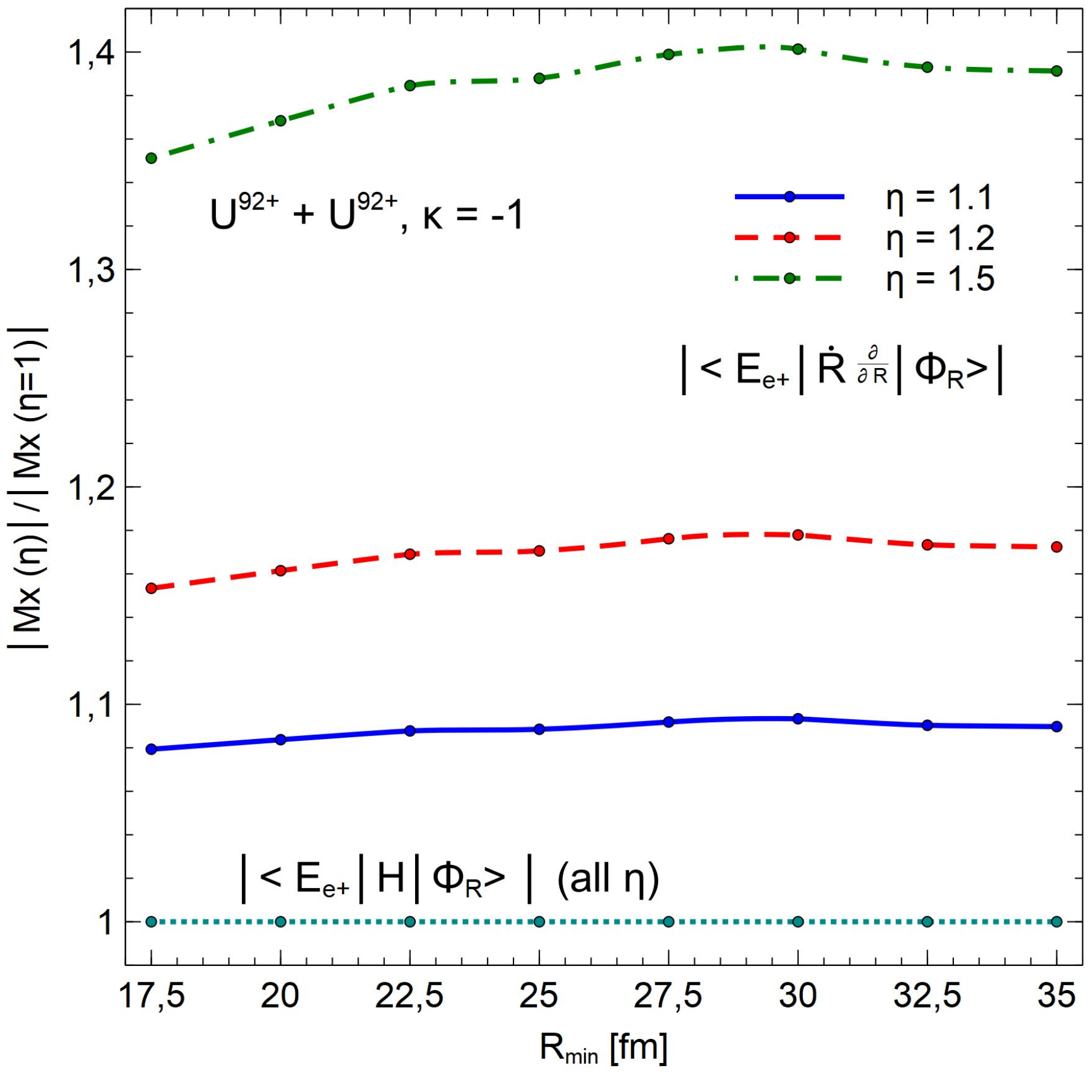

Figure 27: *Similarly to Fig. 22 the ratio of absolute values for matrix elements for $\eta$ >1 are divided by the result for central collisions of U+U for different kinematic parameters $R_{min}$ and $\eta$. The radial coupling matrix elements are dependent on the kinematical parameters, whereas the potential coupling is independent of it.*

| $E_{e^+}$ [keV] | $dP_{e^+}/dE_{e^+}[\frac{1}{keV}]$ | $\lvert a_{1s\sigma,E_{e^+}}\rvert^2$ | $\sum_{n=2}^{8} \lvert a_{ns\sigma,E_{e^+}}\rvert^2$ | $\int \lvert a_{E_{e^-},E_{e^+}}\rvert^2 \, dE_{e^-}$ |
|---|---|---|---|---|
| 337 | $1.191 \cdot 10^{-5}$ | $1.147 \cdot 10^{-5}$ | $3.061 \cdot 10^{-7}$ | $1.337 \cdot 10^{-7}$ |
| 393 | $1.352 \cdot 10^{-5}$ | $1.303 \cdot 10^{-5}$ | $3.448 \cdot 10^{-7}$ | $1.507 \cdot 10^{-7}$ |
| 450 | $1.410 \cdot 10^{-5}$ | $1.359 \cdot 10^{-5}$ | $3.570 \cdot 10^{-7}$ | $1.561 \cdot 10^{-7}$ |
| 511 | $1.384 \cdot 10^{-5}$ | $1.334 \cdot 10^{-5}$ | $3.483 \cdot 10^{-7}$ | $1.525 \cdot 10^{-7}$ |
| 572 | $1.303 \cdot 10^{-5}$ | $1.256 \cdot 10^{-5}$ | $3.268 \cdot 10^{-7}$ | $1.433 \cdot 10^{-7}$ |

Table 3: *Differential emission probability $dP_{e^+}/dE_{e^+}$ for fully ionized $U^{92+} + U^{92+}$- collisions at $E_{Lab} \simeq$ 5.85 MeV/n (2a = 17.5 fm). The composition of the emission probability according to equation (17) in column 2 for 5 different kinetic positron energies $E_{e^+}$ (column 1) around the maximum of the spectrum in Fig. 13 ($\eta$ = 1, $\kappa$ = –1)) by its components in columns 3-5 is shown. The dominant component is $|a_{1s\sigma,E_{e^+}}|^2$.*

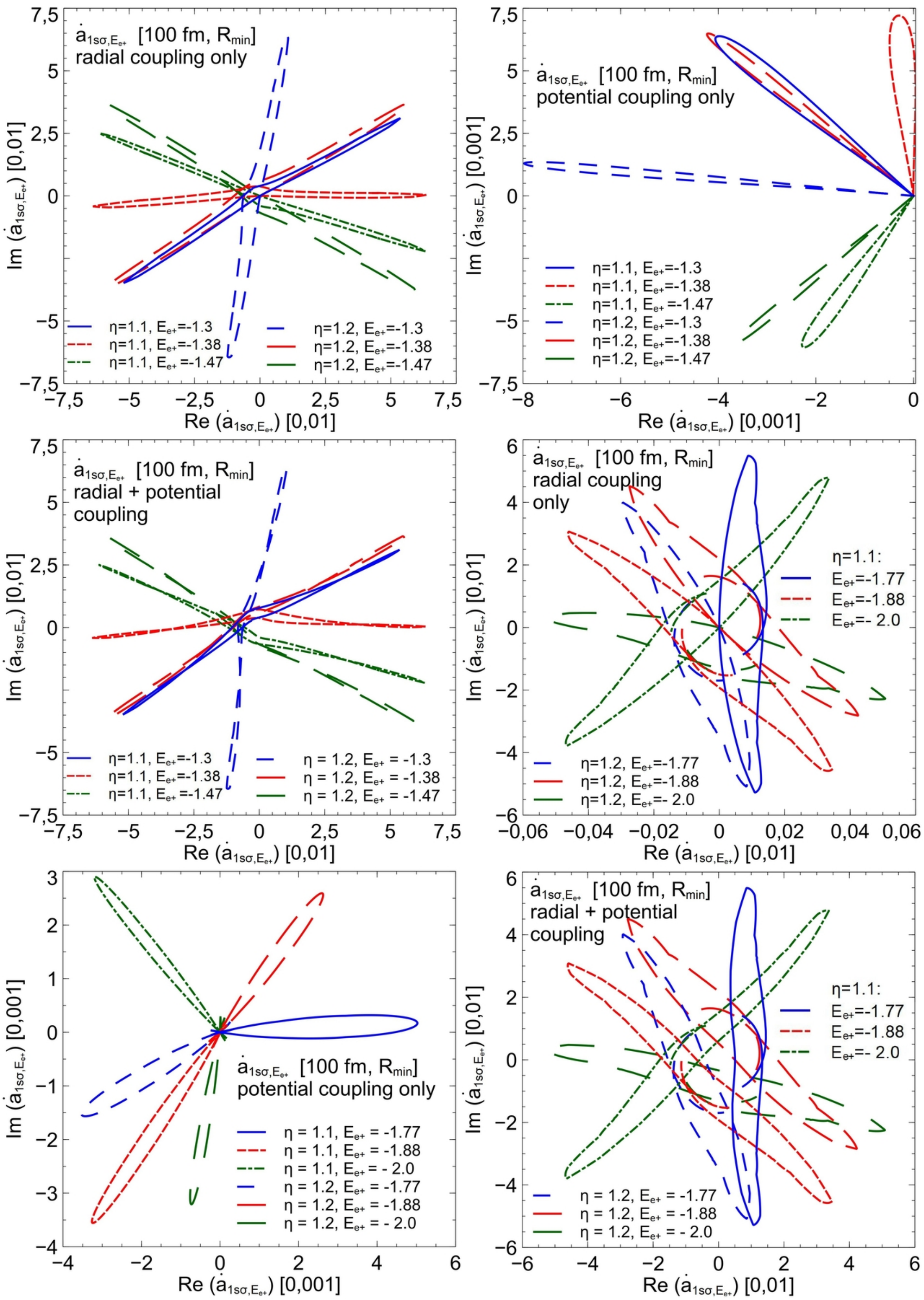

Figure 28: *Top left: Time derivative of* $a_{1s\sigma,E_{e^+}}(t)$ *for the system* $U^{92+} + U^{92+}$ *in collisions with* $R_{min}$ = *17.5 fm (*$F = 0$*), considering only radial coupling for small positron energies. Top right: Same as before considering only potential or spontaneous coupling. Middle left: Complete time derivative of* $a_{1s\sigma,E_{e^+}}(t)$ *(sum of upper figures). Middle right, bottom left and right: Same as above for positron energies* $E_{e^+}$ *located around the maximum in Fig. 13.*

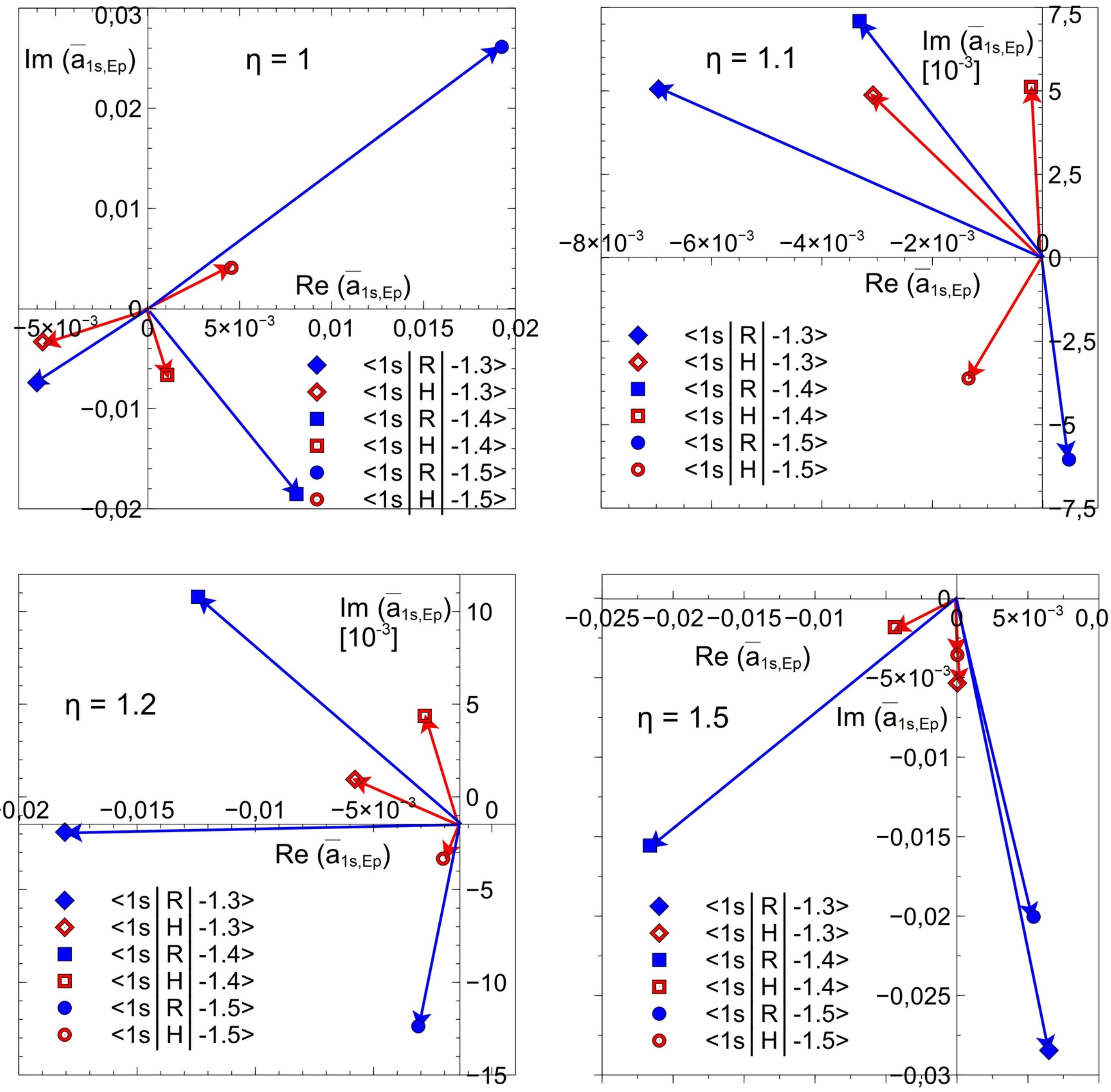

Figure 29: *Approximated occupation amplitudes $\bar{a}_{1s\sigma,E_{e+}}$ for small positron energies $E_{e^+}$ in $U^{92+} + U^{92+}$ collisions at $R_{min} = 17.5$ fm and different parameters of $\eta = 1$, 1.1, 1.3 and 1.5. Open symbols: Contribution of potential coupling, full symbols: Contribution of radial coupling.*

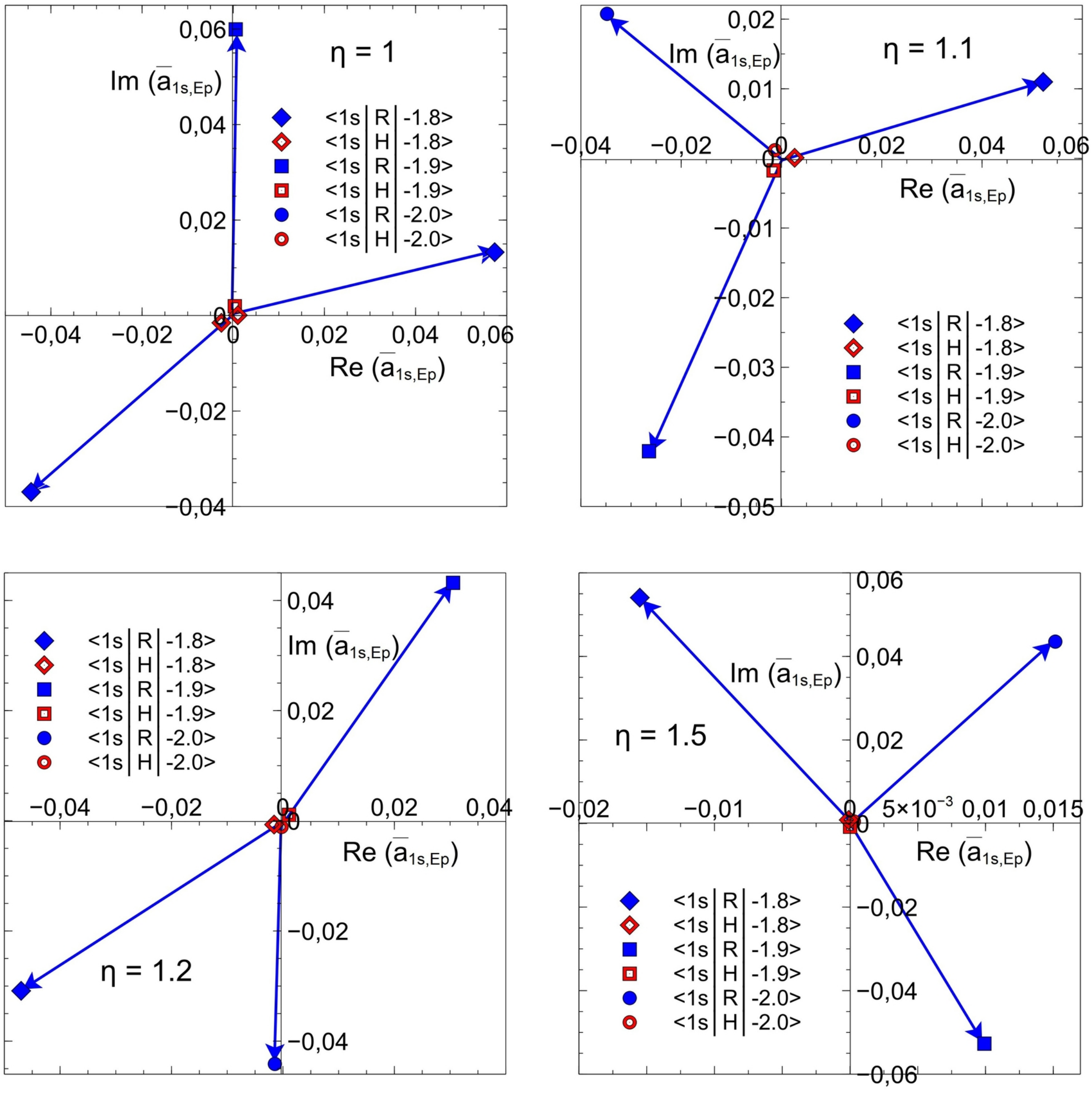

Figure 30: *Same as in Fig. 29, but for positron energies $E_{e^+}$ located around the maximum of the spectrum in Fig. 13.*

### 5.2 Slope of $P_y(\eta)/P_y(1)$ as possible fingerprint of spontaneous positrons

The previous analysis confirms the advantage to study partial spectra as in Fig. 22 (right side), when focusing on spontaneous positron emission in supercritical elastic collisions: Their contribution to the total spectrum is expected to be most pronounced for smaller energies $E_{e^+}$ on the rising edge in the spectrum $dP_{e^+}/dE_{e^+}$ before the maximum. Similar to conventional collisions, induced positron emission is also the dominant channel in the case of bare nuclei and complicates the detection of spontaneous positrons. The increase in the maxima of $dP_{e^+}/dE_{e^+}$ for supercritical collision systems in Figs. 9, 10, 13 (right side) and 14 for decreasing $\eta$- values $\rightarrow$ 1 plus the reversal of the slope for the total positron or pair-creation rates $P_{e^+}$ versus $\eta$ in Figs. 11 (right side) or 15 as discussed in [2] could be interpreted as signal for spontaneous positron emission at first sight. The persistence of this effect in calculations for supercritical collisions, considering only radial couplings in Figs. 23 and 24 in combination with their dominance in coupling strength plus the shortness of the collision time in elastic collisions, suggest that it is not caused uniquely by spontaneous positron emission, but by both mechanisms: induced plus spontaneous positron emission.

Since the previous analysis suggests to focus on smaller positron energies $E_{e^+}$ on the rising edge of the spectra before the maximum, when focusing on spontaneous positrons, the rations $P_x(\eta)/P_x(1)$ are revisited: It seems obvious to investigate partial total positron emission probabilities $P_y$, but restrict the integration of $dP_{e^+}/dE_{e^+}$ in the spectra to energies $E_{e^+} < dP_{e^+}/dE_{e^+} = (1-\text{y})\ (dP_{e^+}/dE_{e^+})_{max}$, however, *left* of the maximum in $dP_{e^+}/dE_{e^+}$for $\eta = 1$, as shown at the right side of Fig. 31.

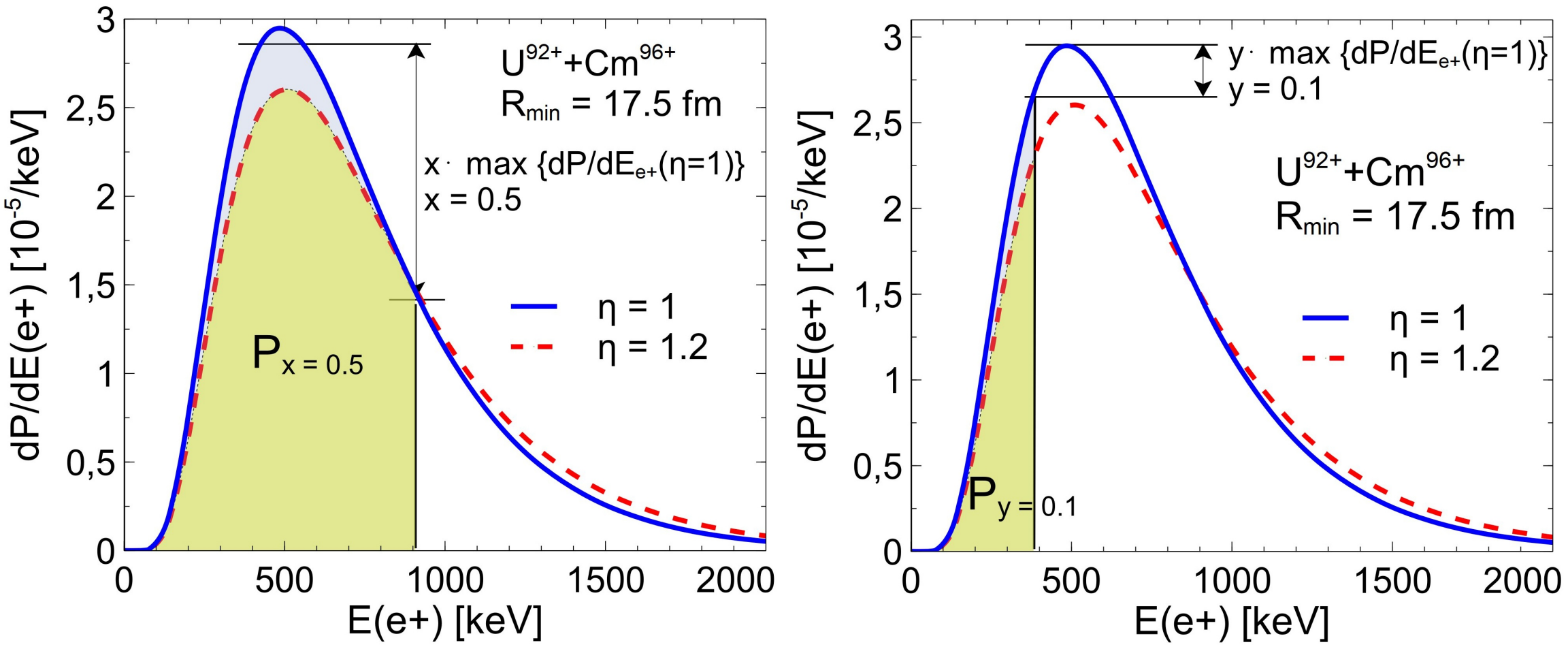

Figure 31: *Left: Definition of partial pair-creation probabilities $P_x$ according to [2], where integration of the $dP_{e^+}/dE_{e^+}$ spectra is restricted to energies $E_{e^+} < dP_{e^+}/dE_{e^+} = (1-x)\ (dP_{e^+}/dE_{e^+})_{max}$ right of the maximum in $dP_{e^+}/dE_{e^+}$for $\eta = 1$. Right: Similarly $P_y$ is defined as integration of $dP_{e^+}/dE_{e^+}$ restricted to energies $E_{e^+} < dP_{e^+}/dE_{e^+} = (1-y)\ (dP_{e^+}/dE_{e^+})_{max}$ left of the maximum in $dP_{e^+}/dE_{e^+}$for $\eta = 1$.*

Calculations of the rations $P_x(\eta)/P_x(1)$ with x = 0.5 and $P_y(\eta)/P_y(1)$ with y = 0.25 for $U^{92+} + U^{92+}$ for different combinations of bombarding energies and impact parameters corresponding to $\eta = 1.1$, 1.2 and 1.5 versus $R_{min}$ are shown in Fig. 32. The left column reflects exact calculations, whereas spontaneous positron emission is switched off in the phenomenological approximation in the right column of the figure. Striking is the change of slope in the bottom right figure (y = 0.25, radial coupling only) for values $R_{min} < R_{min-s} \approx 24$ fm, if no spontaneous positron emission would be present. This effect is once more shown in Fig. 33 in direct comparison between exact calculations and those considering radial coupling only. There is little change, when varying y from 0.1 to 0.25 in the middle and bottom of Fig. 33.

The sign reversal of $P_y(\eta)/P_y(1)$ for decreasing values of $R_{min}$ below $R_{min-s}$ also occurs in the system $U^{92+} + Cm^{96+}$ for $R_{min} < R_{min-s} \approx 30$ fm and values of $\eta = 1.1$ and 1.2 in Fig. 34 when neglecting spontaneous positron emission. The effect is even stronger pronounced and happens at higher values of $R_{min}$, since the critical distance $R_{cr}$ for this collision system is about 40 fm in contrast to $R_{cr} \approx 32.6$ for $U^{92+} + U^{92+}$.

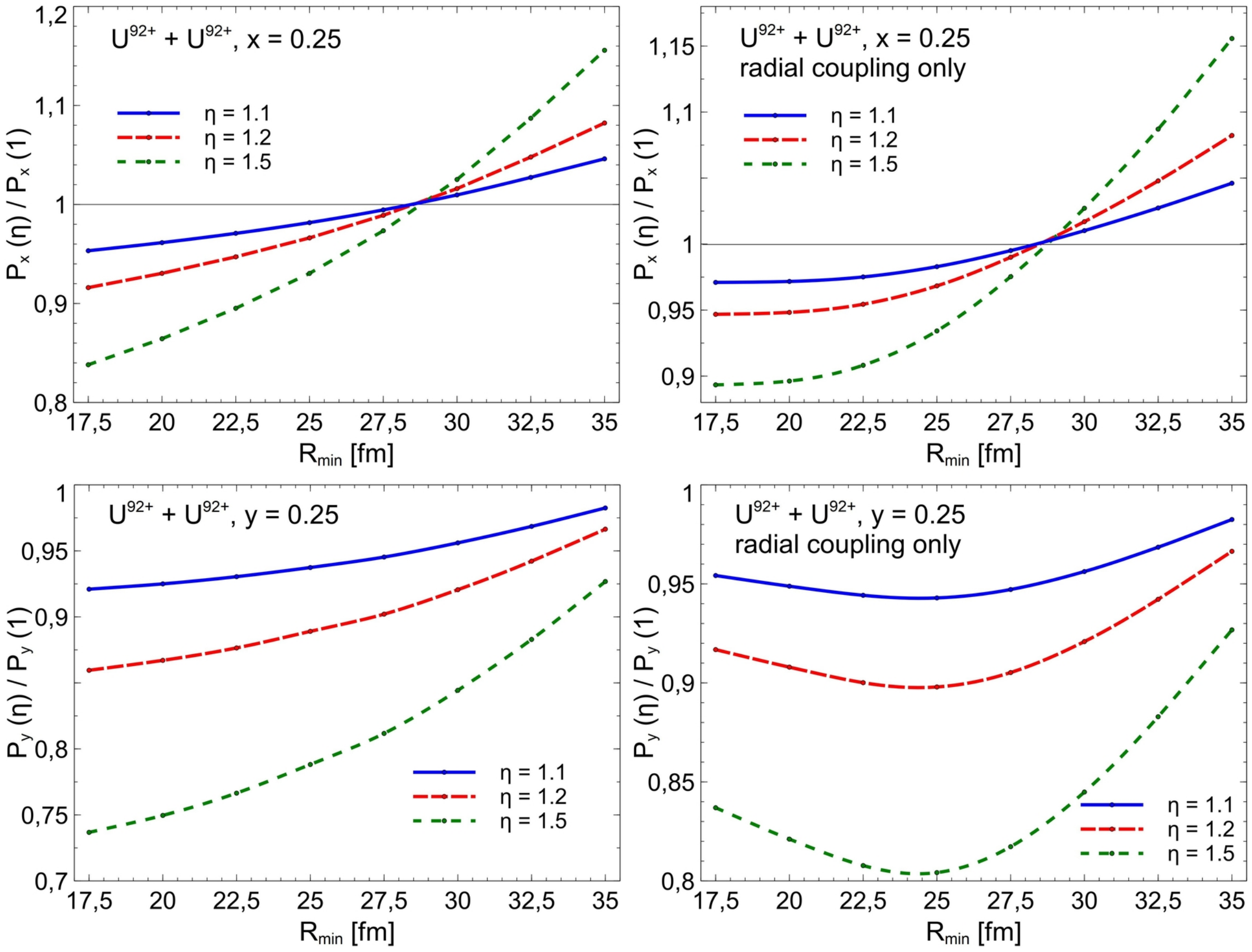

Figure 32: *Ratios $P_x$ and $P_y$ for $U^{92+} + U^{92+}$ collisions versus $R_{min}$ at different values of $\eta$. Left column: Full (exact) calculations, right column: Phenomenological approximations, where only radial coupling is considered.*

The results suggest a possible observability of spontaneous positrons varying y from 0.1 to 0.25 in the middle and bottom of Fig. 33.

The sign reversal of $P_y(\eta)/P_y(1)$ for decreasing values of $R_{min}$ below $R_{min-s}$ also occurs in the system $U^{92+} + Cm^{96+}$ for $R_{min} < R_{min-s} \approx 30$ fm and values of $\eta = 1.1$ and 1.2 in Fig. 34 when neglecting spontaneous positron emission. The effect is even stronger pronounced and happens at higher values of $R_{min}$ since the critical distance $R_{cr}$ for this collision system is about 40 fm in contrast to $R_{cr} \approx 32.6$ for $U^{92+} + U^{92+}$.

The results suggest a possible observability of spontaneous positron emission in elastic collisions of bare heavy ions by analysis of the ratios $P_x(\eta)/P_x(1)$, resp. $P_y(\eta)/P_y(1)$ versus $R_{min}$ in the following sense: If a sign reversal occurs in the slope of $P_y(\eta)/P_y(1)$ for decreasing values of $R_{min}$ below $R_{min-s}$ , or the slope tends towards zero at decreasing values of $P_x(\eta)/P_x(1)$ in the same range of $R_{min}$ , existence of spontaneous positron emission is considered to be unlikely. Conversely, agreement between ratios $P_x(\eta)/P_x(1)$ and $P_y(\eta)/P_y(1)$ obtained from experimental data with results of the exact calculations for values $R_{min} < R_{min-s} \simeq 0.75\ R_{cr}$ could be considered as indirect proof of spontaneous positron emission.

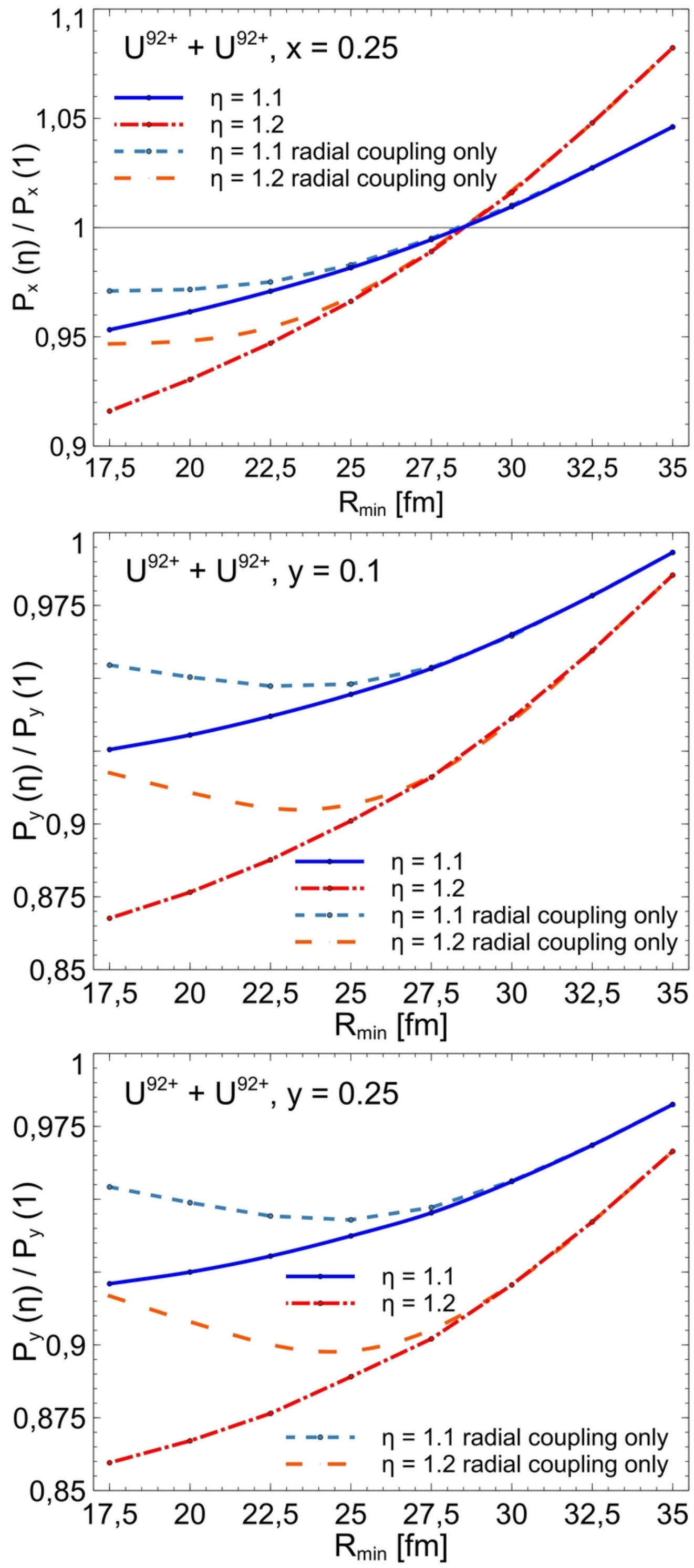

Figure 33: *Ratios $P_x$ and $P_y$ for $U^{92+} + U^{92+}$ collisions as a function of $R_{min}$ for exact calculations and approximations considering radial coupling only in direct comparison for $\eta = 1.1$ and 1.2. The derivative of the ratio $P_y$ changes its sign for $R_{min} < 22.5$ fm, when considering only radial coupling.*

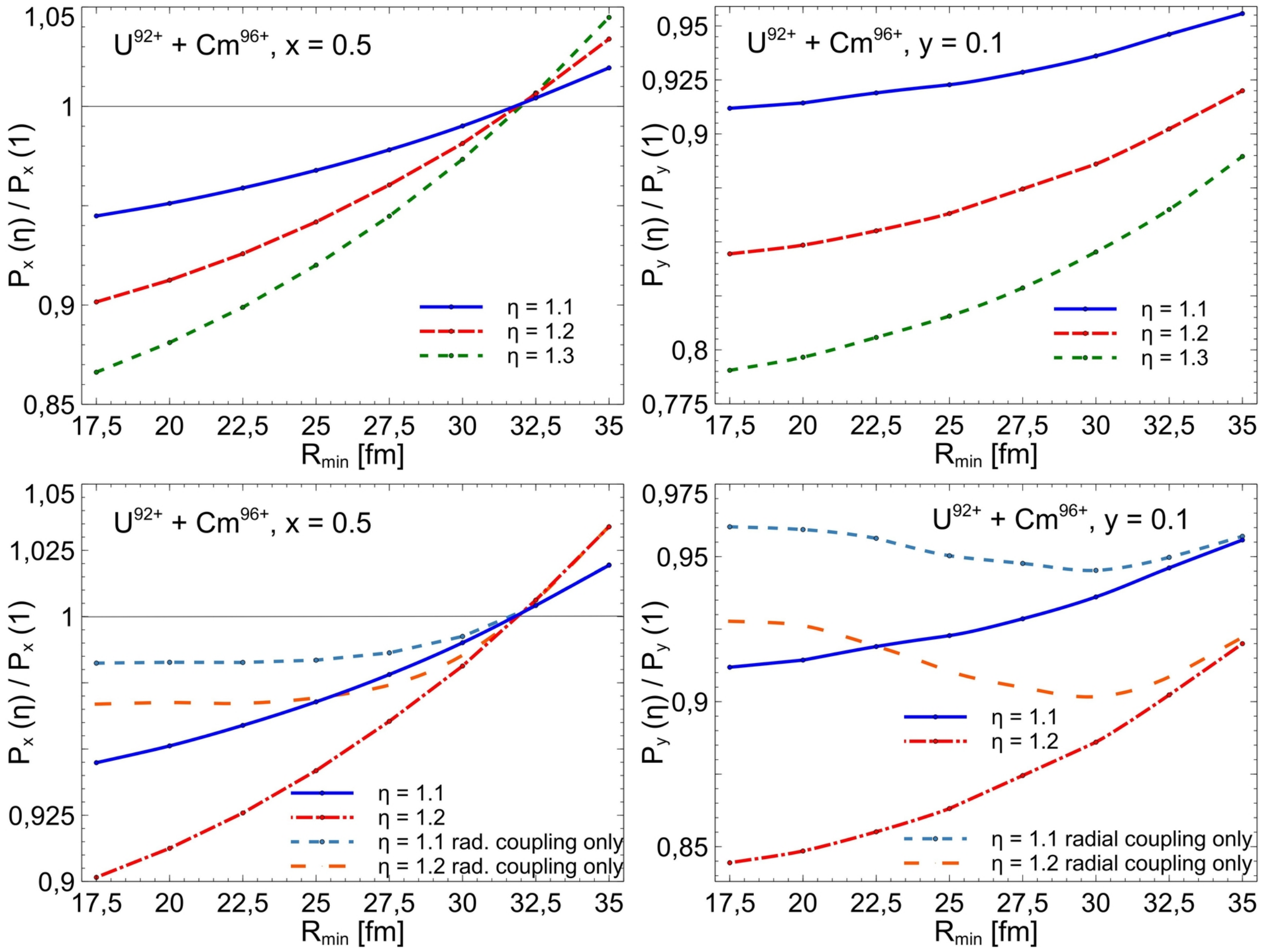

Figure 34: *Top: Ratios $P_x$ and $P_y$ for $U^{92+} + Cm^{96+}$ collisions versus $R_{min}$ for different values of $\eta$ = 1.1, 1.2, 1.3. Bottom: Comparison with results of phenomenological approximation, considering radial coupling only for $\eta$ = 1.1 and 1.2. When neglecting spontaneous coupling, the derivative of $P_y$ changes its sign for $R_{min} < 30$ fm.*

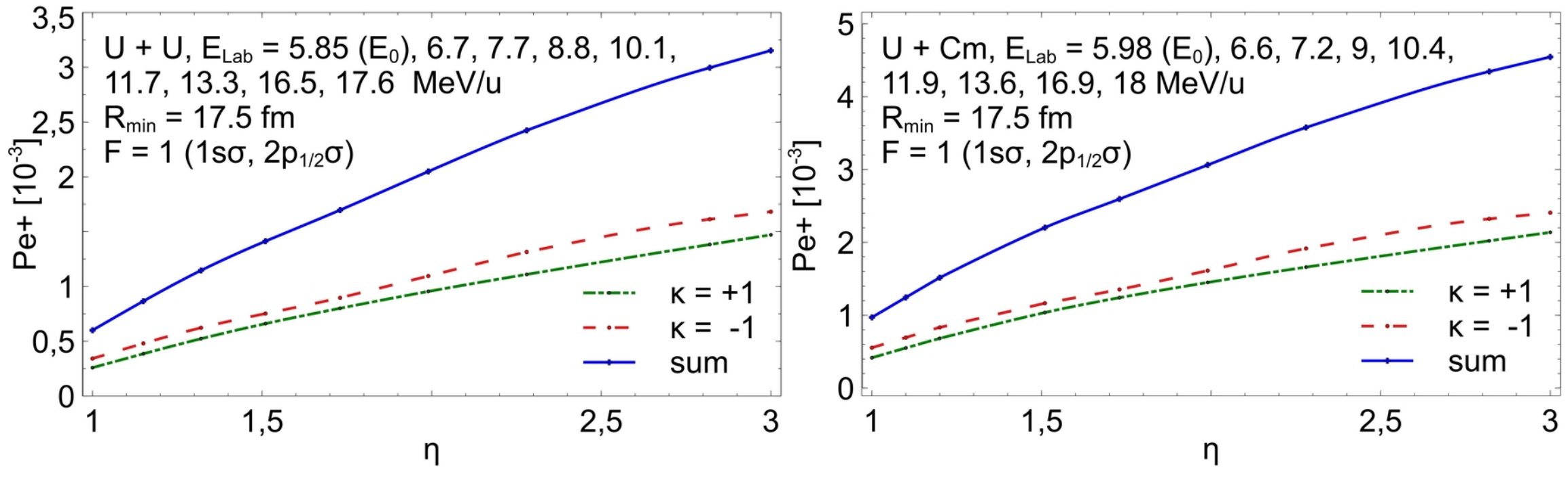

Figure 35: *Total positron emission rates $P_{e^+}$ for the systems U+U (Z = 184) and U+Cm (Z = 188) with a Fermi level F = 1, i.e. only the $1s\sigma$- and $2p_{\frac{1}{2}}$- levels are occupied with electrons. In contrast to Fig. 11 the slope of $dP_{e^+}/d\eta$ remains positive also for values of $\eta \to 1$.*

This indirect proof, however, would only be measurable in collisions of fully ionized heavy ions as Fig. 35 shows: The increase in the pair-production probability at $E/E_0 \to 1$ for a given $R_{min}$ like in Figs. 9 and 10 vanishes as soon as the $1s\sigma$- and $2p_{\frac{1}{2}}\sigma$- levels in U + U collisions are occupied (denoted by a Fermi level F = 1). The effects predicted in Fig. 33 are minimized as Fig. 36 shows, when choosing a Fermi level of F = 1. For this Fermi level the differences between ratios $P_x$ and $P_y$ in U + Cm collisions versus $R_{min}$ in comparison between exact calculations and phenomenological approximations, considering radial coupling only, are still visible, but notably smaller, as the top of Fig. 37 shows. The bottom of Fig. 37 displays the situation for a Fermi level of F = 3 (indicating the levels $1s\sigma$-$3s\sigma$ and $2p_{\frac{1}{2}}\sigma$- $4p_{\frac{1}{2}}\sigma$ to be occupied). This calculation would be used, when bombarding an ionized U- projectile onto a neutral Cm- target. Considering electron screening effects within the adiabatic time dependent Hartree-Fock-approach would a) reduce the critical distance $R_{cr}$ from $\simeq 40$ fm to $\simeq 33$ fm and b) reduce the binding energy of the $1s\sigma$- state by about 120 keV at that point. This means that in the scenario of bombarding a fully stripped U- projectile onto a (neutral) Cm- target, detection of spontaneous positrons in elastic collisions is deemed to be not possible.

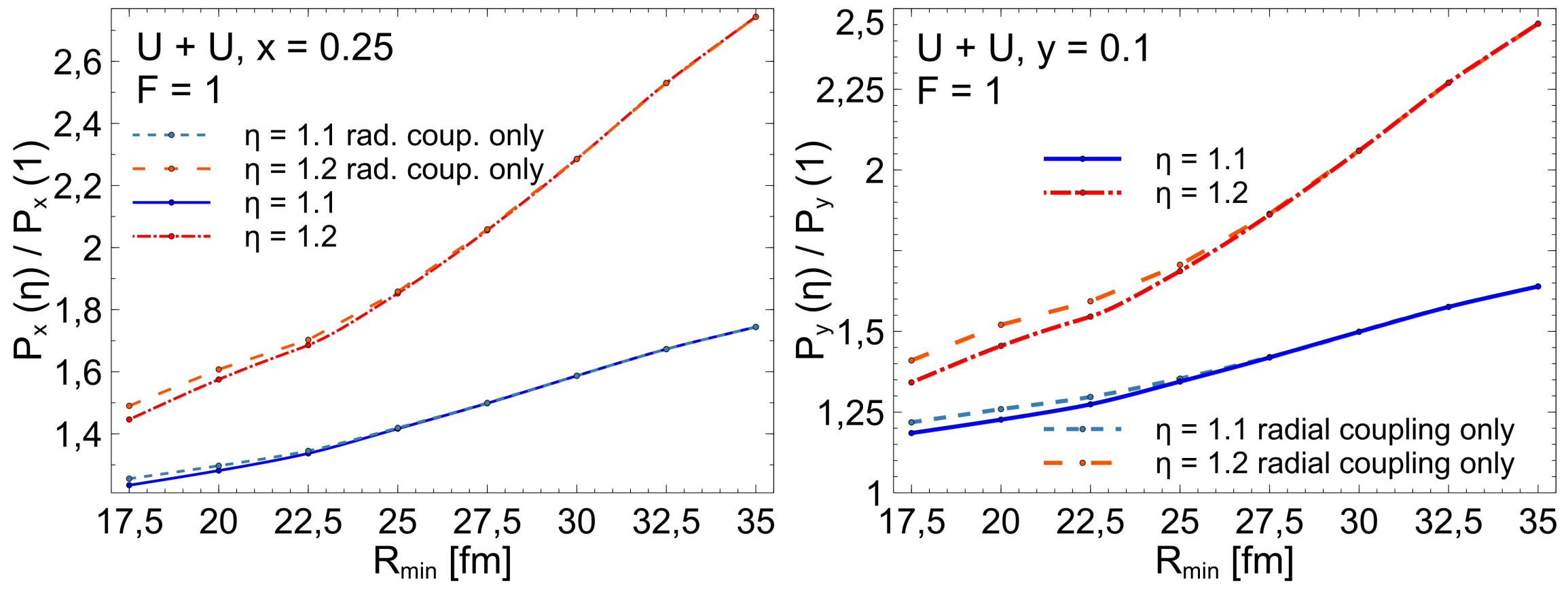

Figure 36: *Ratios $P_x$ and $P_y$ for U + U- collisions as a function of $R_{min}$ for exact calculations and approximations considering radial coupling only in direct comparison for $\eta$ = 1.1 and 1.2. A Fermi level F = 1 is chosen similar to Fig. 35. In contrast to Fig. 33 no significant differences occur between the curves of exact calculations and the approximations considering radial coupling only.*

The calculations using the ratios $P_y(\eta)/P_y(1)$ in comparison with results where the second term in equation (14) is switched off, however, have two shortcomings: Omission of the term $-i\sum_{k\neq j} a_{ik} < \varphi_j|\hat{H}_{TCD}|\varphi_k > \ e^{i(\chi_j-\chi_k)}$ does not yield exact results, but is a phenomenological approximation. An improved ansatz to evaluate the effects of both coupling types on the final coupled channel results for the occupation amplitudes beyond the approximations in this paper would be helpful to assess the validity of the presented results and its implications. Secondly using the ratios $P_y(\eta)/P_y(1)$ for small positron energies decreases the statistics concerning data evaluation obtained from counting rates in experiments and thus will be difficult to measure. Although the emission rates in collisions of fully ionized atoms increase significantly, this complicates an experimental proof.

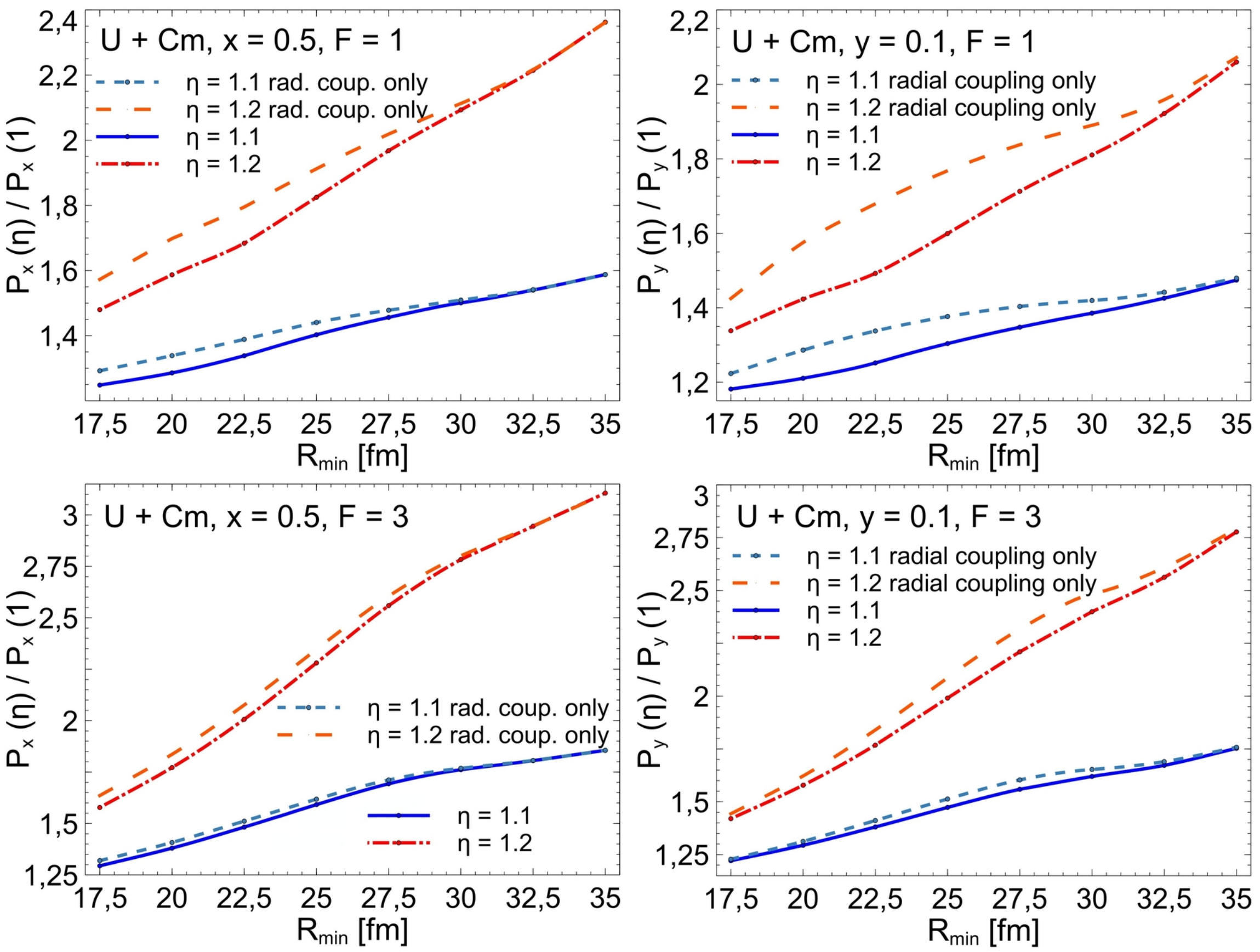

Figure 37: *Top: Ratios $P_x$ and $P_y$ for U + Cm collisions versus $R_{min}$ for $\eta = 1.1$ and 1.2, choosing a Fermi level $F = 1$ in comparison between exact calculations and phenomenological approximations, considering radial coupling only. Bottom: Same as above, however, choosing a Fermi level of $F = 3$.*

## 6 Conclusion

In elastic heavy ion collisions with a combined nuclear charges $> 173$ a continuous transition from the sub- to the supercritical regime takes place, where spontaneous positron emission contributes to the positron spectra. In [1] the idea was developed, to study positron emission rates for elastic heavy ion collisions in dependence on the kinematic parameter $\eta = E_0/E$. The dwell time in the supercritical regime of the quasi-molecule formed in elastic collisions of bare nuclei depends inversely on $\eta$. According to [2] the dependence of the positron emission spectra on $\eta$, i.e. any increase in the pair-production probability at $E/E_0 \rightarrow 1$ for a given $R_{min}$, should indicate the effect of spontaneous positron emission.

The calculations for positron spectra and positron emission rates resulting from elastic collisions of bare nuclei in sub- and supercritical collisions in this paper independently confirm the results achieved by [1, 2], i.e. are in good agreement.

The difference between spectra, emission rates and cross sections for positrons in collisions between a charged projectile onto a neutral target versus fully ionized collision partners is demonstrated for the systems U+Cm and Cm+Cm: The cross sections in collisions of bare nuclei increase by factors of 57 (U+Cm) and 72 (Cm+Cm). In addition emission spectra for $s\sigma$- states show a drastic increase versus those of $p_{\frac{1}{2}}\sigma$ -states. Beyond that the transition of electrons from the negative Dirac continuum into the vacant $1s\sigma$- state delivers the dominant contribution for pair creation in collisions of fully ionized heavy ions.

For an analysis of the contribution of spontaneous positrons in elastic collisions of bare nuclei the strongest contributing matrix elements were investigated, plus approximations for the dominant transition amplitude $|\bar{a}_{1s\sigma,E_{e+}}|^2$ for pair creation. The results suggest a main contribution of spontaneous positrons at small kinetic energies (at $\simeq 230$ keV in U+U) to the spectra but not to its maximum.

For this reason partial spectra $P_x$ and ratios of partial spectra $P_x(\eta)/P_x(1)$ being introduced in [2] are calculated, however, also for small positron energies (denoted as $P_y$). The results of calculations for ratios of partial spectra $P_y(\eta)/P_y(1)$ in Figs. 32-34 show that its slope as a function of $R_{min}$ for values of $R_{min} \lesssim 0.75\ R_{cr}$ could be a fingerprint of spontaneous positron emission. The increase in the pair-production probability at $E/E_0 \rightarrow 1$ for a given $R_{min}$ alone seems to be no sufficient condition to indicate the effect of spontaneous positron emission according to Fig. 23. This increase vanishes as soon as the inner $1s\sigma$- and $2p_{\frac{1}{2}}\sigma$- shells are occupied as Fig. 35 shows.

To detect spontaneous positron emission directly, it seems inevitable to reduce the effect of induced or dynamical positron emission. This leads us again to the proposed scenarios of nuclear sticking [38, 75] or inelastic collisions [39]. Both scenarios switch off ($\dot{R} = 0$ for nuclear sticking) or drastically reduce (small values of $\dot{R}$ for inelastic collisions) the induced or radial couplings (c.f. equ. 21). The effects of nuclear sticking would theoretically cause a pronounced spike in the positron spectrum as in Fig. 38, indicating directly spontaneous positron emission.

The fact, that nuclear sticking in collisions of heavy ions with combined nuclear charges

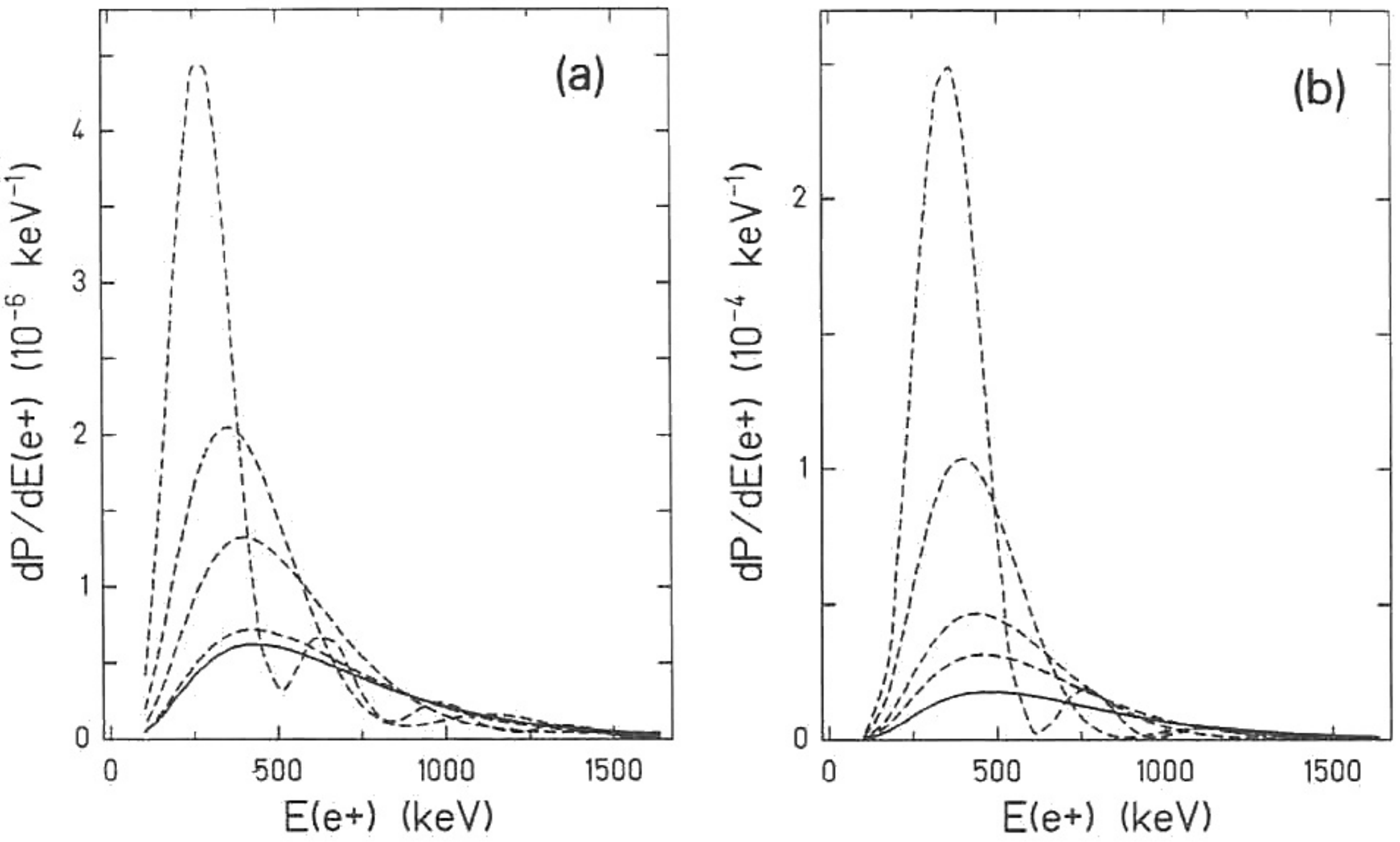

Figure 38: *Positron spectra from central U + U collisions (b = 0) at $E_{c.m.}$ = 740 MeV for elastic Rutherford scattering (solid line) and nuclear time delays of T = 1, 2, 5 and 10 $\cdot 10^{-21}$s (dashed lines) [75]. a) Results for conventional scattering (F=3) and b) for bare nuclei (F=0).*

$>$ 173 has not been detected up to now [30–33], does not mean that it does not exist. It may be, that the share of such reactions is much smaller than expected and therefore could not be detected up to now [87, 88]. Detection of positrons in coincidence with Pb-like fragments might be promising according to this reference, since it could reduce the dynamical or induced positrons by factors up to 1000, so that the spike-like structure indicating spontaneous positron emission could be detectable. In addition spectra from collisions of bare nuclei will significantly increase counting rates and support experimental detection.

Furthermore also inelastic collisions offer a chance to detect signals for spontaneous positron emission (c.f. Fig. 39) by modification of the spectra being analyzed within a small impact parameter range of b [39]. The short time delay in inelastic collisions theoretically causes the effect, that the spectrum shows a significant narrower emission maximum for small impact parameters ($b \leq 4fm$) plus a steeper decrease for positron energies in the range $E_{e^+} = 400 - 800$ keV. Especially the curves for impact parameters $b = 0$, 2 and 4 fm, where comparatively long reaction times occur, contain a considerable

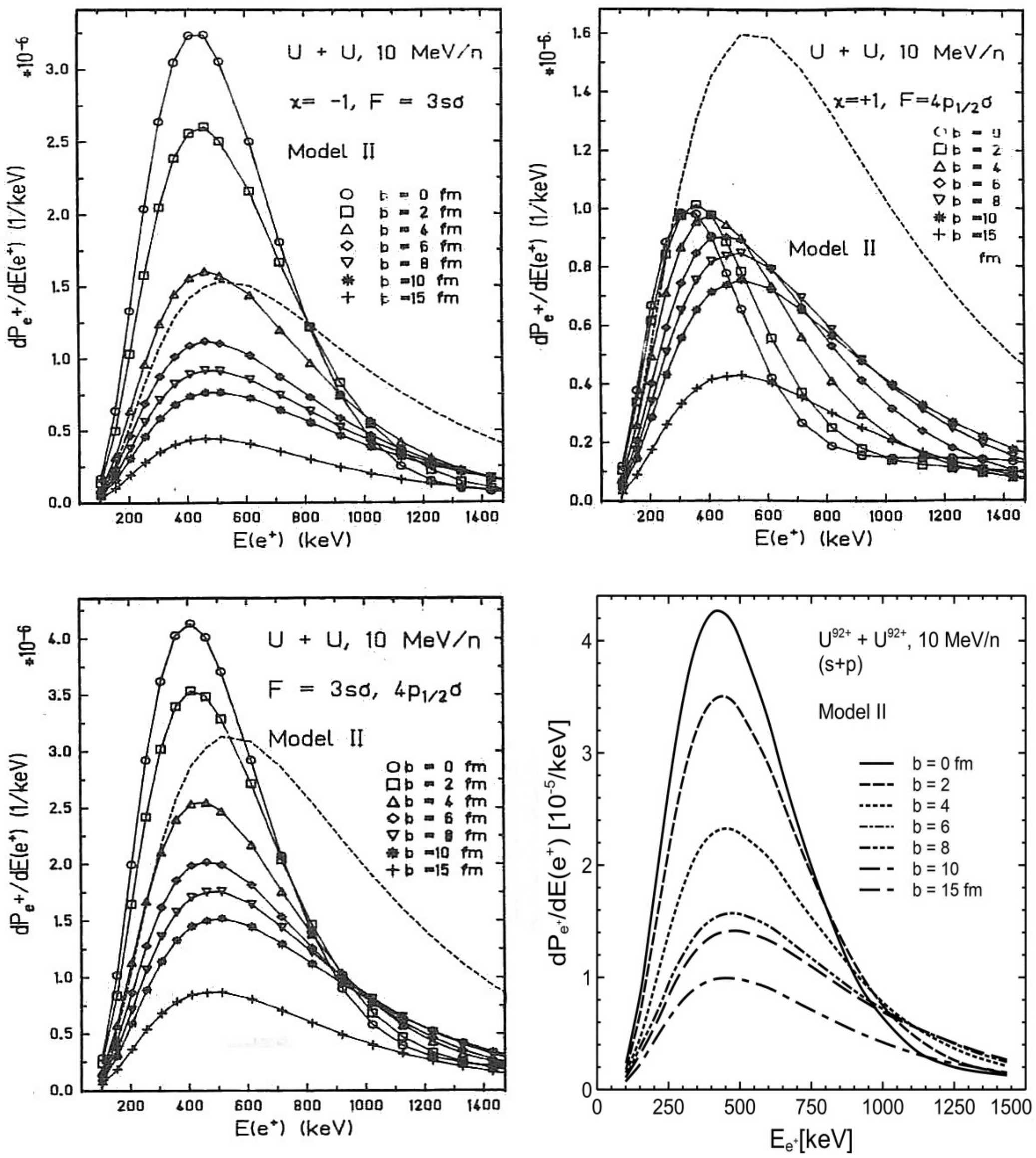

Figure 39: *Positron emission spectra in deep inelastic U + U collisions at $E_{Lab} = 10\,MeV/n$ (F=3). Trajectories according to [89] (Model II) with delay times $\Delta T$ up to 2.3 $\cdot 10^{-21}s$ in central collisions were used [39]. This delay time issues a clear signal for spontaneous positrons in the $\kappa = -1$ spectrum, compared with results of elastic scattering (dashed line). For angular momentum quantum number $\kappa = +1$ destructive interferences lead to reduced emission rates. The sum spectrum (bottom left) shows a significant narrower emission maximum at small impact parameters ($b \leq 10fm$) plus a steeper decrease for positron energies in the range $E_{e^+} = 400 - 800$ keV. Bottom right: Extrapolation of the total spectrum in collisions of bare nuclei (F=0), using scale factors for elastic collisions.*

fraction of spontaneous positrons. The ratio

$$\frac{\int\limits_{b=0}^{b_f} dP_{e^+}^{s+p}\, dE_{e^+} b\, db}{\int\limits_{b=0}^{b_f} dP_{e^+}^{p}\, dE_{e^+} b\, db}$$

is considered for sufficiently small impact parameters. The differential emission probability $dP_{e^+}^{s+p}\, dE_{e^+}$ contains also spontaneous positrons, whereas the same quantity neglecting the vacuum decay is approximated by $2\ dP_{e^+}^{p}\ dE_{e^+}$. For the system U+U shown in Fig. 39 this ratio increases from 1.3 to 2.3 (F=3) for kinetic positron energies in the range $E_{e^+} = 200 - 800$ keV when using $b_f = 4$ fm and offers an alternative tool to detect spontaneous positrons. Time delays in inelastic collisions assumed in the calculations above of about $3 \cdot 10^{-21}$s are quite probable as calculations of [90–92] show, considering even values of $5 \cdot 10^{-21}$s as possible. Conducting this type of experiments with bare nuclei would also significantly support the detectability of these effects.
All possibilities to detect spontaneous positron emission, i.e. in elastic collisions of bare nuclei, in the scenario of nuclear sticking and in deep inelastic collisions, should be considered in future experiments. An intensive collaboration between experimental and theoretical physicists, resp. all those involved in the preparation of upcoming experiments, with exchange of pertinent ideas, information and results are a prerequisite to make progress in this matter. This plus improved accelerator and detector technologies along with adequate data evaluation tools might help to detect the decay of the neutral vacuum finally in supercritical fields of colliding heavy ions.

## 7. Acknowledgments

I am grateful to Christophor Kozhuharov, GSI/Darmstadt and Johannes Kirsch, FIAS/Frankfurt for fruitful discussions, critical reading of the manuscript and valuable proposals concerning structure, contents and references of this paper.
I also thank the members of the Silverfrost Forum - especially mecej4 - for repeated and successful support in modifying legacy Fortran code to run on current compilers.

# References

[1] I.A. Maltsev, V.M. Shabaev, R.V. Popov, Y.S. Kozhedub, G. Plunien, X. Ma, Th. Stöhlker und D.A. Tumakov, Phys. Rev. Lett. **123**, 113401 (2019)

[2] R.V. Popov, V.M. Shabaev, D.A. Telnov, I.I. Tupitsyn, I.A. Maltsev, Y.S. Kozhedub, A.I. Bondarev, N.V. Kozin, X. Ma, G. Plunien, Th. Stöhlker, D.A. Tumakov und V.A. Zaytsev, Phys. Rev. **D102**, 076005 (2020)

[3] W. Pieper und W. Greiner, Z. Phys. **218**, 126 (1969)

[4] S.S. Gershtein und Ya.B. Zel'dovich, Zh. Eksp. Teor. Fiz. **57**, 654 (1969) und Sov. Phys. JETP **30**, 358 (1970)

[5] P.G. Reinhardt, W. Greiner und H. Arenhövel, Nucl. Phys. **A166**, 173 (1971)

[6] V.S. Popov, Yad. Fiz. **12**, 429 (1970) und Sov. J. Nucl. Phys. **12**, 235 (1971)

[7] V.S. Popov, Sov. Phys. JETP **32**, 526 (1971)

[8] V.S. Popov, Sov. J. Nucl. Phys. **15**, 595 (1972)

[9] Ya.B. Zel'dovich und V.S. Popov, Usp. Fiz. Nauk **105**, 403 (1971) und Sov. Phys. Usp. **14**, 673 (1972)

[10] L.P. Fulcher und A. Klein, Phys. Rev. **D8**, 2455 (1973)

[11] J. Rafelski, B. Müller und W. Greiner, Nucl. Phys. **B68**, 585 (1974)

[12] G. Soff, J. Reinhardt, B. Müller und W. Greiner, Phys. Rev. Lett. **38**, 592 (1977)

[13] B. Müller, H. Peitz, J. Rafelski und W. Greiner, Phys. Rev. Lett. **28**, 1235 (1972)

[14] B. Müller, J. Rafelski und W. Greiner, Z. Phys. **257**, 62 (1972)

[15] B. Müller, J. Rafelski und W. Greiner, Z. Phys. **257**, 183 (1972)

[16] J. Rafelski, B. Müller und W. Greiner, Z. Phys. **A285**, 49 (1978)

[17] J. Reinhardt, B. Müller und W. Greiner, Phys. Rev. **A24**, 103 (1981)

[18] E. Berdermann, F. Bosch, M. Clemente, F. Güttner, P. Kienle, W. Koenig, C. Kozhuharov, B. Martin, B. Povh, H. Tsertos, W. Wagner und Th. Walcher, GSI Scientific Report 1980, **GSI-81-2**, 128 (1981) und Verh. Dt. Phys. Ges. (VI) **16**, 606 (1981)

[19] H. Bokemeyer, H. Folger, H. Grein, S. Ito, D. Schwalm, P. Vincent, K. Bethge, A. Gruppe, R. Schulé, M. Waldschmidt, J.S. Greenberg, J. Schweppe und N. Trautmann, GSI Scientific Report 1980, **GSI-81-2**, 127 (1981) und Verh. Dt. Phys. Ges. (VI) **16**, 795 (1981)

[20] H. Backe, P. Senger, W. Bonin, E. Kankeleit, M. Krämer, R. Krieg, V. Metag, N. Trautmann, and J.B. Wilhelmy, Phys. Rev. Lett. **50**, 1838 (1983)

[21] J. Schweppe, A. Gruppe, K. Bethge, H. Bokemeyer, T. Cowan, H. Folger, J.S. Greenberg, H. Grein, S. Ito, R. Schulé, D. Schwalm, K.E. Stiebing, N. Trautmann, P. Vincent und M. Waldschmitt, Phys. Rev. Lett. **51**, 2261 (1983)

[22] M. Clemente, E. Berdermann, P. Kienle, H. Tsertos, W. Wagner, C. Kozhuharov, F. Bosch und W. Koenig, Phys. Lett. **137B**, 41 (1984)

[23] H. Bokemeyer, H. Folger, H. Grein, T. Cowan, J.S. Greenberg, J. Schweppe, A. Balanda, K. Bethge, A. Gruppe, K. Sakaguchi, K.E. Stiebing, D. Schwalm, P. Vincent, H. Backe, M. Begemann, M. Klüver und N. Trautmann, GSI Scientific Report 1984, **GSI-85-1**, 177 (1985)

[24] T. Cowan, H. Backe, M. Begemann, K. Bethge, H. Bokemeyer, H. Folger, J.S. Greenberg, H. Grein, A. Gruppe, Y. Kido, M. Klüver, D. Schwalm, J. Schweppe, K.E. Stiebing, N. Trautmann und P. Vincent, Phys. Rev. Lett. **54**, 1761 (1985)

[25] P. Kienle, Prog. Part. Nucl. Phys. Vol. **15**, 77 (1985)

[26] H. Tsertos, E. Berdermann, F. Bosch, M. Clemente, P. Kienle, W. Koenig, C. Kozhuharov und W. Wagner, Phys. Lett. **162B**, 273 (1985)

[27] J. Schweppe, in: "Electronic and Atomic Collisions", Hrsg.: D.C. Lotents, W.E. Meyerhof und J.R. Peterson, North Holland, Amsterdam, 405 (1986)

[28] E. Berdermann, F. Bosch, S. Huchler, P. Kienle, W. Koenig, C. Kozhuharov, H. Tsertos und W. Wagner, "Positron Lines investigated with the Orange-$\beta$-Spectrometer", GSI Nachrichten, **GSI-01-87**, 7 (1987)

[29] H. Tsertos, F. Bosch, P. Kienle, W. Koenig, C. Kozhuharov, E. Berdermann, S. Huchler und W. Wagner, Z. Phys. **A326**, 235 (1987)

[30] I. Ahmad, S.M. Austin, B.B. Back, R.R. Betts, F.P. Calaprice, K.C. Chan, A.A. Chishti, C. Conner, R.W. Dunford, J.D. Fox, S.J. Freedman, M. Freer, J.S. Greenberg, S. B. Gazes, A.L. Hallin, T. Happ, D. Henderson, N.I. Kaloskamis, E. Kashy, W. Kutschera, J. Last, C.J. Lister, M. Liu, M.R. Maier, D.J. Mercer, D. Mikolas, P.A.A. Perera, M.D. Rhein, D.E. Roa, J.P. Schiffer, T.A. Trainor, P. Wilt, J.S. Winfield, M. Wolanski, F.L.H. Wolfs, A.H. Wuosmaa, G. Xu, A. Young, and J.E. Yurkon, Phys. Rev. **C55**, R2755 (1997)

[31] I. Ahmad, Sam.M. Austin, B.B. Back, R.R. Betts, F.P. Calaprice, K.C. Chan, A. Chishti, C.M. Conner, R.W. Dunford, J.D. Fox, S.J. Freedman, M. Freer, S.B. Gazes, A.L. Hallin, Th. Happ, D. Henderson, N.I. Kaloskamis, E. Kashy, W. Kutschera, J. Last, J.C. Lister, M. Liu, M.R. Maier, D.M. Mecer, D. Mikolas, P.A.A. Perera, M.D. Rhein, D.E. Roa, J.P. Schiffer, T.A. Trainor, P. Wilt, J.S. Winfield, M. Wolanski, F.L.H. Wolfs, A.H. Wuosmaa, A.R. Young und J.E. Yurkon, Phys. Rev. **C60**, 064601 (1999)

[32] S. Heinz, E. Berdermann, F. Heine, O. Joeres, P. Kienle, I. Koenig, W. Koenig, C. Kozhuharov, U. Leinberger, M. Rhein, A. Schröter und H. Tsertos, Eur. Phys. J. **A1**, 27 (1998)

[33] S. Heinz, E. Berdermann, F. Heine, O. Joeres, P. Kienle, I. Koenig, W. Koenig, C. Kozhuharov, U. Leinberger, M. Rhein, A. Schröter und H. Tsertos, Eur. Phys. J. **A9**, 55 (2000)

[34] B. Müller und W. Greiner, Z. Naturforsch. **31a**, 1 (1976)

[35] J. Reinhardt und W. Greiner, Physik in unserer Zeit **6**, 171 (1976)

[36] W. Greiner, B. Müller und J. Rafelski, "Quantum Electrodynamics of Strong Fields", Springer Verlag, Berlin (1985)

[37] J. Reinhardt, U. Müller, B. Müller und W. Greiner, Z. Phys. **A303**, 173 (1981)

[38] U. Müller, G. Soff, T. de Reus, J. Reinhardt, B. Müller und W. Greiner, Z. Phys. **A313**, 263 (1983)

[39] T. de Reus, U. Müller-Nehler, G. Soff, J. Reinhardt, S. Graf, B. Müller, W. Greiner, Phys. Rev. **C40**, 752, (1989)

[40] B. Müller und W. Greiner, Acta Phys. Austriaca, Suppl. **XVIII**, 153 (1977)

[41] G. Soff, U. Müller, T. de Reus, J. Reinhardt, B. Müller und W. Greiner, Nucl. Instr. Meth. **B10/11**, 214 (1985)

[42] T. de Reus, J. Reinhardt, B. Müller, W. Greiner, U. Müller und G. Soff, Z. Phys. **A321**, 589 (1985)

[43] T. de Reus, J. Reinhardt, B. Müller, W. Greiner, U. Müller und G. Soff, Prog. Part. Nucl. Phys. **15**, 57 (1985)

[44] T. de Reus, J. Reinhardt, B. Müller, W. Greiner und G. Soff, Phys. Lett. **169B**, 139 (1986), Erratum: Phys. Lett. **173B**, 491 (1986)

[45] JM. Rhein, R. Barth, E. Ditzel, H. Feldmeier, E. Kankeleit, V. Lips, C. Müntz, W. Nörenberg, H. Oeschler, A. Piechaczek, W. Polai, and I. Schall Phys. Rev. Lett. **69**, 1320 (1992)

[46] M. Rhein, R. Barth, E. Ditzel, H. Feldmeier, E. Kankeleit, V. Lips, C. Müntz, W. Nörenberg, H. Oeschler, A. Piechaczek, W. Polai, and I. Schall Phys. Rev. **C49**, 250 (1994)

[47] T. de Reus, "Innerschalenanregung in Schwerionenstößen bis zu mittleren Einschußenergien", GSI Report 87-8, ISSN 0171-4546 (1987)

[48] H. Backe, L. Handschug, F. Hessberger, E. Kankeleit, L. Richter, F. Weik, R. Willwater, H. Bokemeyer, P. Vincent, Y. Nakayama, J.S. Greenberg, Phys. Rev. Lett. **40**, 1443 (1978)

[49] C. Kozhuharov, P. Kienle, E. Berdermann, H. Bokemeyer, J.S. Greenberg, Y. Nakayama, P. Vincent, H. Backe, L. Handschug und E. Kankeleit, Phys. Rev. Lett. **42**, 376 (1979)

[50] B. Müller, "Die Zweizentren- Dirac- Gleichung", Dissertation, Universität Frankfurt (1973)

[51] B. Müller, Ann. Rev. Nucl. Sci. **26**, 351 (1976)

[52] W. Betz, B. Müller, G. Soff und W. Greiner, Phys. Rev. Lett. **37**, 1046 (1976)

[53] W. Betz, "Elektronen in superschweren Quasimolekülen", Dissertation, Universität Frankfurt (1980)

[54] G. Soff, J. Reinhardt, W. Betz und J. Rafelski, Phys. Scr. **17**, 417 (1978)

[55] T. Tomoda und H.A. Weidenmüller, Phys. Rev. **A26**, 162 (1982)

[56] G. Soff, W. Greiner, W. Betz und B. Müller, Phys. Rev. **A20**, 169 (1979)

[57] S.R. McConnell, A.N. Artemyev, M. Mai and A. Surzhykov, Phys. Rev. **A86**, 052705 (2012)

[58] W.L. Wang und C.M. Shakin, Phys. Lett. **32B**, 421 (1970)

[59] E. Ackad and M. Horbatsch, Phys. Rev. **A78**, 062711 (2008)

[60] J. Reinhardt, "Dynamische Theorie der Erzeugung von Positronen in Stößen sehr schwerer Ionen", Dissertation, Universität Frankfurt (1979)

[61] J.F. Reading, Phys. Rev. **A8**, 3262 (1973)

[62] J.H. McGuire und L. Weaver, Phys. Rev. **A16**, 41 (1977)

[63] G. Soff, J. Reinhardt, B. Müller und W. Greiner, Z. Phys. **A294**, 137 (1980)

[64] E. Kankeleit, Nukleonika, **25**, 253 (1980)

[65] R. Bass, Nucl. Phys. **A231**, 45 (1974)

[66] R. Bass, "Nuclear Reactions with Heavy Ions", Springer (1980)

[67] J.R. Birkelund and J.R. Huizenga, Phys. Rev. Ann. Rev. Nucl. Part. **33**, 265 (1983)

[68] M. Altamira-Arrizabalaga, 'Experimental Methods to Measure Nuclear Radii', Física Nuclear Etsii, 4-17 (1994)

[69] Z. Patyk, A. Baran, J. F. Berger, J. Dechargé, J. Dobaczewski, P. Ring, and A. Sobiczewski, Phys. Rev. C **59**, 704 (1999)

[70] W. M. Seif and H. Mansour, Int. J. Mod. Phys. E Vol. **24**, No. 11, 1550083 (2015)

[71] O.A.P. Tavares, E.L. Medeiros, Int. J. Mod. Phys. E Vol. **28**, No. 4, 1950021 (2019)

[72] H. Hellmann, "Einführung in die Quantenchemie", Franz Deutige Verlag, (1937)

[73] E.R. Davison, in: "Physical Chemistry, An Advanced Treatise", Vol. 3, Hrsg.: D. Henderson, Academic Press, New York, 116 (1969)

[74] U. Müller-Nehler, G. Soff, Phys. Rep. **246**, 101 (1994)

[75] U. Müller, T. de Reus, J. Reinhardt, B. Müller, W. Greiner and G. Soff, Phys. Rev. **A37**, 1449 (1988)

[76] G. Soff, B. Müller und W. Greiner, Phys. Rev. **A299**, 189 (1981)

[77] T. de Reus, W. Greiner, J. Kirsch, B. Müller, J. Reinhardt und G. Soff, in: "Inner-Shell and X- Ray Physics of Atoms and Solids", Hrsg.: D.J. Fabian, H. Kleinpoppen und L.M. Watson, Plenum, New York, 221 (1981)

[78] T. de Reus, J. Reinhardt, B. Müller, W. Greiner G. Soff und U. Müller, J. Phys. **B17**, 615 (1984)

[79] U. Müller, T. de Reus, J. Reinhardt, B. Müller und W. Greiner, Z. Phys. **323**, 261 (1986)

[80] T. de Reus, J. Reinhardt, U. Müller, B. Müller, W. Greiner und G. Soff, Physica **C144**, 237, (1987)

[81] D. Liesen, P. Armbruster, H.H. Behncke, F. Bosch, S. Hagmann, P.H. Mokler, H. Schmidt- Böcking und R. Schuch, in: "Electronic and Atomic Collisions", Hrsg.: N. Oda und K. Takayanagi, North Holland, Amsterdam, 337 (1980)

[82] D. Liesen, Comments At. Mol. Phys. **12**, 39 (1982)

[83] R. Anholt, W.E. Meyerhof, J.D. Molitoris und Xu Jun-Shan, Z. Phys. **A308**, 189 (1982)

[84] S. Ito, P. Armbruster, H. Bokemeyer, F.Bosch, H. Emling, H. Folger, E. Grosse, R. Kulessa, D. Liesen, D. Maor, D. Schwalm, J.S. Greenberg, R. Schulé und N. Trautmann, GSI Scientific Report 1981, **GSI-82-1**, 141 (1982)

[85] M.A. Herath Banda, A.V. Ramayya, C.F. Maguire, F. Güttner, W. Koenig, B. Martin, B. Povh, H. Skapa und J. Soltani, Phys. Rev. **A29**, 2429 (1984)

[86] S. Graf, J. Reinhardt, U. Müller, B. Müller, W. Greiner, T. de Reus and G. Soff, Z. Phys. **A329**, 365 (1988)

[87] V. Zagrebaev and W Greiner, "Giant Nuclear Systems of Molecuar Type", Lecture Notes in Physics Vol. **818**, 267 (2010)

[88] V. Zagrebaev and W Greiner, "Mass transfer and time delay in low energy heavy ion collisions", Symposium on supercritical fields, FIAS Frankfurt, November 5, (2013)

[89] R. Schmidt, V.D. Toneev und G. Wolschin, Nucl. Phys. **A311**, 247 (1978)

[90] C. Golabek, S. Heinz, W. Mittig, F. Rejmund, A. C. C. Villari, S. Bhattacharyva, D. Boilley, G. De France, A. Drouart, L. Gaudefroy, L. Giot, V. Maslov, M. Morjean, G. Mukherjee, Yu. Penionzkevich, P. Roussel-Chomaz and C. Stodel, Eur. Phys. J. **43A**, 251 (2010)

[91] V. I. Zagrebaev, A. V. Karpov, and Walter Greiner, Phys. Rev. **C81**, 044608 (2010)

[92] S. Heinz, "Deep Inelastic Reactions and Time Delay in Collisions of U+U", Symposium on supercritical fields, FIAS Frankfurt, November 5, (2013)